\documentclass[prx,aps,groupedaddress,twocolumn,superscriptaddress,longbibliography]{revtex4-1}

\ifx\pdfsuppresswarningpagegroup\undefined\else\pdfsuppresswarningpagegroup=1\fi\relax
\PassOptionsToPackage{unicode, pdfprintscaling=None, colorlinks}{hyperref}

\usepackage{orcidlink}
\usepackage{multirow}
\usepackage{siunitx}
\usepackage{colortbl}
\usepackage{dcolumn}% Align table columns on decimal point

\usepackage{tikz}
\usetikzlibrary{decorations.markings, arrows, decorations.pathmorphing}

\usepackage{graphicx}
\usepackage{latexsym}
\usepackage{amsmath,amssymb}
\usepackage{amsfonts}
\usepackage{bm}
\usepackage{bbm}
\usepackage{microtype}
\usepackage{xspace}
\usepackage{placeins}
\usepackage{float}
\usepackage{hyperref}
\usepackage[capitalise]{cleveref}
\usepackage{braket}
\usepackage{setspace}
\usepackage{xcolor}
\usepackage{makecell}
\usepackage{ragged2e}
\usepackage{mathtools}

\usepackage{soul}

\usepackage[export]{adjustbox}

\newlength{\FigureWidth}
\usepackage{xcolor}
\usepackage{makecell}

\usepackage{relsize}

\usepackage[acronyms, nohypertypes={acronym}, nopostdot, style=super, nonumberlist, toc]{glossaries}
\setacronymstyle{long-sc-short}
\glsenableentrycount
\makeglossaries  

\definecolor{dkgreen}{rgb}{0.2,0.7,0.4}
\definecolor{dkblue}{rgb}{0.2,0.2,0.7}
\definecolor{dkred}{rgb}{0.8,0,0}
\definecolor{dkgreen}{rgb}{0.2,0.8,0.4}

\begin{document}

%\preprint{APS/123-QED}
\title{Spontaneous symmetry breaking in a $SO(3)$ non-Abelian lattice gauge theory in $2+1$D with quantum algorithms}

\author{Sandip Maiti\,\orcidlink{0000-0002-5248-5316}}
\email{sandip.maiti@saha.ac.in}
\affiliation{Saha Institute of Nuclear Physics, HBNI, 1/AF Bidhannagar, Kolkata 700064, India}
\affiliation{Homi Bhabha National Institute, Training School Complex, Anushaktinagar, Mumbai 400094, India}
\author{Debasish Banerjee\,\orcidlink{0000-0003-0244-4337}}
\email{debasish.banerjee@saha.ac.in}
\affiliation{Saha Institute of Nuclear Physics, HBNI, 1/AF Bidhannagar, Kolkata 700064, India}
\affiliation{Homi Bhabha National Institute, Training School Complex, Anushaktinagar, Mumbai 400094, India}
\affiliation{School of Physics and Astronomy, University of Southampton,  University Road, Southampton, UK}
\author{Bipasha Chakraborty\,\orcidlink{0000-0001-6667-329X}}
\email{B.Chakraborty@soton.ac.uk}
\affiliation{School of Physics and Astronomy, University of Southampton,  University Road, Southampton, UK}
\author{Emilie Huffman\,\orcidlink{0000-0002-4417-338X}}
\email{ehuffman@perimeterinstitute.ca}
\affiliation{Perimeter Institute for Theoretical Physics, Waterloo, ON N2L 2Y5, Canada}

\date{\today}% It is always \today, today,

\begin{abstract}
The simulation of various properties of quantum field theories is rapidly becoming 
a testing ground for demonstrating the prowess of quantum algorithms. Some examples
include the preparation of ground states, as well as the investigation of
various simple wave packets relevant for scattering phenomena. 
In this paper, we study the ability of quantum algorithms to prepare ground states in a
matter-free non-Abelian $SO(3)$ lattice gauge theory in $2+1$D in a phase where the 
global charge conjugation symmetry is spontaneously broken. This is challenging for 
two reasons: the necessity of dealing with a large Hilbert space for gauge theories 
compared to that of quantum spin models, and the closing of the gap between the two 
ground states, which becomes exponentially small as a function of the volume. To deal 
with the large Hilbert space of gauge fields, we demonstrate how the exact imposition 
of the non-Abelian Gauss Law in the rishon representation of the quantum link operator 
significantly reduces the degrees of freedom. Further, to resolve the gap, we introduce 
symmetry-guided ans\"{a}tze in the Gauss-Law-resolved basis 
for trial states as the starting point for the quantum algorithms to prepare the two 
lowest energy states. In addition to simulation results for a range of two-dimensional 
system sizes, we also provide experimental results from the trapped-ion-based 
quantum hardware, IonQ, when working on systems with four quantum links. The 
experimental/simulation results derived from our theoretical developments indicate the
role of metrics--such as the energy and the infidelity--to assess the obtained results.
\end{abstract}

\maketitle

\section{Introduction} \label{sec:intro}
 The success of quantum field theory (QFT) as a paradigm to explain the properties of physical 
systems in Nature has proceeded hand in hand with the development of computational techniques 
in this framework. One of the key advances was the development of renormalized perturbation 
expansion in quantum field theory \cite{Dyson1949}, allowing the computation of quantities which 
could be matched with experiments, and culminating in the resounding success of quantum 
electrodynamics. However, the theory of strong interactions has proven to be a challenge for 
perturbation theory, since the presence of strong interactions between the quarks and gluons 
result in non-perturbative phenomena such as confinement. This necessitated the introduction 
of lattice gauge theory \cite{Wilson1974, Weisz2012}, and the Markov Chain Monte Carlo methods 
for non-perturbative evaluation of physical quantities in QFTs \cite{Creutz:1988mb,Montvay:2007yg}. 

  While there has been considerable improvement in various Monte Carlo techniques, there are 
domains where the role of Monte Carlo as a superior method from existing methods has not been
established. Investigation of matter at finite densities is one prime example, especially in
the case of doped Hubbard model (relevant for high-temperature superconductivity), or quantum
chromodynamics (QCD) at finite baryon density (relevant for equation of state of neutron stars).
Similarly, studies of the real-time dynamics of QFTs or quantum many body systems can hardly
be addressed with Monte Carlo methods. Powerful variational methods \cite{meurice2022tensor,
Banuls2023,Cataldi2024} involving matrix product and tensor network states can address both the above 
problems in lower dimensions, but it is not clear whether these problems can be addressed fully 
in thermodynamically large systems. 

 In this ecosystem, the technological realization of quantum computation, which was theoretically
inspired by \cite{Benioff1980,Feynman1982}, has ushered in a new array of opportunities for the 
development of computational paradigms. Hamiltonians of relevant physical systems
can be designed by controlling various quantum degrees of freedom (such as ions, atoms, or molecules)
in various hardware (ion-traps, optical lattices, superconducting qubits, Rydberg systems) and 
tuning interactions between them \cite{Angelakis2016,Kjaergaard2019,Cheuk2015,Hadzibabic2002, 
Altman2019, Monroe2019, Semeghini2021, Bluvstein2021,Fossfeig2024}. 
In principle, quantum computation may be used for both of the aforementioned difficult 
cases of simulations of matter at finite densities and of real-time dynamics, although 
in reality nontrivial work is necessary to address any physically relevant
system. Currently, efforts are underway to design and test quantum algorithms in toy 
quantum field theories to demonstrate their capabilities of both reproducing and going 
beyond results obtained through well-known classical methods 
\cite{McClean2015,Zhang:2017kde,Kandala2018,Bauer2019,Ciavarella:2020vqm,Hall:2021rbv,
Perez-Salinas:2020nem, Huffman:2021gsi,Kan:2021nyu,Cohen:2021imf,Homeier:2022mkg, 
Fontana:2022dil,Osborne:2022jxq,Bauer2022,Farrell:2024fit}.
Simultaneously, there are also efforts in the development of novel theoretical methods and 
models, which can be seamlessly adapted to the framework of quantum technologies
\cite{Wiese:2013uua,Zohar:2016iic,Bravyi:2000vfj, Paulson:2020zjd,
Davoudi:2020yln,Bhattacharya:2020gpm,Alexandru:2019nsa,Ciavarella:2021nmj,Meurice:2021pvj,
Zache:2021ggw,Liu:2021tef,Gustafson:2021qbt,Kadam:2022ipf,Banerjee:2022jja,Alexandru:2023qzd,
Kadam:2024zkj,Fontana2024}.

As advancements in the controllability of noisy intermediate-scale quantum (NISQ) 
\cite{Preskill:2018jim} computers have emerged, there is a growing focus on variational quantum 
simulation (VQS). The main objective of VQS involves using variational algorithms, such 
as variational quantum eigensolvers (VQEs) \cite{Tilly:2021jem}, to estimate the ground-state 
spectrum of a quantum Hamiltonian. At the core of VQEs lies the development of parametrized 
quantum circuits. \cite{Kandala2017,Sim:2019yyv}
As an example, a specific VQE variant, utilizing the hardware-efficient 
ansatz consisting of parametrized single-qubit rotation gate layers and non-parametrized 
entangling gate blocks, has been employed to address the ground-state energy of a quantum 
many-body system. An extension of VQE, known as variational quantum deflation (VQD) 
\cite{Higgott:2018doo}, allows for the computation of excited state spectra by incorporating 
overlap terms into the optimization function. This procedure comes at almost no extra cost. 

  In contrast to the VQE utilizing the hardware-efficient ansatz, another well-known type of
variational algorithm is the quantum approximate optimization algorithm (QAOA) 
\cite{Farhi:2014ych, Zhou:2018fwi}, where the circuit ansatz is referred to as the Hamiltonian 
variational ansatz, and the design of the quantum circuit is intricately linked to the problem 
Hamiltonian. It was initially designed for solving combinatorial minimization 
problems like the Max-Cut problem \cite{Crooks:2018vud}.  As is sometimes expected from large
multidimensional variational problems, one can run into barren plateaus. While barren plateaus 
are present in the optimization landscape of both VQEs using the hardware-efficient ansatz 
and the QAOA, the QAOA has been developed in part to reduce the probability of encountering 
such plateaus, and in both cases sometimes minor adaptions in a particular ansatz may 
eliminate them \cite{Larocca:2021ksf}.

 In addition to the general simulation issues described above, there are also symmetry-based 
 issues that may arise in studying particular phases of physical systems. Symmetries play 
 a crucial role in modern physics in the context of classifying various 
phases of matter. The Ginzburg-Landau paradigm \cite{Hohenberg2015} of classifying phases 
and phase transitions has largely governed numerous theoretical and experimental explorations 
both in classical and quantum physics. Consequently, the idea that symmetries can be 
spontaneously broken, especially at low temperatures or at finite densities, has facilitated
the identification of phases present in systems of physical interest. In fact, the spontaneous 
breaking of chiral symmetry in quantum chromodynamics (QCD) is responsible for the mass
of visible matter (such as protons and neutrons) around us, while the spontaneous breaking 
of a global $U(1)$ symmetry is responsible for superconductivity in a theory of 
(weakly-interacting) fermions. Given the importance of spontaneous symmetry breaking (SSB) to 
physically-relevant systems, it is natural to develop quantum algorithms suited for the 
preparation of these symmetry-broken ground states.  In a given system, the phenomenon of 
SSB indicates the presence of multiple ground states $\ket{\psi_i}$ (where $i$ labels the 
different symmetry broken ground state), which transform into each other by the 
action of a global symmetry operator $U$. In the scenario where SSB does not occur, then
the ground state is unique, and has even quantum numbers corresponding to all symmetries.

 However, the relevant theoretical setup for (classical or quantum) numerical studies is 
a finite box with a lattice structure, such that both ultraviolet and infrared fluctuations are 
regulated. In such a finite volume setup, the ground state is not degenerate, but gapped. Moreover,
the gap decreases exponentially with increase in the volume. Therefore, it is relevant
to ask how would a variational algorithm, especially realized with quantum hardware, fare 
when asked to prepare the ground state(s) of such a phase. Note that no problems are expected 
when such a study is undertaken for the ground state of a gapped theory: the separation 
between the ground state energy $E_0$ and the first excited state $E_1$ is typically of 
the energy scale of theory: $\Delta E \sim J$, where $J$ is the energy scale associated 
with the Hamiltonian. In contrast, for SSB one has $\Delta E \sim \exp(- c V)$, thus 
challenging the gap extraction using variational methods (where $c$ is a constant number, 
and $V$ is the physical volume). 

 Our primary goal is to address these challenges and demonstrate SSB within a pure gauge 
theory using variational quantum algorithms. The impracticality of directly implementing 
the Wilsonian version of the theory (commonly used in classical computation) on a quantum 
computer arises from the infinite-dimensional Hilbert space associated with each gauge link. 
One direction to proceed is to truncate the local infinite-dimensional Hilbert space,
leading one to deal with breaking of gauge invariance appropriately. A viable alternative is
to explore a different framework within gauge theory referred to as quantum link models (QLM)
\cite{Chandrasekharan:1996ih}, where each gauge link is replaced by a finite-dimensional 
Hilbert space while preserving the local gauge invariance, and rendering it suitable for 
quantum computer implementation. Thus, it is possible to ensure that gauge symmetry is 
preserved throughout the quantum simulation. Abelian formulations have already been 
extensively explored, and we proceed to non-Abelian gauge theories while treading the 
road to quantum chromodynamics in the long-term. We concentrate on a theory characterized by 
local symmetries of $SO(3)$ and investigate its representations across various lattice 
geometries, including bubble, triangular, and square lattice structures. The model has 
been previously investigated \cite{Rico:2018pas}, particularly regarding spontaneous symmetry 
breaking (SSB) phenomena through the use of exact diagonalization (ED). However, as the 
computational demands in ED grow exponentially with system volume, exploring significantly 
larger systems becomes impractical. Additionally, the Monte Carlo method becomes difficult, 
mainly due to the sign problem within the chosen basis of the Hilbert space. Consequently, 
it would be beneficial to employ quantum computing to study the model and demonstrate SSB 
phenomena in larger systems. In illustrating SSB, we employ a range of quantum algorithms 
to calculate both the ground state and a subset of excited state spectra.

In this article, we thus propose and benchmark a class of quantum algorithms to extract 
the low-energy spectrum of a $SO(3)$ non-Abelian lattice gauge theory without matter. 
Typically, gauge theories have many more degrees of freedom than a corresponding spin 
or fermionic model, and are thus more resource expensive to simulate on a quantum platform. 
Moreover, only gauge-invariant degrees of freedom contribute to the dynamics, and thus 
mapping all degrees of freedom of the original model onto the quantum computer is not 
very useful. We show that in the quantum link formulation it is possible to impose Gauss' 
Law analytically, and reformulate the model entirely in terms of gauge-invariant degrees 
of freedom. The other question which we address in this paper is the efficacy of the 
various variational quantum algorithms to capture the ground state and the mass gap of 
the theory, which in turn depends on the global symmetries of the Hamiltonian, and whether 
they are broken or not. 

The rest of the paper is arranged as follows: in Sec.~\ref{sec:model}, we describe the model and
its local gauge invariance, and formulate it in a gauge-invariant way; in Sec.~\ref{sec:methods}, 
we provide a comprehensive description of the quantum algorithms used in the investigation of 
symmetry breaking physics. Sec.~\ref{sec:results1} is dedicated to the discussion of our results: 
first we discuss the VQE methods on real hardware (before imposing gauge invariance)
and display our results; then we discuss our attempts to study SSB phenomena using quantum algorithms 
on classical hardware, up to 12 qubits. In Sec.~\ref{sec:results2}, we compare our results to that 
obtained for the transverse field Ising model (TFIM) in the SSB phase to demonstrate the difficulty 
of simulating a full-fledged gauge theory as opposed from a spin model. We conclude our discussion 
in Sec.~\ref{sec:conclusion}, summarizing the main results and providing an outlook for the research 
direction which this work inspires.

\section{Model, Symmetries, and Gauge Invariant States} \label{sec:model}
\begin{figure}[h] 
\begin{center}
\includegraphics[width=1.2\linewidth]{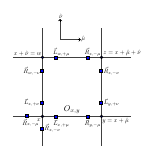}
\caption{The plaquette in a two-dimensional spatial lattice. The gauge field operators are denoted as $O^{ab}_{xy}$ where $a$, $b$ are the color indices and $x$,$y$ are the ends of the links on which the gauge field is defined. The non-Abelian electric fields are $L^a_{x,+\mu}$ and $R^a_{y,-\mu}$, and are defined on the left and right side of the link joining sites $x$ and $y$ respectively.}
\label{fig:latt}
\end{center}
\end{figure}

 Here we discuss the model with a local $SO(3)$ gauge invariance, operators corresponding
to microscopic gauge fields, and the appropriate gauge symmetries. Readers familiar with the
structure of quantum link models can skip this section, an almost equivalent description is provided
in \cite{Rico:2018pas}. The basic degrees of freedom are the $SO(3)$ matrix-valued gauge fields: 
$O^{ab}_{xy}$ (where $a,b\in 1,2,3$). Each element of the gauge field is a Hermitian operator 
$O^{ab \dagger} = O^{ab}$ which lives on the link joining the lattice sites $x$ and $y=x+\mu$. 
We denote the unit vectors in the positive direction as $+\mu, +\nu, \cdots$, while the unit 
vectors in the negative direction are $-\mu, -\nu, \cdots$. This notation is useful since we
will define operators which live on the left and right (top and bottom) positions of a link.
The canonically conjugate momenta are the matrix-valued left and right electric fields, denoted 
as $L^a_{x,+\mu}$ and $R^a_{x+\mu,-\mu}$ respectively (also Hermitian), and shown in 
\cref{fig:latt}. The non-Abelian electric fields at different links always commute with each 
other. However, for a specified link, while $L^a$ and the $R^a$ commute with each other, 
$[L^a, R^b] = 0$, the others satisfy the following commutation relations:
\begin{equation}\label{eq1}
[L^a, L^b] = 2 i \varepsilon^{abc} L^c,~~~[R^a, R^b] = 2 i \varepsilon^{abc} R^c,
\end{equation}
where $\varepsilon^{abc}$ is the usual Levi-Civita symbol.

 Just like position and momentum operators, the electric and the gauge field operators on the same
link satisfy certain commutation relations among themselves (while those associated with different
links commute):
\begin{equation}\label{eq2}
    [L^a, O^{bd}] = 2 i \varepsilon^{abc} O^{cd};~~~ [R^a, O^{bd}] = -2 i O^{bc} \varepsilon^{acd},
\end{equation}
and similarly, due to their non-Abelian nature the different elements of the $O^{ab}$ satisfy the 
following commutation rules:
\begin{equation}\label{eq3}
    [O^{ab}, O^{cd}] = 2 i \delta^{ac} \varepsilon^{ebd} R^e 
    + 2 i \delta^{bd} \varepsilon^{eac} L^e.
\end{equation}

Using these operators, we can now construct the Hamiltonian operator. A generic Hamiltonian for a (lattice) gauge theory has terms containing the electric field energy and the magnetic field energy, 
${\cal H} = {\cal H}_E + {\cal H}_B$. In terms of the electric field operators, the first term is ${\cal H}_E = \frac{g^2}{2} \sum_{x, \mu} \left( L^a_{x,+\mu} L^a_{x,+\mu} 
+ R^a_{x+\mu,-\mu} R^a_{x+\mu, -\mu}\right)$. The magnetic term is the plaquette term, defined as a product of the four oriented links around the smallest square loop on the lattice, ${\cal H}_B = -\frac{1}{4g^2}\sum_{\Box} {\rm Tr} {\cal O}_{\Box}$,
where ${\cal O}^{ab}_{\Box} = O^{am}_{xy} O^{mn}_{yz} O^{np}_{zw} O^{pb}_{wx}$, and $x,y,z,w$ label the four corners of the plaquette $\Box$ starting from bottom left and moving anticlockwise. Since the operator is already Hermitian, the conjugate is unnecessary.

Hamiltonians with this structure are invariant under a larger class of local 
transformations, often called gauge symmetries. These transformations are generated by the local Gauss Law, which is the non-Abelian analogue of $\nabla \cdot E = 0$, 
\begin{equation}\label{eq4}
    G^a_x = \sum_{\mu} (L^a_{x,+\mu} + R^a_{x,-\mu}), ~~ [G^a, G^b] = 2 i \varepsilon^{abc} G^c,
\end{equation}
where the various components of the Gauss Law do not commute. Moreover, the electric field
operators which appear in the Gauss Law are schematically shown in \cref{fig:latt}.

Typically, if one is working in a computational basis diagonal in the electric field, it is 
non-trivial to form totally gauge invariant states. In this paper we will take a different route: 
for our chosen operators, we first construct a basis which directly projects to the $\vec{G} = 0$ 
sector, and then construct the Hamiltonian in this gauge invariant basis. Under a generic gauge 
transformation $V = \prod_x {\rm exp}(i \alpha^a_x G^a_x)$, quantum link operators transform as:
\begin{equation}\label{eq5}
    {O'}^{ab}_{xy} = (V^\dagger O_{xy} V)^{ab} = 
    \big[{\rm e}^{i\alpha^{m}_x {t^m}}\big]^{ac} O^{cd}_{x,y} \big[{\rm e}^{-i\alpha^{p}_y {t^p}}\big]^{db},
\end{equation}
and $t^{a}_{bc}=-\varepsilon^{abc}$ are the generators of the $SO(3)$ group. A sketch of
the operators on the lattice is shown in \cref{fig:latt}. The detailed transformation is 
provided in the \cref{sec:gtrafo} as a reference. 

\vspace{0.3cm}
{\bf Representations of field operators:}
 As the next step, we need to choose concrete representations for the operator structures 
discussed above. There is a simple method to construct such representations following 
\cite{Brower_1999,Banerjee:2012xg}. We first note that in order to represent $O^{ab}$, 
we need $N^2$ Hermitian operators, for each of $L^a, R^a$, we need $N$ hermitian operators, 
and thus a total of $N^2 + 2N$ operators. With $N=3$, this gives 15 hermitian operators,
and this can be represented by the 15 elements of the $so(6)$ algebra, linearly independent 
by construction. The $so(6)$ forms the embedding algebra for this model. 

 The simplest representation for the operators is to have a spin-$\frac{1}{2}$ bilinear 
operator to represent the gauge and the electric fluxes as follows:
\begin{equation}\label{eq6}
\begin{split}
   O^{ab}_{xy}  & = \sigma^a_{x,+\mu} \otimes \sigma^b_{x+\mu,-\mu},\\ 
   L^a_{x,+\mu} & = \sigma^a_{x,+\mu} \otimes \mathbb{I},~~
R^a_{x+\mu,-\mu} = \mathbb{I} \otimes \sigma^a_{x+\mu,-\mu}.
\end{split}
\end{equation}
 Each operator in the bilinear is called a rishon.
 Note that we have explicitly chosen the smallest representation possible here, the 
spin-$\frac{1}{2}$, and the generators are then simply the tensor products of the
Pauli operators. In general, it is also possible to choose a spin-1,
or any other allowed representations. Typically, it is expected that with
integer-valued spins one obtains theories whose ground states behave qualitatively
similar to that of the corresponding Wilson formulation of the theory 
\cite{Zache:2021ggw}. On the other hand, choice of a half-integer spin gives rise 
to a novel phases, often relevant in the context of non-trivial $\theta$-terms
\cite{Banerjee:2013dda,Banerjee:2023pnb}.

 This representation was also the subject of \cite{Rico:2018pas}, where the 
physics in $(1+1)$-d dimension was studied in the presence of dynamical fermionic 
fields. In this article, we extend the studies to two spatial dimensions,
and inclusion of fermions in the two-dimensional model is underway.
A key feature of the spin-$\frac{1}{2}$ representation is that the electric field energy
term does not explicitly appear in the Hamiltonian, since both the fields square
to yield a constant. However, the fields still remain dynamical, influencing the
theory by choosing the physical Hilbert space through the Gauss' Law. The magnetic 
field is fully non-trivial, and in terms of the chosen operators we have:
\begin{equation}\label{eq7}
\begin{split}
    {\rm Tr} {\cal O}_\Box &= (\sigma^a_{x,+\mu} \otimes \sigma^b_{y,-\mu}) \otimes  
    (\sigma^b_{y,+\nu} \otimes \sigma^c_{z,-\nu}) \\
    &\qquad \otimes (\sigma^c_{z,-\mu} \otimes \sigma^d_{w,+\mu}) 
    \otimes (\sigma^d_{w,-\nu} \otimes \sigma^a_{x,+\nu} ).
\end{split}
\end{equation}
 The location of the operators are shown in \cref{fig:latt}, and the trace on the 
left-hand side is implemented on color indices, as can be seen explicitly in the above
equation. The local plaquette is clearly a 256-dimensional matrix. 

\vspace{0.3cm}
{\bf Gauss Law:} A general gauge transformation is given by 
$V = \prod_x \exp(-i \alpha^a_x G^a_x)$. Using equation (\ref{eq7}), the Gauss Law 
(in the absence of any matter field) is 
\begin{equation}\label{eq8}
G^a_{x}  = \sum_{\mu} \bigg( {\sigma}^a_{x,+\mu} + {{\sigma}^a}_{x,-\mu}\bigg).
\end{equation}
 Demanding a physical state to be gauge-invariant is equivalent to selecting states
according to the condition $G^a_x \ket{\psi} = 0$.

\vspace{0.3cm}
{\bf Construction of gauge-invariant states:} Now, we discuss the construction 
of singlet states 
under gauge transformations in both one and two spatial dimensions. These states
are sometimes called \emph{glueball} states \cite{Banerjee:2012xg}. In one spatial
dimension, there are two links touching a site $x$, and a gauge invariant state 
can be easily constructed as follows:
\begin{equation}\label{eq9}
\begin{aligned}
    & \ket{\psi_s}_{x,+\mu,-\mu}\\
     &\;\;\; = \frac{1}{\sqrt{2}} \bigg( \ket{\uparrow}_{x,+\mu}\ket{\downarrow}_{x,-\mu} 
     - \ket{\downarrow}_{x,+\mu}\ket{\uparrow}_{x,-\mu} \bigg)\textcolor{blue}{.}
     \end{aligned}
\end{equation}
The state $\ket{\psi_s}_x$ is gauge-invariant, which means $G^3_x \ket{\psi_s}_x = 0$, 
$G^+_x \ket{\psi_s}_x = 0$, and $G^-_x \ket{\psi_s}_x = 0$, where the operators $G^+_x$ and $G^-_x$ are defined as 
$G^+ = \frac{1}{\sqrt{2}} ( G^1 + i G^2)$, $G^- = \frac{1}{\sqrt{2}} ( G^1 - i G^2)$ respectively.
We can construct a triplet state at site $x$ as follows
\begin{equation}\label{eq10}
\begin{split}
 \ket{\psi_1}_{x,+\mu,-\mu} &= \ket{\uparrow}_{x,+\mu}\ket{\uparrow}_{x,-\mu}\\
 \ket{\psi_2}_{x,+\mu,-\mu} &= 
 \frac{1}{\sqrt{2}} \bigg( \ket{\uparrow}_{x,+\mu}\ket{\downarrow}_{x,-\mu} 
\\&\qquad\qquad\qquad\qquad+ \ket{\downarrow}_{x,+\mu}\ket{\uparrow}_{x,-\mu} \bigg)\\
 \ket{\psi_3}_{x,+\mu,-\mu} &= \ket{\downarrow}_{x,+\mu}\ket{\downarrow}_{x,-\mu}.
\end{split}
\end{equation}
The triplet states represent external static charges and are useful to track
the total number of allowed states with the chosen representation.

\begin{figure}[h] 
\begin{center}
\includegraphics[trim=5pt 4pt 6pt 3pt, clip, width=0.49\linewidth]{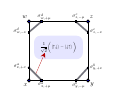}
\includegraphics[trim=5pt 4pt 1pt 1pt, clip, width=0.49\linewidth]{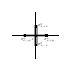}
\includegraphics[trim=5pt 4pt 1pt 1pt, clip, width=0.6\linewidth]{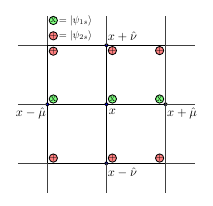}
\caption{(Top left): Gauge-invariant states for the plaquette can be constructed by creating singlets of each pair of spins at the corners. The figure illustrates how the singlets are constructed at each corner. (Top right): The location of the four spins relative to a lattice site, which is used in the construction of gauge-invariant states for a 2D lattice. Four spin-$\frac{1}{2}$ are considered, and as explained in the text, two singlets can be constructed. (Bottom): The figure shows one of the states contributing to the ground state, labeled as $\ket{0011}$, on a $2\times2$ lattice with periodic boundary conditions (PBC) where $\ket{0}$ represents $\ket{\psi_{1s}}$ and $\ket{1}$ represents $\ket{\psi_{2s}}$.} 
\label{fig:plaq-GIS}
\end{center}
\end{figure}

Next in complexity, consider a single plaquette state. In this case also, there are two
links touching a site, but in orthogonal directions. As before, we can build four singlet
states using two spins touching each corner. Labelling the corner sites as $x,y,z,w$, the
single gauge-invariant state in this case can be represented as
\begin{equation}\label{eq11}
\begin{aligned}
\ket{\psi_s}_{\Box}=&\ket{\psi_s}_{x,+\mu,+\nu} \ket{\psi_s}_{y,-\mu,+\nu} \\
& 
  \ket{\psi_s}_{z,-\mu,-\nu} \ket{\psi_s}_{w,+\mu,-\nu}\textcolor{blue}{.}
  \end{aligned}
\end{equation}
Since each singlet state is gauge-invariant separately, the state $\ket{\psi_s}_{\Box}$ is 
also gauge-invariant trivially. Further, it is trivial to compute the ground state energy 
for this state. Noting that each singlet contributes $-\frac{3}{4}$, while there is an 
additional factor of $2^8$ for defining the Hamiltonian via the Pauli $\vec{\sigma}$, instead
of the usual $\vec{S}$ operators. This normalization is better suited to studies of the model
on quantum computers. The ground state energy for the state $\ket{\psi_s}_{\Box}$ is thus
\begin{equation*}
    \mathcal{H} \ket{\psi_s}_{\Box} = -\frac{256}{4g^2} \big(-\frac{3}{4}\big)^4 
    \ket{\psi_s}_{\Box} = -\frac{81}{4g^2} \ket{\psi_s}_{\Box} \textcolor{blue}{.}
\end{equation*}

Let us give an example of how to track the total number of states separately in different
Gauss' Law sectors. Because every link operator consists of two spin-$\frac{1}{2}$s, there are four possible states for each link, so a square plaquette has a total of $4^4 (=256)$ possible states. As we argued before, only the one such state remains invariant under gauge transformation, which corresponds to the tensor product of pairwise singlets as in \cref{eq10}. The other states correspond to different charge insertions on the lattice sites. We can decompose the 256 states into different gauge sectors as
\begin{equation}\label{eq12}
    256 = 1 \oplus 12 \oplus 54 \oplus 108 \oplus 81,
\end{equation}
where the $1$ is the full gauge invariant state (singlets at all corners), and $81$ is the number of states at each sites with triplet charges $3^4 = 81$. With a single triplet charge on any lattice site, one has $^4 C_1 \cdot 3 = 12$ states, and with three triplet charges, one obtains $^4 C_3 \cdot 3^3 = 108$ states, and finally the $54$ corresponds to the situation when any two of the sites have triplet charges.

\vspace{0.3cm}
{\bf Gauge-invariant states for four spins:} Once the pattern of building singlets to 
impose the Gauss Law is understood, it is straightforward to push the construction for
a large lattice in higher dimensions. We restrict to two space-dimensional plaquettes 
in this article. For a square lattice, there are four links which touch a single site,
and we need to count how many singlets can be constructed with four spin-$\frac{1}{2}$s.
Clearly, since two spin-$\frac{1}{2}$s give a singlet and a triplet, $1 \oplus 3$, 
with four spin-$\frac{1}{2}$s, we get $(1 \oplus 3) \otimes (1 \oplus 3) 
= 2 \cdot 1 \oplus 3 \cdot 3 \oplus 5$, which means that there are two singlets, 
three triplets and a single quintet, giving a total of 16 states, as expected. 

Consider two spin singlet states at site $x$ given by 
\begin{equation}\label{eq13}
 \ket{\psi_s}_{x,+\mu,-\mu} , \ket{\psi_s}_{x,+\nu,-\nu}.
\end{equation}
These two states correspond to four spins, and we can create gauge-invariant singlet 
states for the four spins in two ways. The first one is given by
\begin{equation}\label{eq14}
\begin{split}
 \ket{\psi_{1s}}_x &= \ket{\psi_s}_{x,+\mu,-\mu} \otimes \ket{\psi_s}_{x,+\nu,-\nu}.
\end{split}
\end{equation}
Using \cref{eq12} by combining two triplets, we can construct another
gauge-invariant spin singlet state at site $x$ with the linear combination
\begin{equation}\label{eq15}
\begin{split}
  \ket{\psi_{2s}}_x =&\; a \ket{\psi_1}_{x,+\mu,-\mu} \ket{\psi_3}_{x,+\nu,-\nu}\\
    &+ b \ket{\psi_2}_{x,+\mu,-\mu} \ket{\psi_2}_{x,+\nu,-\nu}\\
    & + a \ket{\psi_3}_{x,+\mu,-\mu} \ket{\psi_1}_{x,+\nu,-\nu}\textcolor{blue}{.}
\end{split}
\end{equation}
We find the constants $a=-\frac{1}{\sqrt{3}}$ and $b=\frac{1}{2\sqrt{3}}$ by demanding 
the state $\ket{\psi_{2s}}$ to be normalized and annihilated by $G^+_x$ or $G^-_x$.

It is then possible to use a reduced Hilbert space to study the gauge invariant sector 
that consists of the two singlet states per site, $\ket{\psi_{1s}}_x$ and 
$\ket{\psi_{2s}}_x$. The Hamiltonian can be expressed in a gauge-invariant way as
follows (see \cref{fig:plaq-GIS} for the site indices, and \cref{sec:GIHam}
for a derivation of the same):
\begin{equation}\label{eq16}
\begin{split}
  H_{\mathrm{inv}} &= -\frac{1}{4g^2}\prod_{i=x,z}
  \left(\frac{1}{4}\left(\tau^3_i -\mathbbm{1}_i\right) + \frac{\sqrt{3}}{4} \tau^1_i \right) \\ 
   & \qquad\qquad\times \prod_{i=y,w}
  \left(\frac{1}{4}\left(\tau^3_i -\mathbbm{1}_i\right) - \frac{\sqrt{3}}{4} \tau^1_i \right)\textcolor{blue}{.}
\end{split}
\end{equation}
In moving to the gauge-invariant basis we have reduced our Hilbert space to $2$ 
states per site rather than $4$ states per link. 
In addition, for clarity, we have used the $\vec{\tau}$ operators to 
describe the action of the Hamiltonian within the two-dimensional gauge-invariant Hilbert
space spanned by $\ket{\psi_{1s}}$ and $\ket{\psi_{2s}}$.
Thus, for a general 2D lattice with extent $L_x \times L_y$, instead of $4^{2 \cdot L_x \cdot L_y}$, one
has to work with $2^{L_x \cdot L_y}$ states. In terms of actual numbers, for the $2 \times 2$
system, one can get away by diagonalizing a $16 \times 16$ matrix instead of a
$65536 \times 65536$ one. In actual calculations, we have always used the normalization 
$\frac{1}{4 g^2}=1$.

Note that $G^z_x$ is zero on these states by construction. The generator for gauge 
transformations used in this case can be expressed as:
\begin{equation}\label{eq17}
\begin{aligned}
 G^a_x  = \sigma^a_{x,+\mu} + \sigma^a_{x,-\mu} + \sigma^a_{x,+\nu} + \sigma^a_{x,-\nu},
\end{aligned}
\end{equation}
where $a = +,-,z$ correspond to the three Gauss' Laws. A visual representation of
this is presented in \cref{fig:plaq-GIS}.

\begin{figure}[ht]
\begin{center}
\includegraphics[trim=5pt 4pt 6pt 3pt, clip, width=0.83\linewidth]{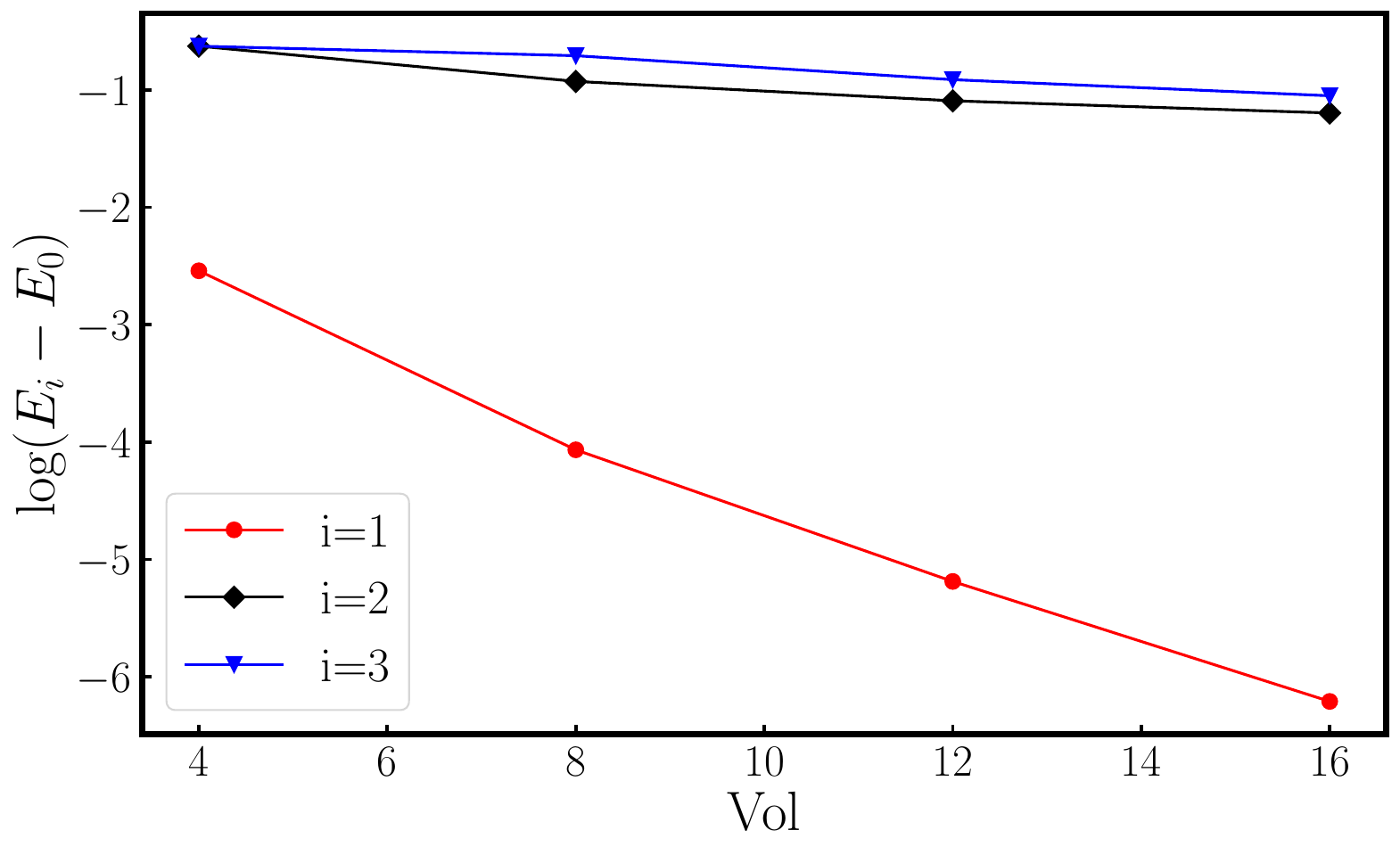}
\includegraphics[trim=5pt 4pt 6pt 3pt, clip, width=1.0\linewidth]{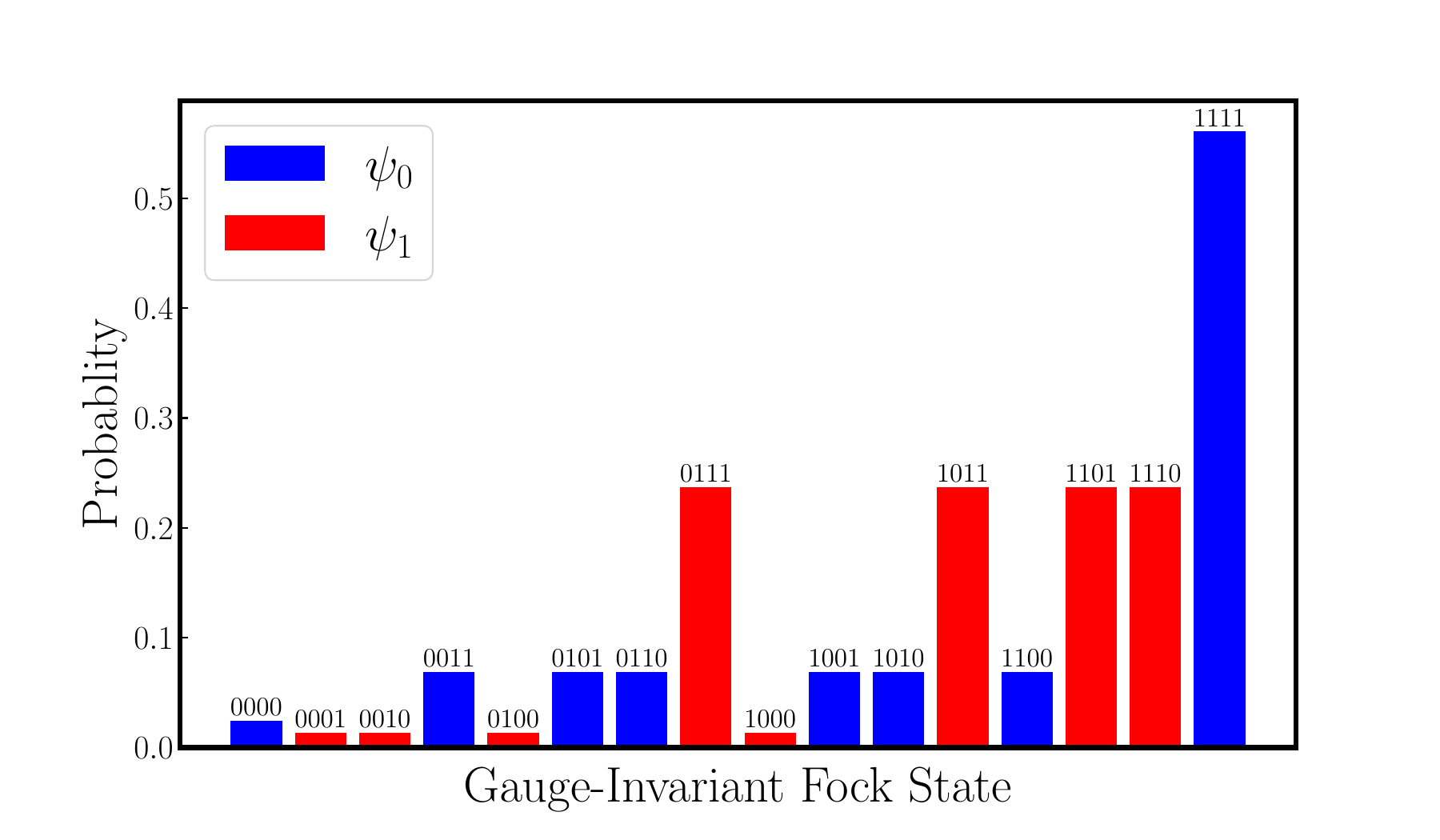}
\caption{\label{dE-gap} (Top): Plot of the energy difference within the pure $SO(3)$ QLM 
in (2 + 1)-d. If a discrete symmetry breaks spontaneously, the smallest mass gap 
becomes exponentially small with increase in volume. However, higher energy gaps are 
insensitive to this. The ED results are consistent with this hypothesis.
(Bottom): The probabilities from the wavefunction for the ground state (blue) and first excited state (red) for the 2$\times$2 (Vol=4) system expressed in the gauge invariant basis.}
\end{center}
\end{figure}

It turns out that the ground state of this model breaks the lattice translation
symmetry by a single lattice spacing spontaneously, which is actually identified
with charge conjugation \cite{Rico:2018pas}. This definition of charge conjugation 
ensures a smooth integration with staggered fermions, which we are addressing in a
future work. The physical translation operator is equivalent to two lattice spacings.
From exact diagonalization (ED), the lowest energy gap exponentially decreases with volume, as 
$\Delta E \sim \exp{(-\alpha V)}$. This is the telltale signature of discrete symmetry 
breaking in a finite volume, given that the ground state has $C = +1$, while the
first excited state has $C = -1$, where $C$ represents the charge conjugation quantum 
number. This behaviour is illustrated in \cref{dE-gap} (top panel), 
where the energy difference between the ground state and the first excited state becomes 
exponentially smaller as a function of volume. But the higher energy gaps ($E_2 - E_0$ and 
$E_3 - E_0$) are insensitive to the volume. The bottom panel of \cref{dE-gap} shows
the ground state and the first excited state wavefunctions, where the symmetry breaking
is evident.

\section{Method} \label{sec:methods}
 In this section, we describe quantum algorithms to target the low-lying energy states on
quantum computers. Since most of the quantum algorithms for this purpose use variational 
methods, our results indicate how robustly the exponentially small gap in an SSB phase can be 
extracted using quantum algorithms. 

\subsection{Variational Quantum Algorithms}
 It is well-known from basic quantum mechanics that for a given system described by a 
quantum Hamiltonian ($H$), we can estimate the ground state using variational principles.
This is directly used in the variational quantum eigensolver (VQE) algorithm, where the
following cost function is minimized with respect to the different parameters 
represented by $\Vec{\theta} = \{\theta_1, \cdots, \theta_N\}$ (assuming $N$ variational 
parameters):
\begin{equation} \label{eq:18}
    E(\Vec{\theta}) = \braket{\psi(\Vec{\theta}) | H | \psi(\Vec{\theta})},
\end{equation}
 where $\psi(\Vec{\theta})$ is a parameterized ansatz for the real parameters $\Vec{\theta}$. 
Finding the ground state energy of $H$ is equivalent to minimizing the cost function, 
$E(\Vec{\theta})$. Additionally, we can compute the excited state energies using a variational 
algorithm known as the variational quantum deflation (VQD) algorithm \cite{Higgott:2018doo}, 
which is an extension of the VQE algorithm. The primary idea is to iteratively remove the 
influence of the previously found states from the Hamiltonian to find higher excited states. 
In VQD, the cost function often involves terms that ensure orthogonality to previously found 
states to prevent overlap. To find the $k$-th excited state we minimize the cost function:
\begin{equation}\label{eq:19}
\begin{split}
F(\vec{\theta}_k) &= \braket{ \psi(\vec{\theta_k}) |H| \psi(\vec{\theta_k})} 
 + \sum_{i=0}^{k-1} \beta_i |\braket{ \psi(\Vec{\theta_k}) | \psi(\Vec{\theta_i})}|^2\\
  &= E(\vec{\theta}_k) + \sum_{i=0}^{k-1} \beta_i |\braket{ \psi(\vec{\theta}_k) | \psi(\vec{\theta}_i)}|^2,
\end{split}
\end{equation}
where the first term can be calculated using the same method as VQE, while the second part 
acts as a penalty, ensuring that the current state is orthogonal to all the previously 
optimized ones. In practice, the ansatz state $\ket{\psi(\vec{\theta_k})}$ may not be 
perfectly orthogonal to the previously found states $\ket{\psi_0}$, $\ket{\psi_1}$,
$\cdots$, $\ket{\psi_{k-1}}$ during the optimization process. The penalty terms help 
the optimization to enforce orthogonality. For example, if we already found the ground 
state $\ket{\psi_0}$ and the first excited state $\ket{\psi_1}$, the cost function for 
the second excited state $\ket{\psi_2}$ would look like:

\begin{equation}\label{eq:20}
\begin{split}
    F(\theta) = \braket{ \psi_2( \vec{\theta}) | H | \psi_2 (\vec{\theta}) } 
    + \beta_0 \left| \braket{ \psi_2(\theta) | \psi_0 } \right|^2 \\
    + \beta_1 \left| \braket{ \psi_2(\theta) | \psi_1 } \right|^2,
\end{split}
\end{equation}
where $\beta_0$ and $\beta_1$ are penalty coefficients, and 
$\left| \braket{ \psi_2(\vec{\theta}) | \psi_0 } \right|^2$ and 
$\left| \braket{ \psi_2(\vec{\theta}) | \psi_1 } \right|^2$ represent the overlaps of 
$\ket{\psi_2(\vec{\theta})}$ with $\ket{\psi_0}$ and $\ket{\psi_1}$ respectively. During 
optimization, the penalty terms $\beta_0 \left| \braket{\psi_2(\vec{\theta}) | \psi_0 }\right|^2$ 
and $\beta_1 \left| \braket{\psi_2(\vec{\theta}) | \psi_1 } \right|^2$ penalize 
overlaps with the states with lower energy. If $\ket{\psi_2(\vec{\theta})}$ has 
a non-zero overlap with $\ket{\psi_0}$ or $\ket{\psi_1}$, these terms increase 
the cost function value, discouraging the optimizer from selecting parameters 
that result in such overlaps.

Even though we want to find a state that is orthogonal to the previously found states, 
achieving perfect orthogonality through the optimization process can be difficult due 
to the circuit complexity and the higher dimensional parameter space. The penalty terms 
provide an effective way to ensure the ansatz state becomes orthogonal by the end of the 
optimization process. As the optimization progresses, the penalty terms actively reduce 
any overlap with previously found states. The choice of $\beta_i$ depends on the specific 
system and ansatz used. For example, if $\beta_i$ values are too high, the optimization 
process will prioritize orthogonality over minimizing the energy. This can lead to a state 
that is highly orthogonal to previous states but may not represent the true $k$-th excited 
state in terms of energy. Conversely, if $\beta_i$ values are too low, the optimization 
might not sufficiently enforce orthogonality, leading to an overlap with lower energy states, 
which could result in an incorrect excited state. These coefficients balance the cost function 
between lowering the energy and keeping the new state orthogonal to previously found states.
In the particular case of the $SO(3)$ model, we start with small values of $\beta$, and
increase it while keeping track of the dependence of converged energy $E_1$ with $\beta$.
An optimal value is chosen from the plateau where $E_1$ is stable. For the $2 \times 2$ lattice, 
this is $\beta=1$, and for $2 \times 4$ lattice, we chose $\beta=5$. 

\begin{figure}
\includegraphics[width=8cm]{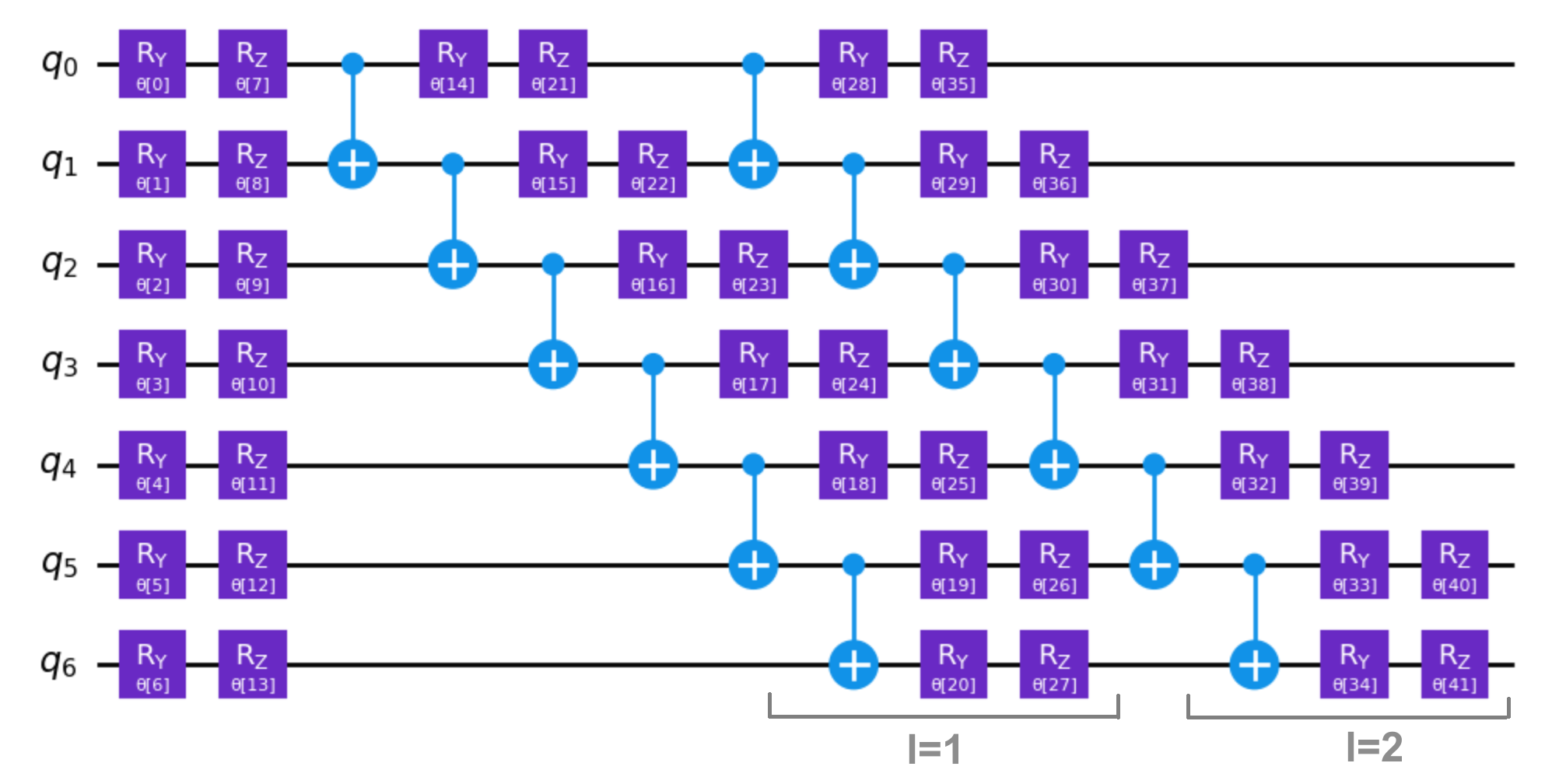}
\caption{The variational ansatz for seven qubits and two layers, which alternates 
two-qubit CNOT gates with single-qubit rotational gates in each layer.}\label{var}
\end{figure}

There are several options for choosing the initial variational states. For our calculations, 
we used the variational ansatz with linear connectivity. The basic structure in each layer 
consists of a sequence of qubits, connected to each other with 2-qubit CNOT gates, and have 
rotation gates in the $Y$ and $Z$ spin components. The full circuit has several layers of 
such gates to increase its expressivity. Mathematically, this can be written as:
\begin{equation} \label{eq:21}
\begin{aligned}
&\ket{\psi(\theta)} =\\&\prod_{l=1}^{N_l}\left( \prod_x \exp(-i ({\theta'}_{l,x}/2) Q^3_x) 
 \prod_x \exp(-i (\theta_{l,x}/2) Q^2_x)\right.\\&\left. \qquad\prod_{x} 
 \exp( i\pi/4(I_x-Q^3_x) \otimes (I_{x+1}-Q^1_{x+1})) \right)  
 \\ & \prod_x \exp(-i ({\theta'}_{0,x}/2) Q^3_x)\prod_x \exp(-i (\theta_{0,x}/2) Q^2_x)
 \ket{0}.
 \end{aligned}
\end{equation}
\cref{var} illustrates one such ansatz for seven qubits, and for two layers $N_l=2$. 
Note that we have called the qubit degrees of freedom to be generically $Q^i$. 
In the electric flux basis, they are the same operators as $\sigma^i$, while in the gauge
invariant basis they are represented through $\tau^i$.

\subsection{Quantum Adiabatic Algorithm and the QAOA}
 As mentioned in the introduction, the QAOA 
\cite{Lloyd:2018fsu, morales2020universality, Crooks:2018vud} is a variational quantum 
algorithm which exploits the form of the quantum Hamiltonian as well as the quantum 
adiabatic theorem in order to approximate the ground state of the Hamiltonian. According 
to the quantum adiabatic theorem, if one starts from the ground state of a (simple) 
Hamiltonian, and adds a coupling which varies with time, then the final state will be 
(arbitrarily) close to the eigenstate of the final Hamiltonian, provided the variation 
is done slowly, and assuming non-degenerate initial and the final states \cite{Farhi:2000ikn}. 
Using this idea, to use the QAOA algorithm, the Hamiltonian is divided into $N_\alpha$ Trotter-inspired parts, $H_\alpha$, such that
\begin{equation}\label{eq:22}
H = \sum_{\alpha=1}^{N_\alpha} H_\alpha \textcolor{blue}{,}
\end{equation}
where each $H_\alpha$ consists of a sum of terms that commute with one another.
This facilitates the choice of an initial state which is the ground state of the starting 
Hamiltonian. 

Then the QAOA ansatz is of the form
\begin{equation}
\ket{GS}_{\mathrm{QAOA}} = \prod_{k=1}^{p} \prod_{\alpha=1}^{N_\alpha} 
 e^{iC_{\alpha,k} H_{\alpha}} \ket{\psi_A},
 \label{qaoagen}
\end{equation}
where $\ket{\psi_A}=\ket{\psi_0}$ is the ground state of a portion of the Hamiltonian, 
which (without loss 
of generality) we set to be the first term in the sum, $H_{\alpha=1}$. In the limit of infinite 
layers, there will be a set of $C_{\alpha,k}$ variational parameters that yield the ground state. 
QAOA as an algorithm works by approximating the ground state for a finite number of parameters, 
and we can make the ansatz more expressive by increasing $p$, the number of QAOA layers.

For the $SO(3)$ model in particular, when we break down the Hamiltonian into separate parts, we do so such that one of these terms is a magnetic field in the $z$-direction, which we set to be our $H_{\alpha=1}$ due to its trivial ground state 
(note that since the gauge invariant basis is used here, we use the $\tau^i$ 
operators):
\begin{equation}
    \begin{aligned}
        H_{\alpha=1} &= J\sum_{x} \tau^3_x ,
        \qquad \ket{\psi_{A=0}}  = \ket{\uparrow\uparrow... \uparrow}.
        \label{qaoags}
    \end{aligned}
\end{equation}
For the $2\times2$ lattice, the Hamiltonian given by \cref{eq16} consists of four spins, 
and after expanding, we divide it into nine pieces for QAOA:
\begin{equation}\label{eq:25}
\begin{aligned}
H_1 = &\frac{1}{4^3}\sum_{x=1}^4 \tau^3_x,\\
H_2 = & -\frac{1}{4^3}\sum_{x\neq y} \tau^3_x \tau^3_y
    + \frac{1}{4^3}\sum_{x\neq y\neq z} \tau^3_x \tau^3_y \tau^3_z \\
      &-\frac{1}{4^3} \tau^3_1 \tau^3_2 \tau^3_3 \tau^3_4, \\
H_3 =& -\frac{3}{4^3} \tau^1_1 \tau^1_2 \left(-\tau^3_3\tau^3_4 
    + \tau^3_3 + \tau^3_4 - I\right),\\
H_4 =& -\frac{3}{4^3} \tau^1_3 \tau^1_4 \left(-\tau^3_1\tau^3_2 
    + \tau^3_1 + \tau^3_2 - I\right),\\
H_5 =& -\frac{3}{4^3} \tau^1_2 \tau^1_3 \left(\tau^3_1\tau^3_4 
    - \tau^3_1 - \tau^3_4 + I\right),\\
H_6 =& -\frac{3}{4^3} \tau^1_1 \tau^1_3 \left(-\tau^3_2\tau^3_4 
    + \tau^3_2 + \tau^3_4 - I\right),\\
H_7 =& -\frac{3}{4^3} \tau^1_2 \tau^1_4 \left(-\tau^3_1\tau^3_3 
    + \tau^3_1 + \tau^3_3 - I\right),\\
H_8 =& -\frac{3}{4^3} \tau^1_1 \tau^1_4 \left(\tau^3_2\tau^3_3 
    - \tau^3_2 - \tau^3_3+I\right),\\
H_9 =& -\frac{3^2}{4^3} \tau^1_1 \tau^1_2 \tau^1_3 \tau^1_4 .
\end{aligned}
\end{equation}

Note that in the typical spin-models considered in the literature,
the use of the QAOA form would require at most a few terms in the Hamiltonian. The
additional complexity in our case is the due to non-Abelian nature of the gauge
theory under consideration. The decomposition of Hamiltonian for the $2\times 4$ lattice
can be found in the \cref{sec:trot}. The periodic boundary conditions on 
ladder systems give rise to more cancellations than are possible for a 
square geometry. The key point in all of these decompositions is the presence 
of a term $H_1$ of the form $H_1=\sum_{x=1}^N \tau^3_x$. 

In order to approximate the first excited state, we use a QAOA-inspired ansatz 
that makes use of symmetry to ensure it is orthogonal to the ground state (which
is then in a different symmetry sector). In analogy to $\ket{\psi_0}$, we define 
$\ket{\psi_{A}}$ as
\begin{equation}
\ket{\psi_{A=1}} = \ket{\uparrow\uparrow... \downarrow},
\label{qaoaes}
\end{equation}
where we have flipped the last spin. This state is the one of the degenerate 
first excited states $\ket{\psi_1}_{i} = \tau^x_i \ket{\psi_0}$ of $H_{\alpha=1}$, 
and because the Hamiltonian only flips an even number of spins at a time, it 
is impossible to get to the ground state by adiabatic evolution. We thus use this 
state to approximate the first excited state. 
In fact, \cref{qaoags} and \cref{qaoaes} are not the only 
possible initial states for the ground and first excited states. If we start with 
any state from the same symmetry sector as an initial state, the final state after 
optimization will always stay in that sector. Moreover, in terms of circuit depth 
and convergence, the optimization with different initial states is comparable. 

\section{Results for the SO(3) QLM} \label{sec:results1}
  We have applied the different quantum algorithms described in \cref{sec:methods} on the
pure gauge $SO(3)$ model as described in \cref{sec:model}. We have used both classical and 
quantum hardware platforms in order to benchmark the performance of the algorithms. 
The quantum hardware of choice was the trapped-ion quantum computer, IonQ. With the resources 
available to us, we have been able to study the problem on a small quantum hardware of 
four qubits. The classical simulation results are for systems with up to 12 qubits.

\subsection{Simple VQE Ans\"atze with Real Hardware}

First, we explore how effective the quantum algorithms perform on NISQ 
hardware to find the ground state, when working in the gauge field basis where gauge 
invariance is not imposed. The results can then be compared with exact solutions. We
consider three simple systems: the \emph{bubble} plaquette (consisting of two links), the 
 \emph{triangular} plaquette (consisting of three links), and the \emph{square} plaquette 
 (consisting of four links). The IonQ quantum hardware was used to obtain the results only 
 in the first case, the bubble plaquette. 

\begin{widetext}
\begin{figure*}
  \centering 
  \includegraphics[trim=1pt 5pt -5pt 3pt, clip, scale=3.5]{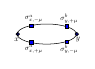}
  \includegraphics[trim=1pt 5pt -5pt 3pt, clip, scale=1.8]{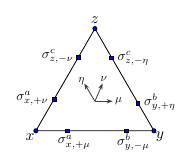}
  \includegraphics[trim=1pt 5pt 4pt 3pt, clip, scale=2.56]{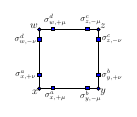}
  \caption{\label{FH-space}(Left): A plaquette with two links, the bubble. 
  (Middle): A plaquette with three links, the triangle. (Right): A plaquette 
  with four links, the square.}
\end{figure*}
\end{widetext}

{\bf Bubble plaquette:} Consider the simplest system first, the bubble plaquette, composed of
only two links (\cref{FH-space}, left). From the expression for the link from \cref{eq6}, we obtain
\begin{equation}\label{eq:27}
    \mathcal{O}^{ab}_{xy} = \sigma^a_{x,+\mu} \otimes \sigma^b_{y,-\mu}\textcolor{blue}{,}
\end{equation}
and the Hamiltonian for the bubble plaquette is then given by
\begin{equation}\label{eq:28}
  \mathcal{H}_{\rm bub} = -J ( {{\sigma}^a}_{x,+\mu} {{\sigma}^b}_{y,-\mu} ) 
                   ( {{\sigma}^b}_{y,+\mu} {{\sigma}^a}_{x,-\mu} ).
\end{equation}

 The two links are physically distinct, and so the operators do not act on the same point. 
A pictorial representation of this is presented in \cref{FH-space} (left), where the blue 
squares indicate the rishon sites. In the spin $S = \frac{1}{2}$ representation
(we use Pauli matrices at the expense of a factor of 2), there are 
two spin halves on each of the two rishon sites on a link, and therefore naively each link has four states. Moreover, the use of the $\sigma^i$ operators make it
clear that we are in the electric flux basis, which is not explicitly gauge invariant.

The analytic argument to obtain gauge singlets is simple: each lattice site connects two 
rishon sites, one to the immediate left and the other to the immediate right. The 
two spins on the rishon sites can form a spin singlet, and a spin triplet. The spin-triplet 
transforms as a charged operator under the Gauss law and thus lies in a high-energy 
manifold. This is true for both the lattice sites. The total gauge-invariant state for the bubble 
plaquette is then obtained by a tensor product of the two singlets, one situated at 
site $x$ and the other at site $y$. Mathematically, the gauge-invariant ground state 
can be represented as: 
\begin{equation}\label{eq:29}
 \ket{\psi_s}_{\rm bub} = \ket{\psi_s}_x \otimes \ket{\psi_s}_y,
\end{equation}
where $\ket{\psi_s}_x$ is the singlet state formed at site $x$ as defined in \cref{eq9}. 
The energy for the bubble plaquette is the product of the energy of two independent singlets, 
\begin{equation}\label{eq:30}
\mathcal{H}_{\rm bub} \ket{\psi_s}_{\rm bub} = 
 -J (-3)^2 \ket{\psi_s}_{\rm bub} = -9 J \ket{\psi_s}_{\rm bub}.
\end{equation}

The VQE used to optimize the parameters of the ansatz is defined as:
\begin{equation}\label{bubansz}
\begin{aligned}
    \ket{\psi(\theta)}_{\mathrm{bub}}&=\mathrm{CNOT}_{01}
    \cdot R_z(\theta)_0 \cdot H_0\\& 
    \qquad\times\mathrm{CNOT}_{23}
    \cdot R_z(\theta)_2 \cdot H_2 \ket{0101}, 
\end{aligned}
\end{equation}
where $i=0,1,2,3$ denote the sites ($x,+\mu$),  ($x,-\mu$),  ($y,-\mu$), and  
($y,+\mu$) respectively and $H_i$ are the Hadamard operators acting on qubit $i$.

\begin{figure}[ht]
\includegraphics[width=0.8\linewidth]{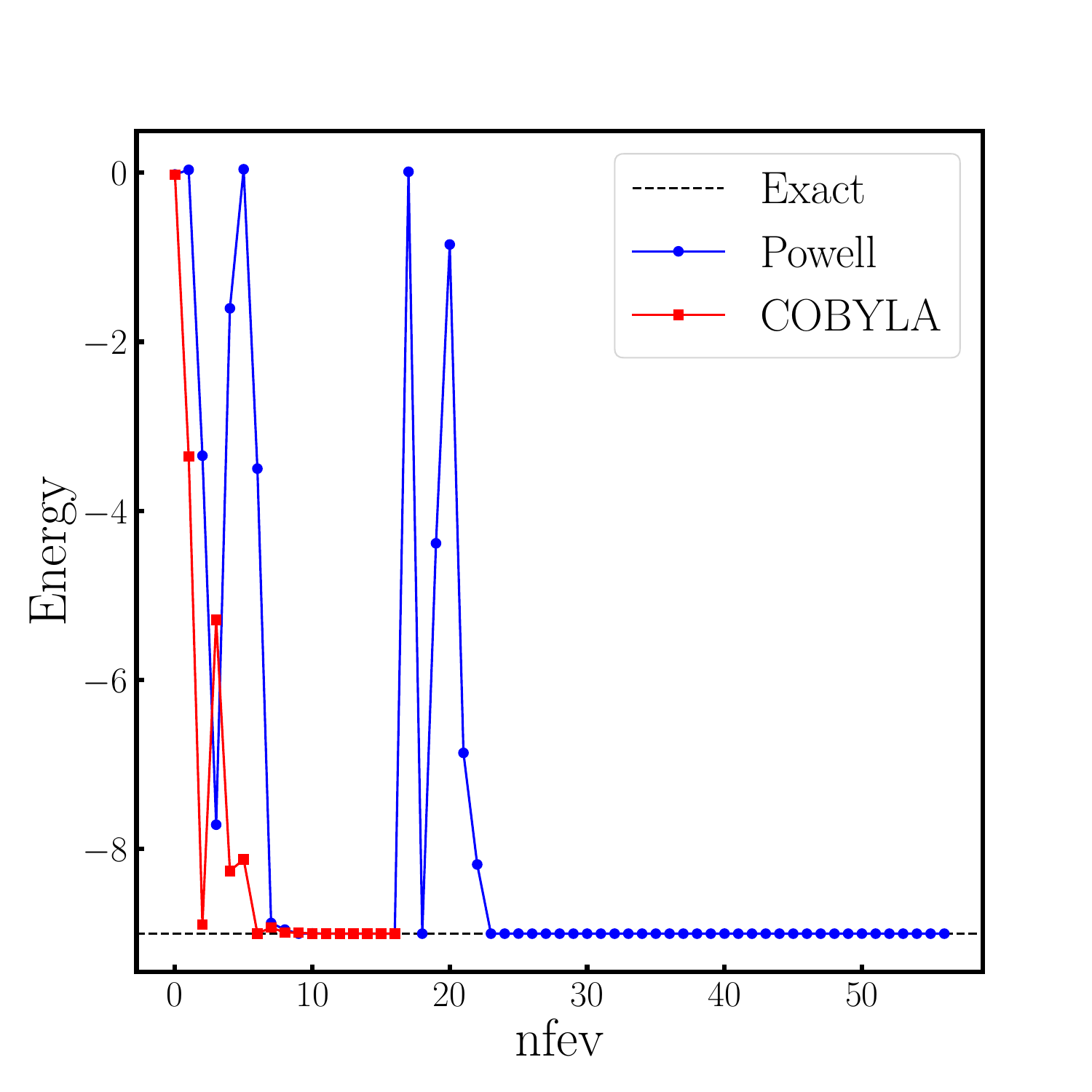}
\caption{
 \label{VQE-ansz}   
 The convergence of the energy with the number of 
 function evaluations (denoted as nfev on the x-axis) using the classical simulator for two different classical optimizers. In most cases, the COBYLA optimizer performs 
 significantly better than the Powell one.}
\end{figure}

A single layer of quantum gates with a single variational parameter 
is sufficient for this example, which is optimized by running the VQE algorithm on 
an exact quantum simulator. This gives the optimal value of $\theta = \pi$, and the 
corresponding optimized energy (shown on the right  of \cref{VQE-ansz}) converges to the exact 
value of $-9.0$ (as given by \cref{eq:30}) very rapidly. The optimized wavefunction 
is obtained by using $\theta = \pi$ in \cref{bubansz} and consists of four Fock
states, which can be written as (up to an overall phase)
\begin{equation} \label{bubGS}
  \ket{\psi}_{\rm GI} = \frac{1}{2} (\ket{1010} - \ket{0110} -\ket{1001} + \ket{0101}).
\end{equation}

\begin{widetext}
\begin{figure*}[ht]
\begin{center}
 \includegraphics[width=0.3\linewidth]{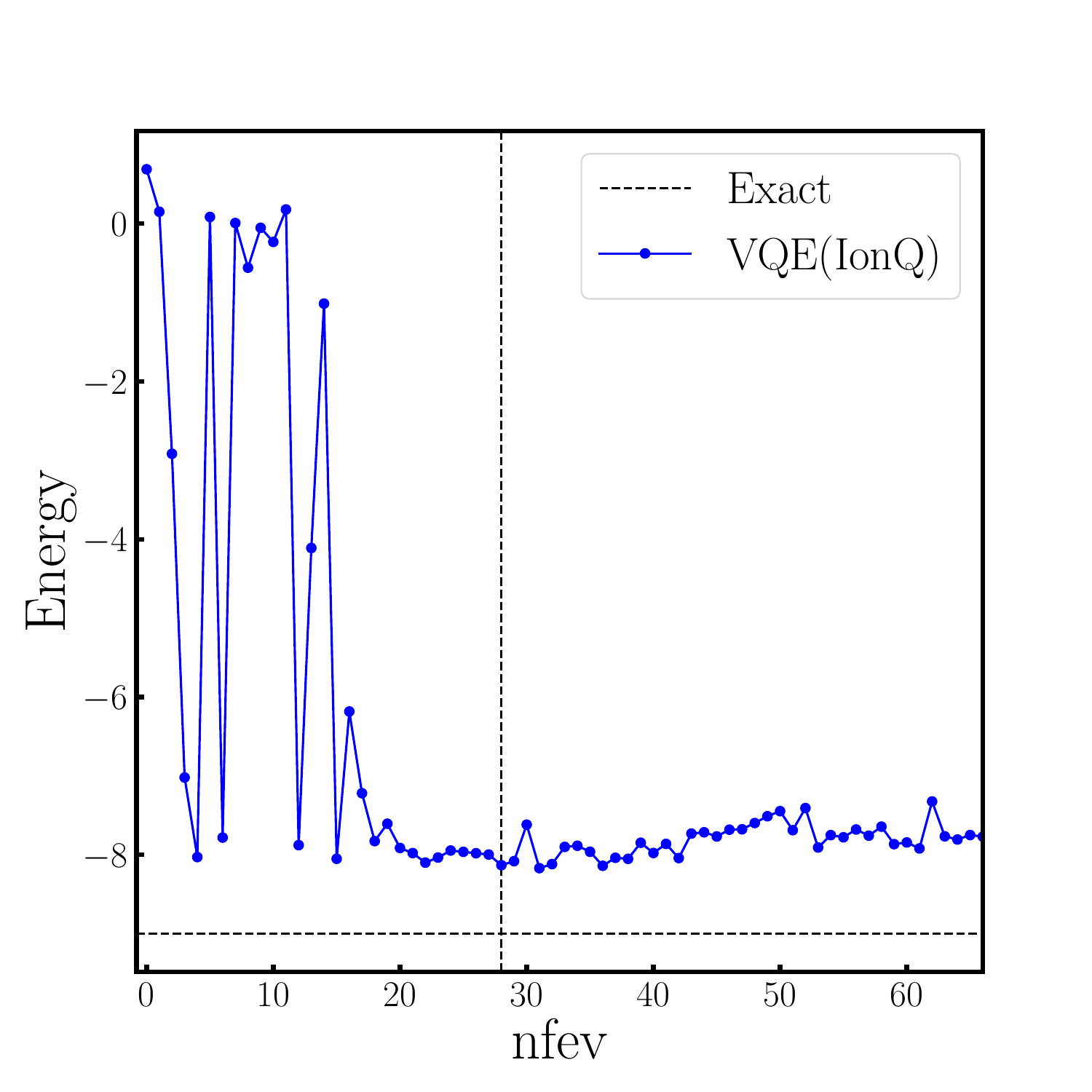}
 \includegraphics[width=0.3\linewidth]{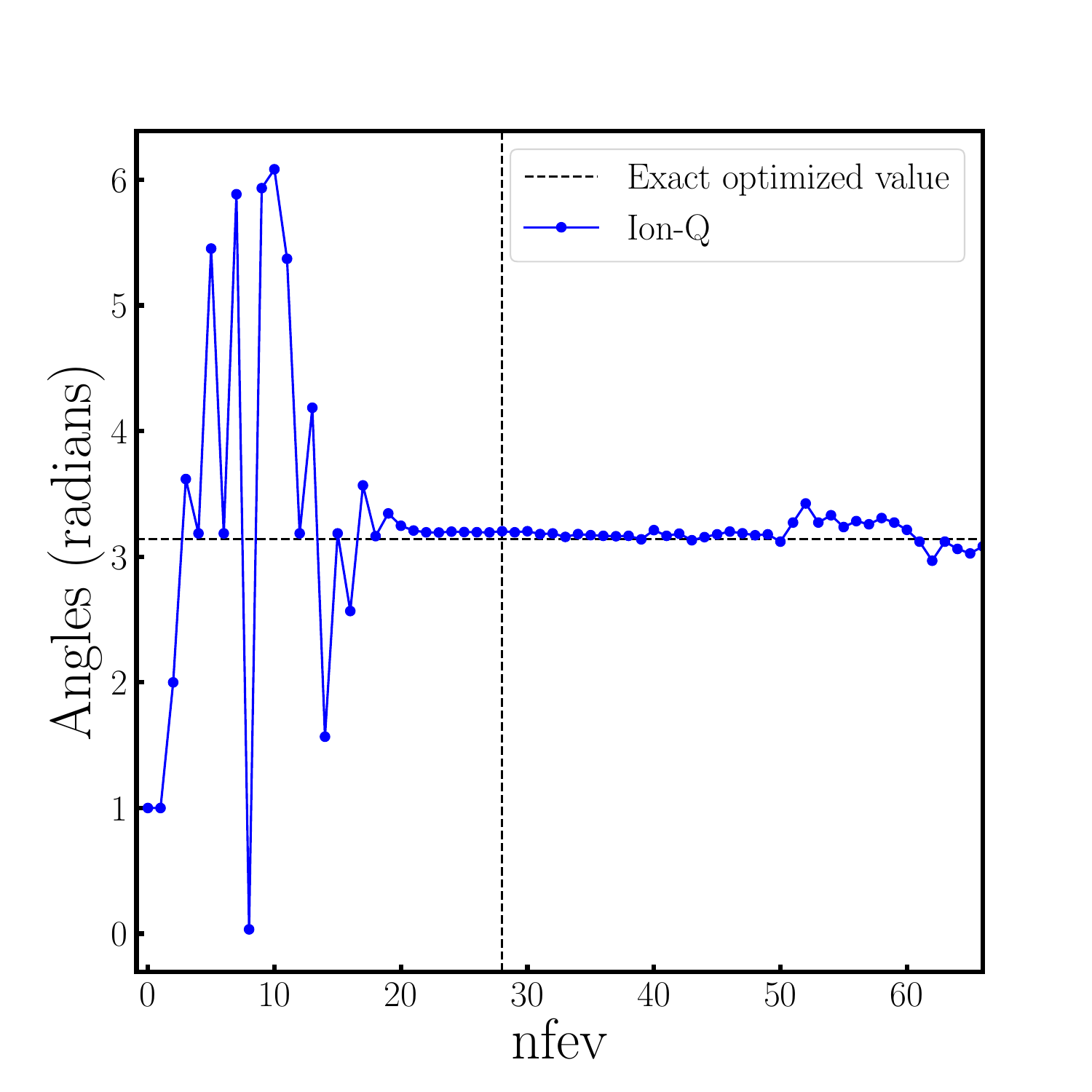}
 \includegraphics[width=0.3\linewidth]{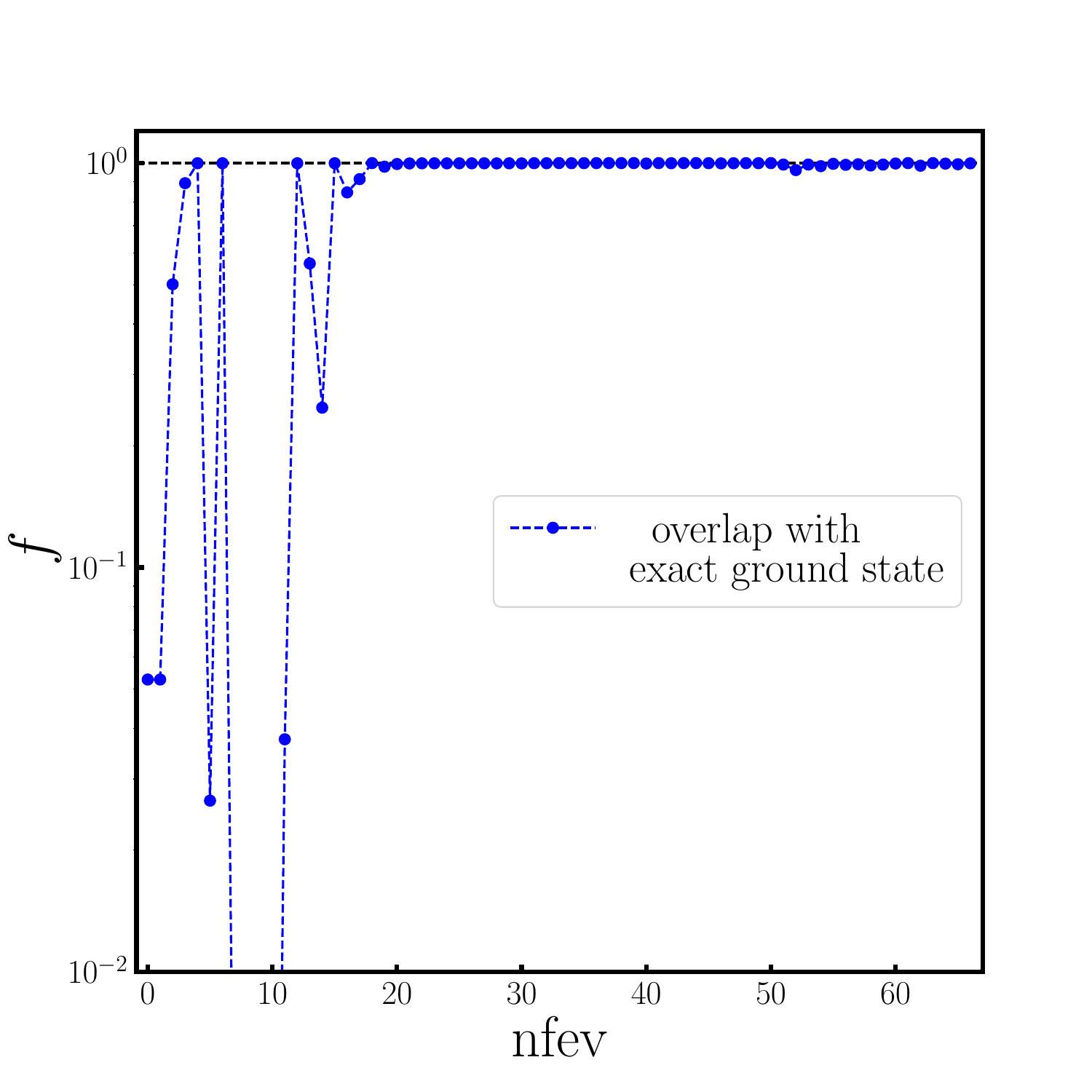}
\caption{\label{bubb-Expt} Experimental results for the bubble plaquette: (left) 
Plot of the energy with the number of function evaluations using the quantum hardware (IonQ 
trapped ions). The deviation from the exact result is commented upon in the text.
(middle) The estimate of the variational parameter $\theta$ at each step of the
optimizer. (right) Measure of the fidelity of the variational wavefunction with the
exact wavefunction at each step in the optimization process. The dashed value indicates
the analytical result.}
\end{center}
\end{figure*}
\end{widetext}

For this example, results using IonQ hardware to obtain the ground state,
as shown in \cref{bubb-Expt}. The left subplot shows the energy at each step of the optimization process, while the middle one shows the estimate of the variational parameter, and the right panel plots the fidelity. The fidelity, defined as the overlap between two wavefunctions is commonly used figure-of-merit to judge their equivalence, and is mathematically defined as $f = \left| \braket{\psi_0 | \psi_1} \right|^2$. 
The ground state energy found by the real hardware is around $-8.0$, 
significantly larger than the exact ground state energy of $-9.0$, even though the optimizer for the wavefunction reaches the correct optimal value, $\theta=\pi$. \cref{bubb-Expt} shows that the energy estimates stabilize after about the first third of the plotted optimization steps. We have put a mark at the 28-th step to note the apparent equilibration of results and use the wavefunction obtained at this step to compare with the exact wavefunction. It is important to note that the optimizer is unaware of the gauge invariance of the ground state. 
The noisy hardware generates spurious states, instead of only the four states 
expected from \cref{bubGS}. \cref{bub-comp-TAB} lists the proportions of all the states 
obtained using the quantum hardware at the $28$th step of the optimization and compares 
them with the exact classical result. While we do see the expected four states that form the gauge-invariant ground state wavefunction dominating (each with probability approximately $\sim 0.25$), there are 11 other states with probabilities two orders of magnitude smaller than the dominant states. These states will contribute to the wave-function generated by the hardware and will raise the measured energy to $-8.1318 $ instead of $-9$. Note that we only have direct access to the measured probabilities of the experimental wave function, and not to the relative signs of the basis states. However, since the optimization of $\theta$ parameter of $3.203$ is much closer to the analytic value $\pi$ (the energy deviation is about $\sim 10\%$ while the parameter deviation is about $\sim 2 \%$), the fidelity between the ansatz wave function at each variational step is much closer to the analytical answer.
This is precisely because the construction of the wave function is gauge-invariant 
(making singlets at each site) and thus the fidelity calculation (shown in \cref{bubb-Expt} right) 
is significantly less contaminated by the gauge non-invariant states than the energy. 

\begin{table*}[htbp]
\begin{tabular}{|c|c|c|c|}
    \hline
    \textbf{States} & \makecell{\textbf{Probablities from } \\ \textbf{exact States}} & 
    \makecell{\textbf{Probablities from } \\ \textbf{classical simulator}} & 
    \makecell{\textbf{Probablities from} \\ \textbf{IonQ hardware}} \\
    \hline
    0000  & 0.0   &  0.0  &  0.0      \\
    0001  & 0.0   &  0.0  &  0.0064   \\
    0010  & 0.0   &  0.0  &  0.0061   \\
    0011  & 0.0   &  0.0  &  0.0002   \\
    0100  & 0.0   &  0.0  &  0.0082   \\
    0101  & 0.25  &  0.25 &  0.2436   \\
    0110  & 0.25  &  0.25 &  0.2386   \\
    0111  & 0.0   &  0.0  &  0.0066   \\
    1000  & 0.0   &  0.0  &  0.0048   \\
    1001  & 0.25  &  0.25 &  0.2364   \\
    1010  & 0.25  &  0.25 &  0.2303   \\
    1011  & 0.0   &  0.0  &  0.0061   \\
    1100  & 0.0   &  0.0  &  0.0004   \\
    1101  & 0.0   &  0.0  &  0.005    \\
    1110  & 0.0   &  0.0  &  0.003    \\
    1111  & 0.0   &  0.0  &  0.0002   \\
    \hline
\end{tabular}
\caption{\label{bub-comp-TAB} Comparison of results between the exact state, classical simulator, 
and the quantum hardware for the bubble plaquette. While the classical simulator gives exact 
zeroes for the probabilities of 12 out of the 16 states, there are small nonzero 
``leakage'' probabilities for 11 of these states in the real quantum hardware.}
\end{table*}

{\bf Triangular and square plaquettes:} The calculations done for the triangular 
and the square plaquettes (with three and four links respectively) are very similar 
to that of the bubble plaquette. The geometry is illustrated in (\cref{FH-space} 
(middle and right) respectively, where the rishon sites are also indicated). The 
Hamiltonians for the two systems are:
\begin{align}\label{triSqr}
  \mathcal{H}_{\rm tri} &= -J ~~ {\rm Tr} (\mathcal{O}_{\triangle} ) \nonumber \\
    &= -J ( {{\sigma}^a}_{x,+\mu} {{\sigma}^b}_{y,-\mu} )  
     ( {{\sigma}^b}_{y,+\eta}{{\sigma}^c}_{z,-\eta})\nonumber\\ &\qquad\qquad
     ( {{\sigma}^c}_{z,-\nu} {{\sigma}^a}_{x,+\nu}), \\
  \mathcal{H}_{\rm plaq} &= -J ~~ {\rm Tr} (\mathcal{O}_{\Box} ) \nonumber \\
    &= -J \hspace{0.2cm} (\sigma^a_{x,+\mu} \sigma^b_{y,-\mu})  
    (\sigma^b_{y,+\nu} \sigma^c_{z,-\nu}) \nonumber  \\
    &\qquad\qquad (\sigma^c_{z,-\mu} \sigma^d_{w,+\mu}) (\sigma^d_{w,-\nu} \sigma^a_{x,+\nu} ).
\end{align}

Let us consider the triangle first, where it is not possible to build the ground state 
by forming singlets at each site due to the frustrated nature of the lattice. Instead, 
the ground state is not gauge-invariant, but has two singlets at two sites, and a triplet 
on the third site. The corresponding energy is then $E =- (-3)^2 \cdot 1 = -9$. The wavefunction 
is a linear combination of three terms where the triplet can be located at the three possible 
sites,
\begin{equation} \label{eq:GStri}
\ket{\psi_G} = \frac{1}{\sqrt{3}} \left[ 
   \ket{\psi_s}_{x} \ket{\psi_s}_{y} \ket{\psi_2}_{z} + \rm{2~permutations}
          \right]\textcolor{blue}{,}
\end{equation}
where the states $\ket{\psi_s}$ and $\ket{\psi_2}$ are described in \cref{eq9} and
\cref{eq10}. 

The square plaquette is made up of four links and eight spin-1/2 particles (\cref{FH-space}). 
The ground state is gauge-invariant and is the product of four two-spin singlets as described in 
\cref{eq9}, with energy $E = -(-3)^4 = -81$. 

\begin{figure}[ht]
\begin{center}
\includegraphics[width=0.49\linewidth]{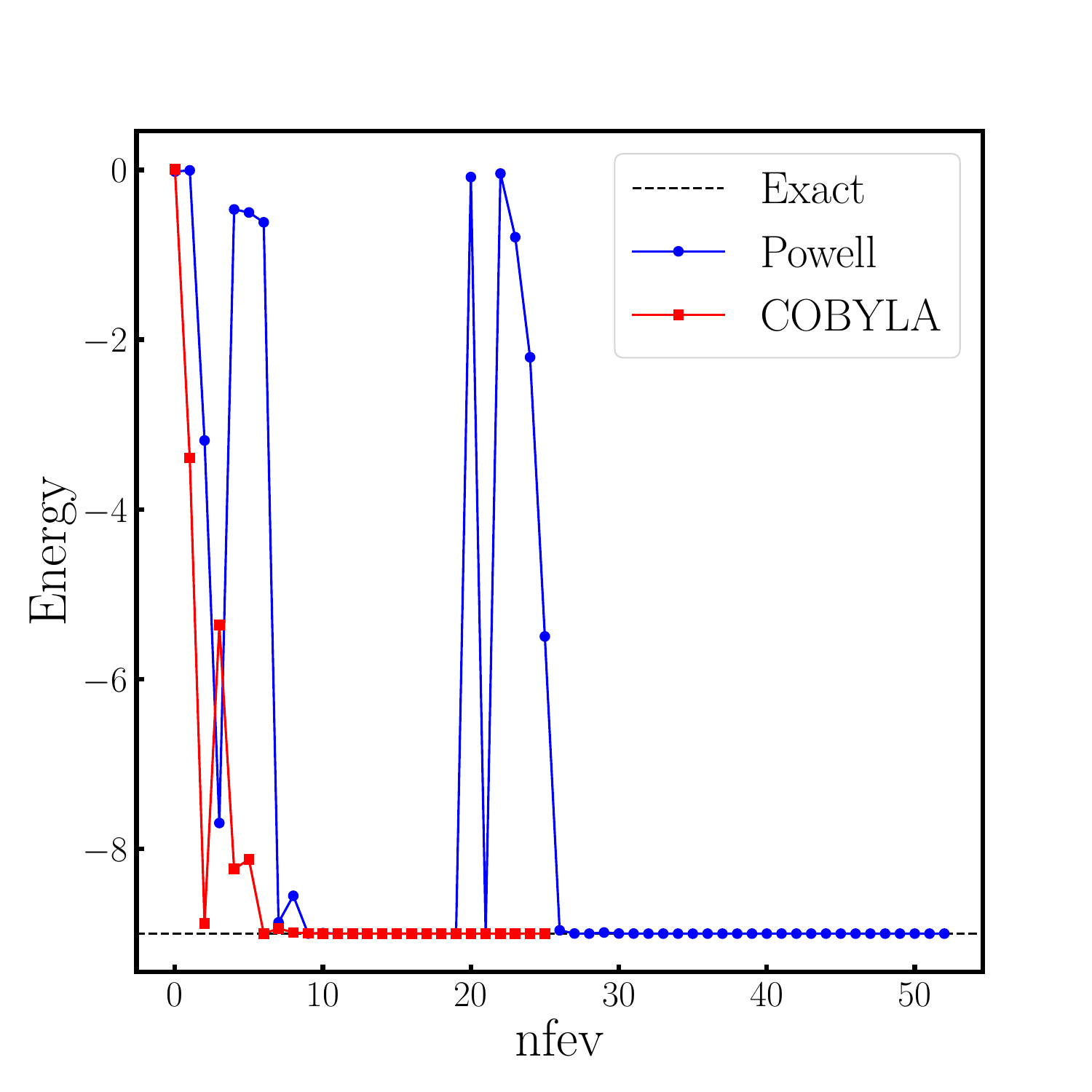}
\includegraphics[width=0.49\linewidth]{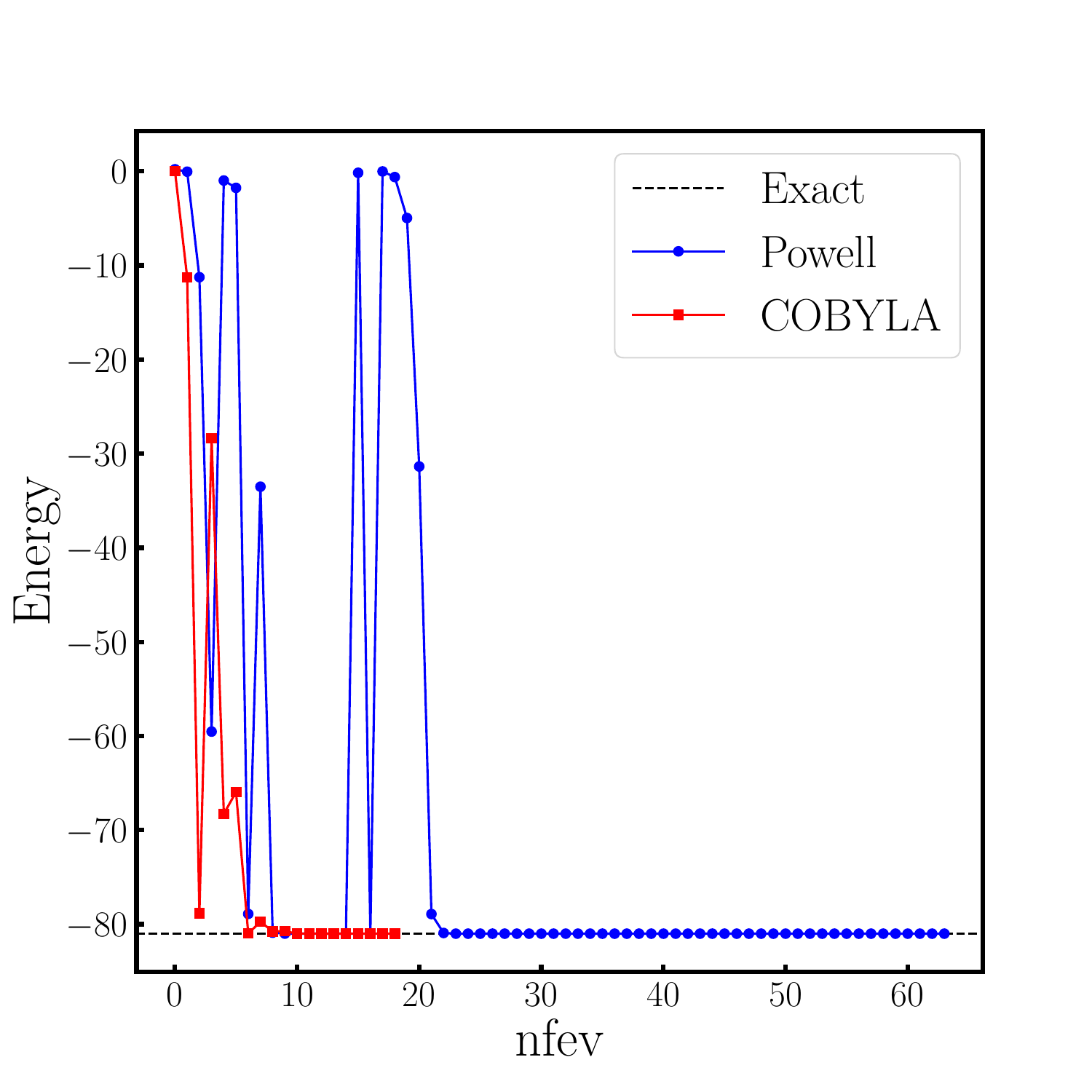}
\caption{\label{triSqr-nit}Plots of the energy with the number of function evaluations for the 
triangular plaquette (left) and the square plaquette (right).}
\end{center}
\end{figure}

We use a VQE algorithm on an exact simulator to check against analytical results 
for both the triangular and square plaquettes. Their variational ans\"{a}tze are given by
\begin{equation}\label{trisquareansz}
\begin{aligned}
    \ket{\psi(\theta)}_{\mathrm{tri}} =&\;\mathrm{CNOT}_{01}
    \cdot R_z(\theta)_0 \cdot H_0\\&\times 
    \mathrm{CNOT}_{23}
    \cdot R_z(\theta)_2 \cdot H_2 \\&\times 
    \mathrm{CNOT}_{45}
    \cdot R_z(\theta)_4 \cdot H_4  \ket{000101},\\ 
    \ket{\psi(\theta)}_{\mathrm{square}} &=\\&\;\mathrm{CNOT}_{01}
    \cdot R_z(\theta)_0 \cdot H_0\\&\times 
    \mathrm{CNOT}_{23}
    \cdot R_z(\theta)_2 \cdot H_2  \\&\times 
    \mathrm{CNOT}_{45}
    \cdot R_z(\theta)_4 \cdot H_4\\& 
    \times 
    \mathrm{CNOT}_{67}
    \cdot R_z(\theta)_6 \cdot H_6\ket{01010101}.
\end{aligned}
\end{equation}
As in the bubble case, a single variational parameter is sufficient to parameterize the states.
In \cref{triSqr-nit}, we present the convergence of VQE, demonstrating that the result converges 
to the exact analytical results. The convergence behavior is shown for two different classical 
optimizers, COBYLA and Powell, and in both cases COBYLA clearly outperforms the Powell method, 
consistent with what we observed for the bubble plaquette in \cref{VQE-ansz}. We compare both 
the ED results and the results from the VQE algorithm in \cref{FHilbert-TAB}, using the simple  
VQE Ans\"atze defined in \cref{trisquareansz}. 
%\cref{VQE-ansz} and \cref{VQE-ansz1}. 
The calculation for the VQE ansatz involves minimization of a single parameter, and could be done exactly. 

\begin{table*}[htbp]
\renewcommand{\arraystretch}{1.5}
\centering
\begin{tabular}{|c|c|c|c|c|c|c|}
    \hline
    \makecell{\textbf{Lattice} \\ \textbf{Geometry}} & \makecell{\textbf{N-terms} \\ \textbf{in} \(\mathcal{H}\)} & \textbf{N-qubits} & \multicolumn{2}{c|}{\makecell{\textbf{Circuit Depth} \\ \textbf{VQE}}} & \multicolumn{2}{c|}{\makecell{\textbf{Ground State} \\ \textbf{Energy}}} \\
    \cline{4-7}
    & & & \textbf{CNOT} & \textbf{p} & \textbf{ED} & \textbf{VQE} \\
    \hline
    Bubble    & 9   & 4  & 1  & 1 & -9.0  & -9.0  \\
    Triangle  & 27  & 6  & 1  & 1 & -9.0  & -9.0  \\
    Square    & 81  & 8  & 1  & 1 & -81.0 & -81.0 \\
    \hline
\end{tabular}
\caption{\label{FHilbert-TAB}Computational Resource for Bubble, Triangular, and Square plaquettes using simple VQE Ans\"atze.}
\end{table*}

\begin{table*}[htbp]
\centering
\resizebox{\textwidth}{!}{
\renewcommand{\arraystretch}{2.0}
\begin{tabular}{|c|c|c|c|c|c|c|c|c|c|c|c|c|c|}
    \hline
    \textbf{Lattice} & \makecell{\textbf{N-terms} \\ \textbf{in} $\mathcal{H}$} & \makecell{\textbf{No. of} \\ \textbf{Qubits}} & \multicolumn{2}{c|}{\textbf{CNOT Depth}} & \multicolumn{2}{c|}{\textbf{ED}} & \multicolumn{2}{c|}{$\textbf{VQE}_{\mathrm{lin}}$} & \multicolumn{2}{c|}{$\textbf{QAOA}_{\mathrm{shots}}$} & \multicolumn{2}{c|}{$\textbf{QAOA}_{\mathrm{EX}}$} \\
    \cline{4-13}
    & & &  $\textbf{VQE}_{\mathrm{lin}}$ & \textbf{QAOA} & $\textbf{E0}$ & $\textbf{E1}$ & $\textbf{E0}$ & $\textbf{E1}$ & $\textbf{E0}$ & $\textbf{E1}$ & $\textbf{E0}$ & $\textbf{E1}$ \\
    \hline
    $2\times 2$ & 41  & 4   & 15(p=5)    & 665(p=5)     & -0.6745  & -0.5957   & -0.6745  & -0.5957  & -0.6745 & -0.5957  & -0.6745  & -0.5957   \\
    $2\times 4$ & 164 & 8   & 385(p=55)  & 5592(p=12)   & -1.2809  & -1.2638   & -1.2809  & -1.2638  & -1.2809 & -1.2637  & -1.2809  & -1.2638   \\
    $2\times 6$ & 246 & 12  & 440(p=40)  & 12690(p=18)  & -1.9118  & -1.9062   & -1.9051  & -1.9044  & -1.9042 & -1.9019  & -        & -         \\
    $2\times 8$ & 328 & 16  & 
    %15(p=1) 
    - & 
    %7596(p=8)   
    -& -2.5464  & -2.5444   & -        & -        & %-2.3647 
    - & %-2.3363 
    - & -        & -         \\
    \hline
\end{tabular}
}
\caption{\label{SO3-comp-TAB}Computational resource and algorithms comparison for the pure $SO(3)$ QLM. In addition to QAOA calculations using a quantum circuit simulator, which we called $\mathrm{QAOA}_\mathrm{shots}$, we also have included a column that we used to provide checks on the QAOA where we used exact matrix multiplications to compute each parameter-dependent energy, which we called $\mathrm{QAOA}_{\mathrm{EX}}$. Since this operation scales exponentially with the volume, it is not possible to do this calculations without access to large memory nodes for the $2 \times 6$ and the $2 \times 8$ systems. 
}
\end{table*}

\subsection{Spontaneous Symmetry Breaking with VQE, VQD and QAOA}
 The last section demonstrated the extraction of the ground state (GS) in the electric
 basis, but since the basis is not gauge-invariant, noisy hardware leads to contributions
 from other Gauss Law sectors. To completely eliminate any traces of gauge-variant states, here we adopt the gauge-invariant basis described in \cref{sec:model}. 
 Using variational techniques in this basis, we aim at recovering not 
 only the GS, but also the lowest-lying excitation. The ground state of the model breaks 
 the $\mathbb{Z}_2$ charge conjugation symmetry spontaneously, leading to a scenario where the GS and the first excited state are identified with $\mathbb{Z}_2$-even and 
 $\mathbb{Z}_2$-odd quantum numbers respectively, and the gap between them closes exponentially with the volume. 
 
 We show that spontaneous symmetry breaking can be detected with two specific variational 
 techniques: the first uses a generic linearly connected VQE ansatz, and the second involves 
 QAOA-inspired ans\"{a}tze for both the GS and first excited state (see \cref{sec:methods}). 

\begin{figure}[h]
\begin{center}
\includegraphics[width=0.92\linewidth]{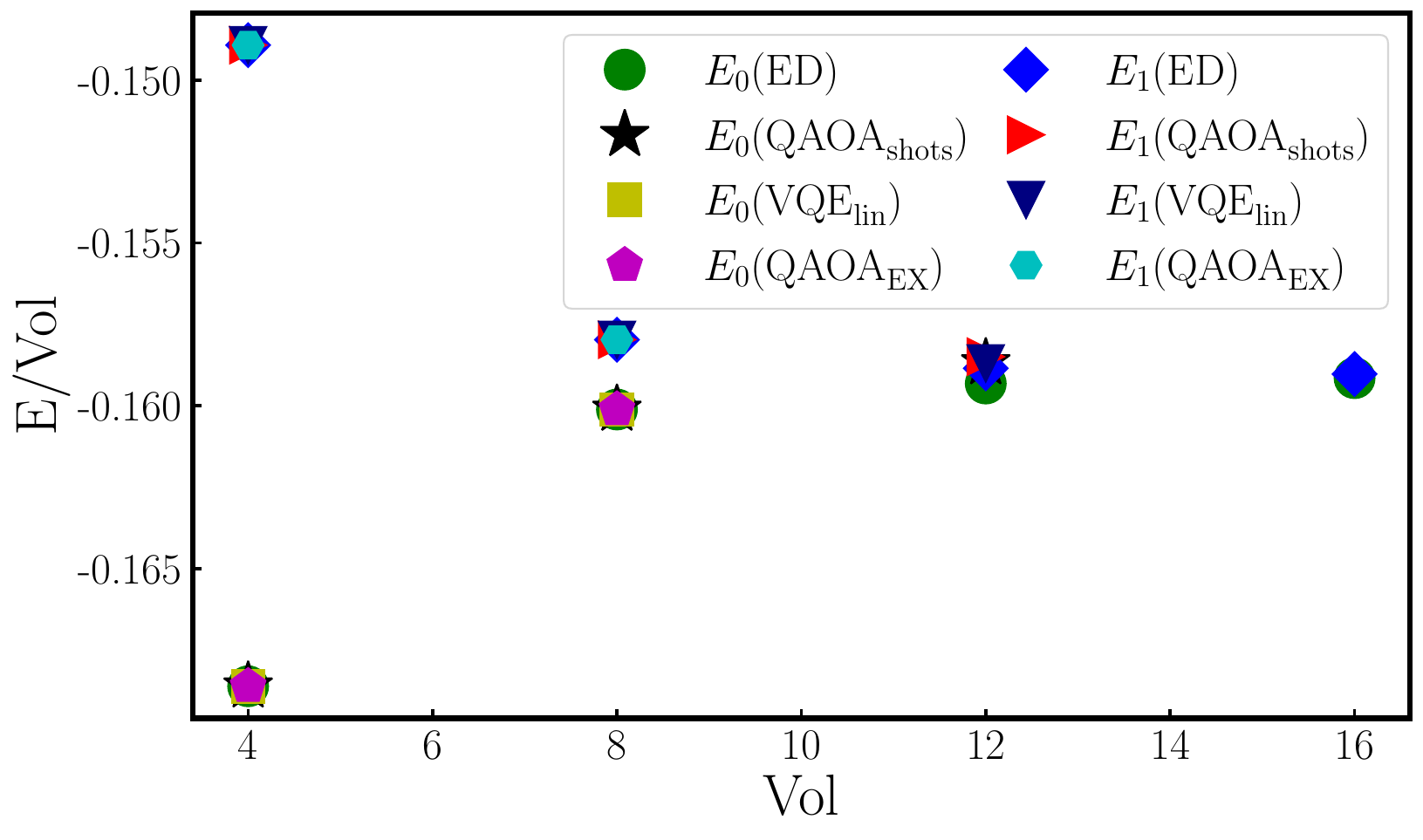}
\includegraphics[width=0.9\linewidth]{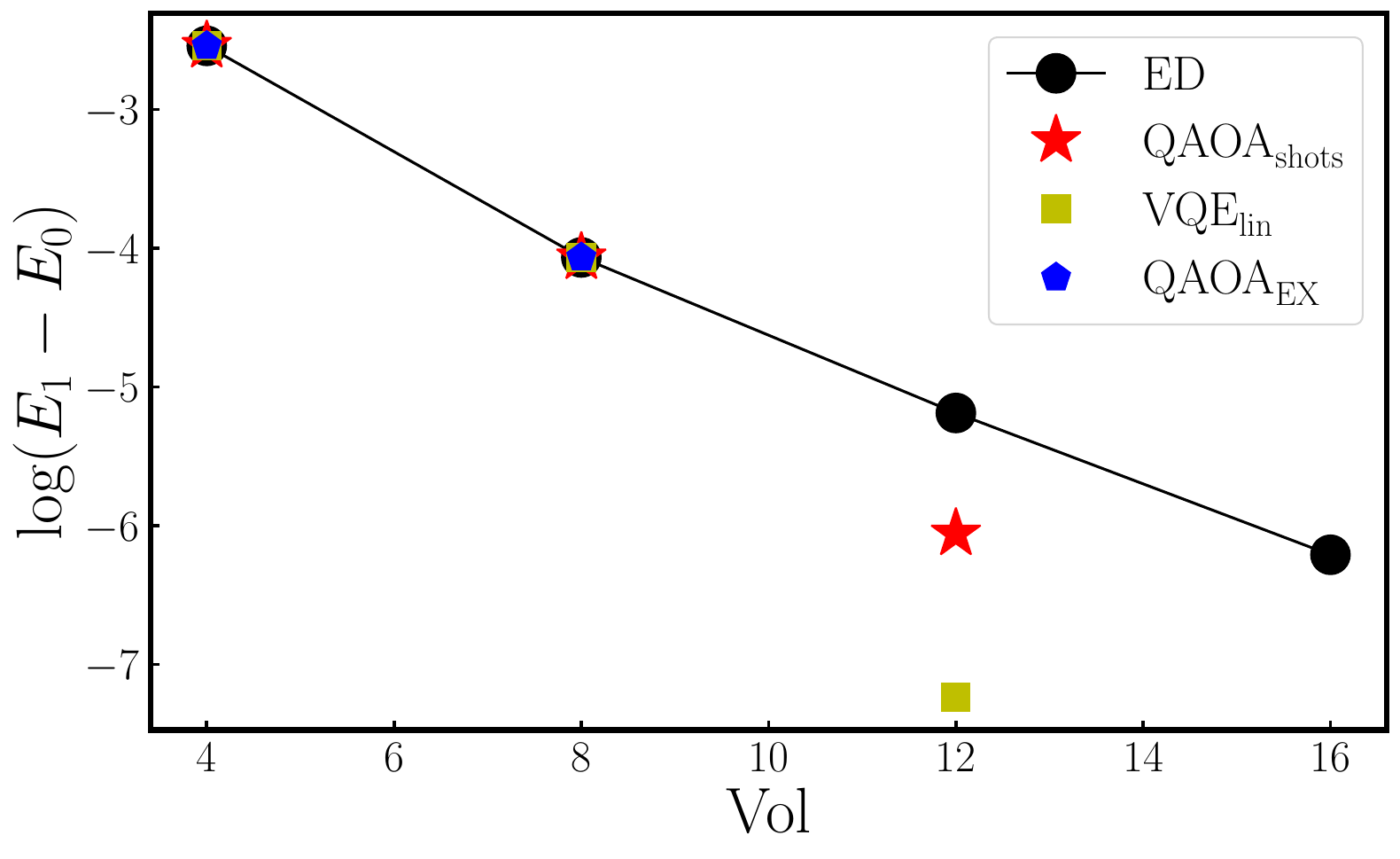}
\caption{\label{ed-vqd} (Top): Comparison of the results of ED, $\mathrm{QAOA}_{\mathrm{shots}}$, 
$\mathrm{VQE}_{\mathrm{lin}}$, and $\mathrm{QAOA}_{\mathrm{EX}}$ by calculating $E_0$ and $E_1$. (Bottom): The smallest energy gap ($E_1-E_0$) for different system sizes in the pure $SO(3)$ QLM in 2-d. Note that although the extraction of the energy looks good, extraction of the gap exposes the difficulty of the problem. Both the $\mathrm{VQE}_\mathrm{lin}$ and the $\mathrm{QAOA}_\mathrm{shots}$ have difficulty in convergence on classical hardware with the stated circuit depth on the 
$2 \times 6$ lattice. }
\end{center}
\end{figure}

{\bf $SO(3)$-symmetric gauge theory:} For the $SO(3)$ model, we compute the two lowest-lying 
energies, $E_0$ and $E_1$, at four lattice sizes: $2\times 2, 2\times 4, 2\times 6$, and 
$2\times 8$, in order to understand the behavior of the system in the thermodynamic limit. We 
use the gauge-invariant basis in order to study the system at these larger lattices, and we 
get the $E_0$ and $E_1$ using two different variational techniques. The first technique employs 
the linearly connected VQE ansatz defined in \cref{eq:21} to get the ground state, and then uses 
VQD with the optimized VQE state in order to get the first excited state. We use the SLSQP 
optimizer for these variational methods. The second technique uses the QAOA ansatz defined in 
\cref{qaoagen} and \cref{qaoags} to get the ground state, and then for the first excited state 
it uses a QAOA-inspired ansatz with \cref{qaoagen} and \cref{qaoaes}. We used the L-BFGS-B optimizer. 

\cref{ed-vqd} summarizes the results, with the top panel showing $E_0$ and $E_1$ computed using 
the two variational methods as well as the ED results, and the bottom panel showing the energy gap 
$E_1-E_0$ computed using these methods as a function of the volume in a semi-log plot. We use 
$\mathrm{VQE}_\mathrm{lin}$ to denote simulator linearly connected variational results, 
$\mathrm{QAOA}_\mathrm{shots}$ to denote simulator QAOA results, and $\mathrm{QAOA}_\mathrm{EX}$ 
to denote QAOA results where each parameter-dependent energy is computed exactly rather than with 
simulator shots. For the VQE linearly connected ansatz and QAOA approaches, we have completed calculations 
on $2\times 2$, $2\times 4$, and $2\times 6$ lattices. On the top panel of \cref{ed-vqd}, we note 
good visual agreement between the methods with the data points on top of each other. 
\cref{SO3-comp-TAB} gives the numerical values: even in the most difficult case of the $2 \times 6$ 
lattice, the energy results agree to better than 0.1\%. By plotting $\mathrm{log}(E_1-E_0)$ as a function 
of volume, the bottom panel of \cref{ed-vqd} shows a gap that closes exponentially with the system volume. 
With the data from three lattice sizes, we thus have evidence of spontaneous symmetry breaking in the 
ground state by making use of quantum-circuit-friendly variational ans\"{a}tze. Here, we note that in 
ED calculations, the computational time grows exponentially with the volume of the system. However, 
in the quantum algorithms, the computational scaling is determined by the number of layers necessary 
to make the ansatz sufficiently expressive to capture the ground state. While we cannot make a statement 
about the scaling of the number of layers necessary to capture this gapless physics, we see from the 
top of \cref{ed-vqd} that the QAOA energies are close to the ED energy with linear scaling in the number 
of layers (for VQE we are able to achieve the same results without scaling up the layers between 
$2\times4$ and $2\times 6$). Thus, at least for these system sizes, we are able to get a qualitative sense 
that the system will be gapless in the thermodynamic limit without using an ansatz with depth that 
scales exponentially with the volume. This would put the scaling in line with quantum Monte Carlo 
algorithms when there is no sign problem, such as projector quantum Monte Carlo.

\begin{figure}[!htb]
\begin{center}
\includegraphics[width=0.98\linewidth]{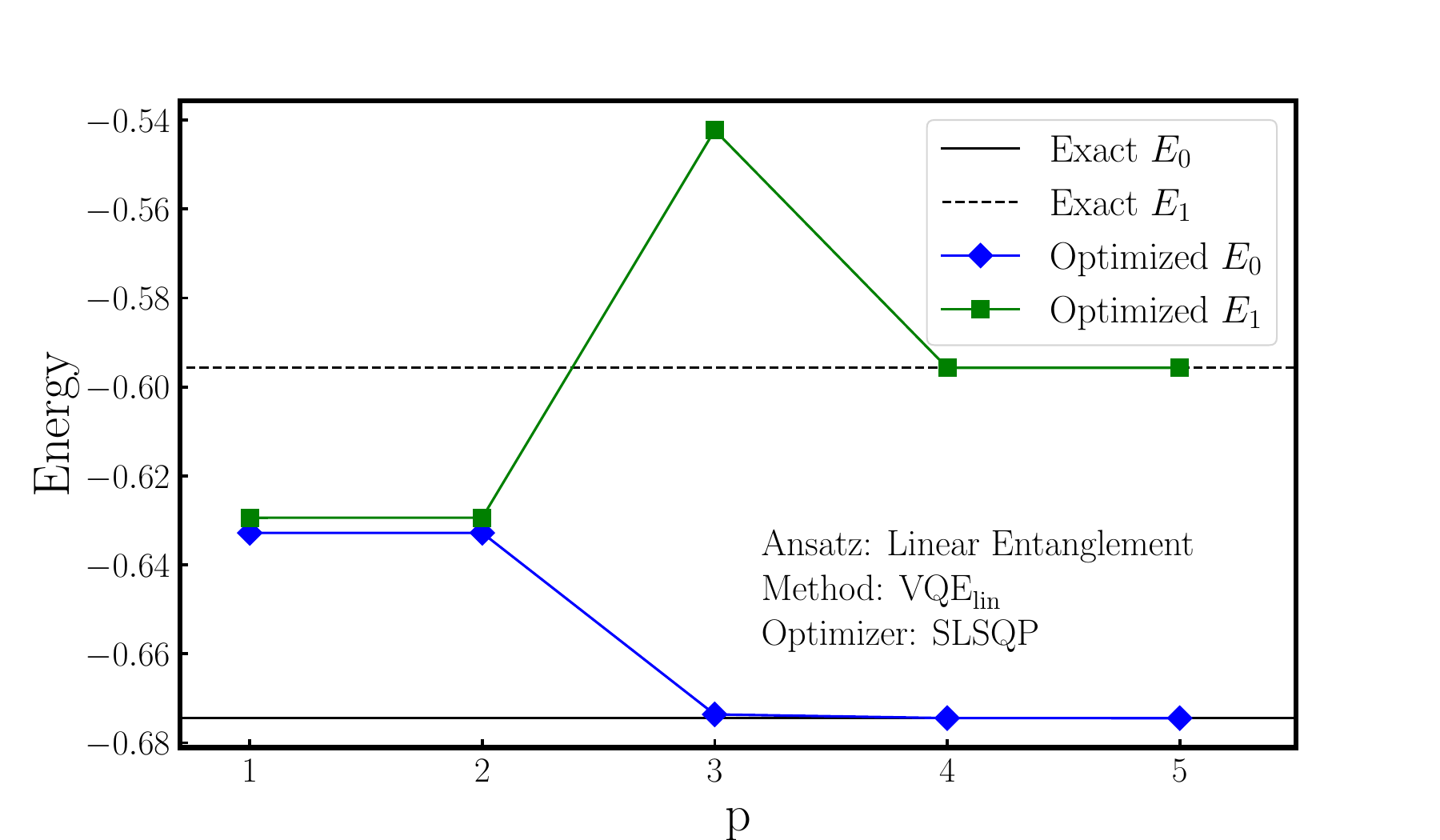}
\includegraphics[width=0.98\linewidth]{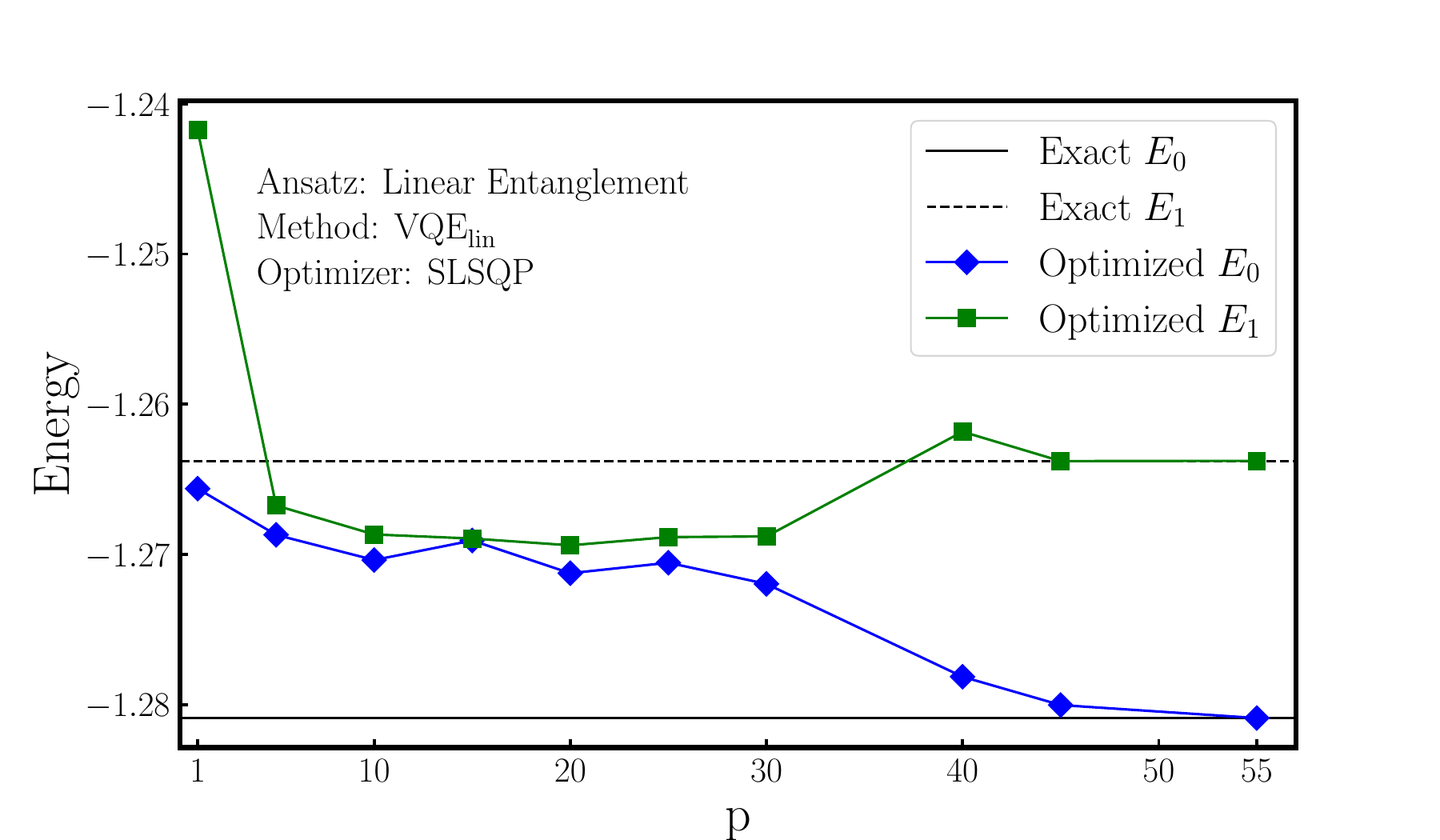}
\caption{\label{vqd-so3} (Top): Convergence of the ground state energy as a 
function of the circuit depth ($p$) on a $2\times 2$ lattice using the VQD 
algorithm ($\mathrm{VQE}_\mathrm{lin}$). If the starting state is not created 
in a symmetry resolved way, the ground and the first excited states mix for 
lower circuit depths. However, with an increase in the circuit depth, the VQD 
optimizer can resolve the symmetry between these states, allowing the ground 
state energy to converge properly.
(Bottom): The convergence of the first excited state energy for the $2\times 4$ 
lattice using the VQD algorithm. The convergence occurs only at a much larger 
circuit depth compared to the $2\times 2$ lattice.}
\end{center}
\end{figure}

For the VQE algorithm with the linearly-entangled ansatz, it is difficult to 
design a parameterized ansatz that respects the different symmetries of the 
ground state and the first excited state. In \cref{vqd-so3}, we show the 
energy convergence as a function of circuit depth ($p$) for both the ground 
state and the first excited state energy for $2\times2$ and $2\times4$ 
lattices. For the $2\times2$ lattice, the ground state energy gets closer to 
the exact value as the circuit depth increases. For the first excited state, 
we see large fluctuations up until $p=4$. This happens because, with smaller p, 
the VQD optimizer tends to mix the ground state and the first excited state, 
causing an overlap that may push $E_1$ above the exact value. As the circuit 
depth increases, the optimizer is able to better resolve the symmetries 
leading to a good convergence. We observe a similar behavior for the 
$2 \times 4$ lattice, where the convergence requires a much larger 
circuit depth ($p$).

From \cref{SO3-comp-TAB} we also note the significantly larger circuit 
depth necessary for QAOA versus the linearly-entangled VQE ansatz. This 
is a disadvantage of the QAOA as we can see it is still possible to resolve 
the energy gap with the simpler linearly entangled VQE ansatz. However, 
an advantage that the QAOA-inspired ans\"{a}tze offer over VQE/VQD is 
that the variational algorithm to find the first excited state is 
independent from the estimation of that of the ground state, which 
removes a potential source of error inherent to VQE/VQD for low-lying 
excitations.

\begin{widetext}    
\begin{figure*}[ht]
\begin{center}
\includegraphics[width=0.325\linewidth]{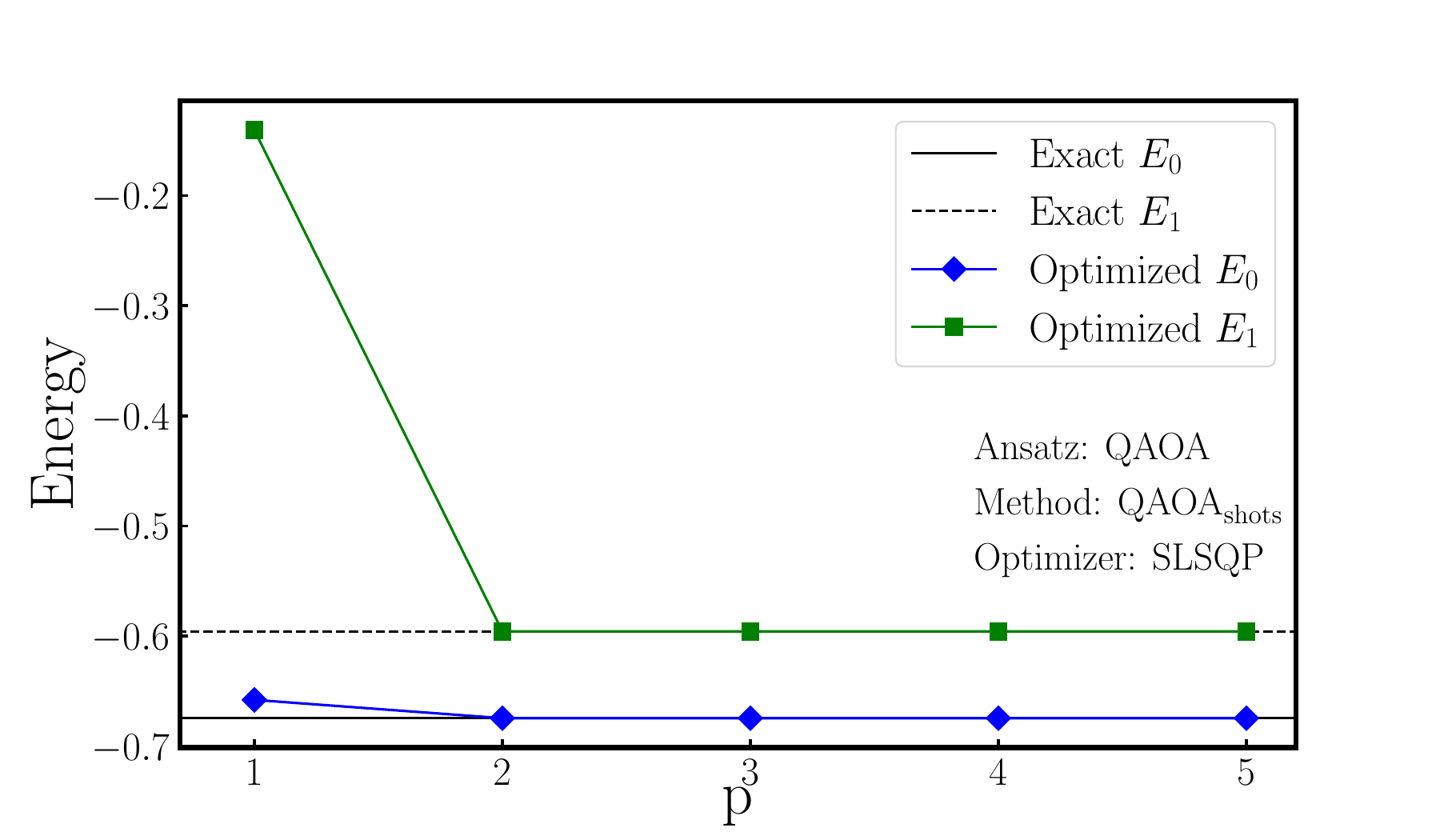}
\includegraphics[width=0.325\linewidth]{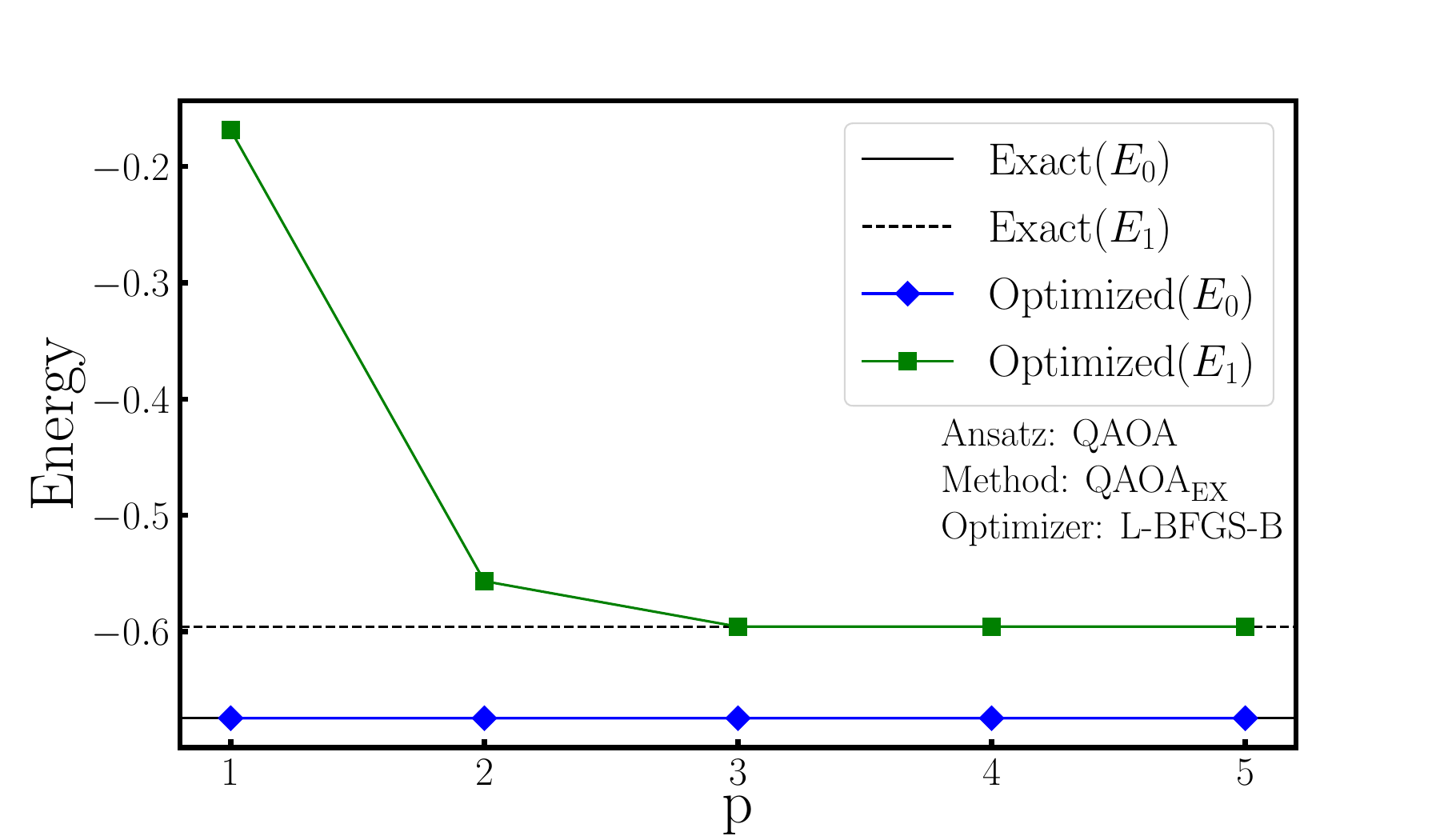}
\includegraphics[width=0.325\linewidth]{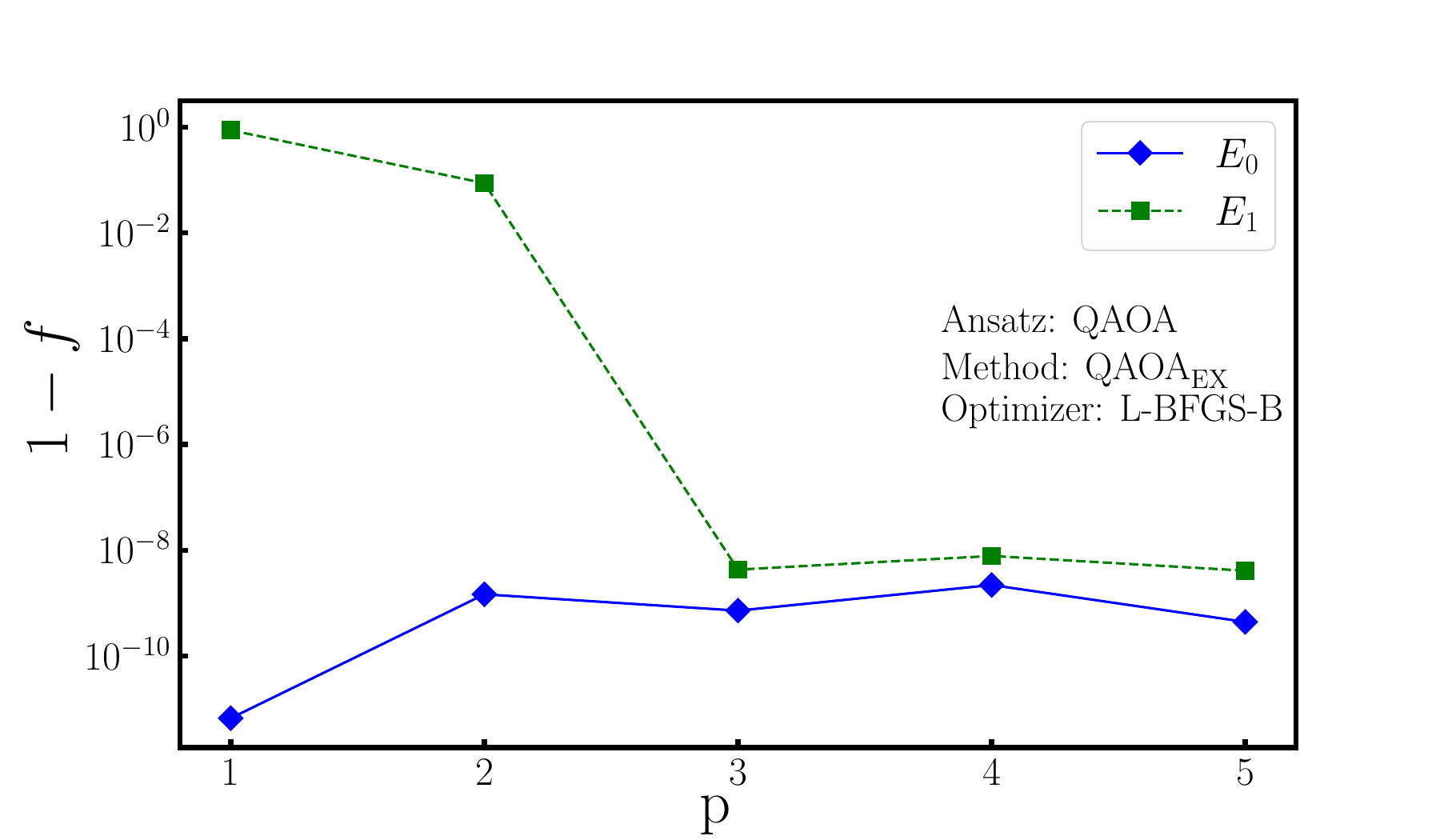} 
\includegraphics[width=0.325\linewidth]{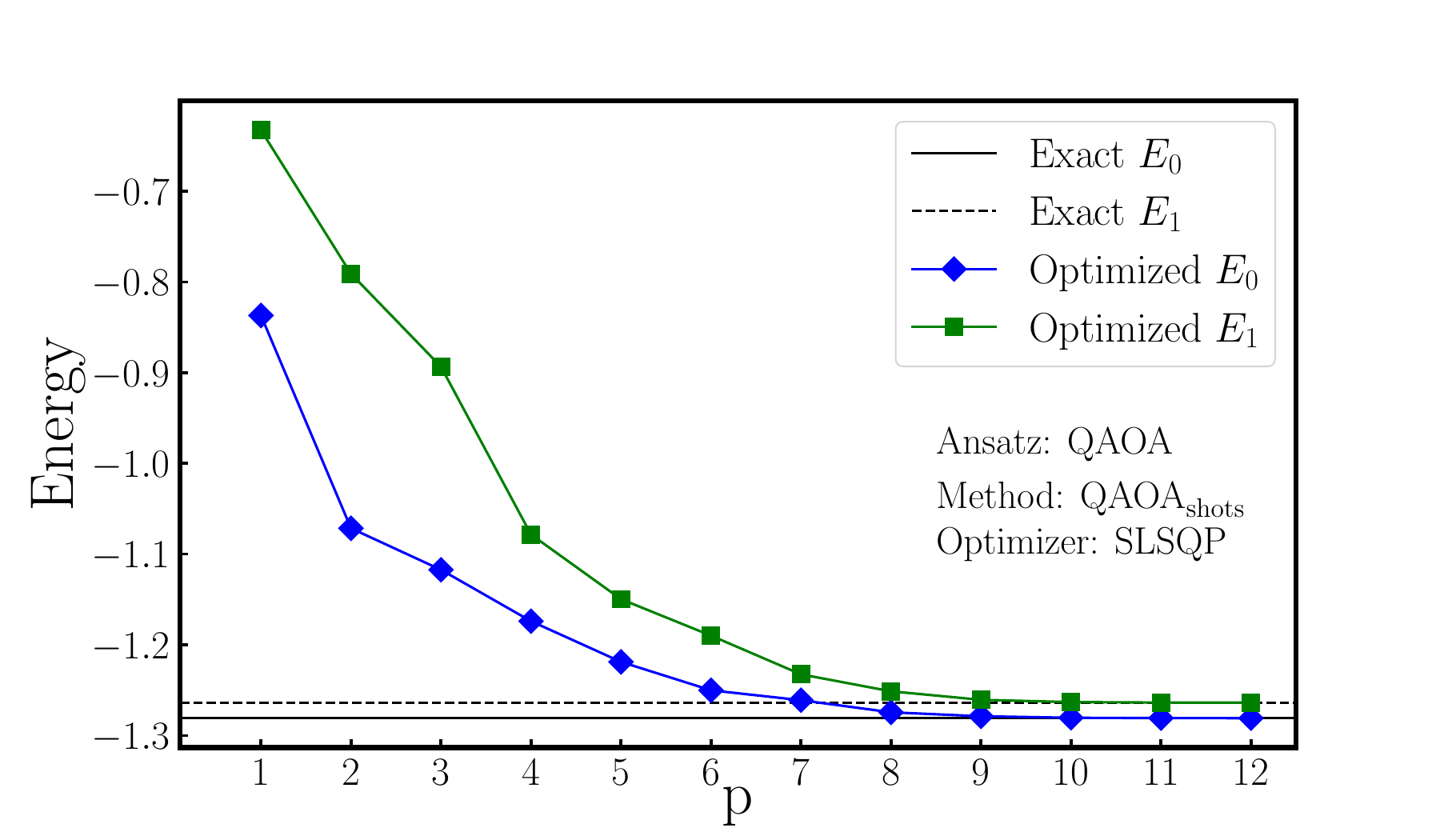}
\includegraphics[width=0.325\linewidth]{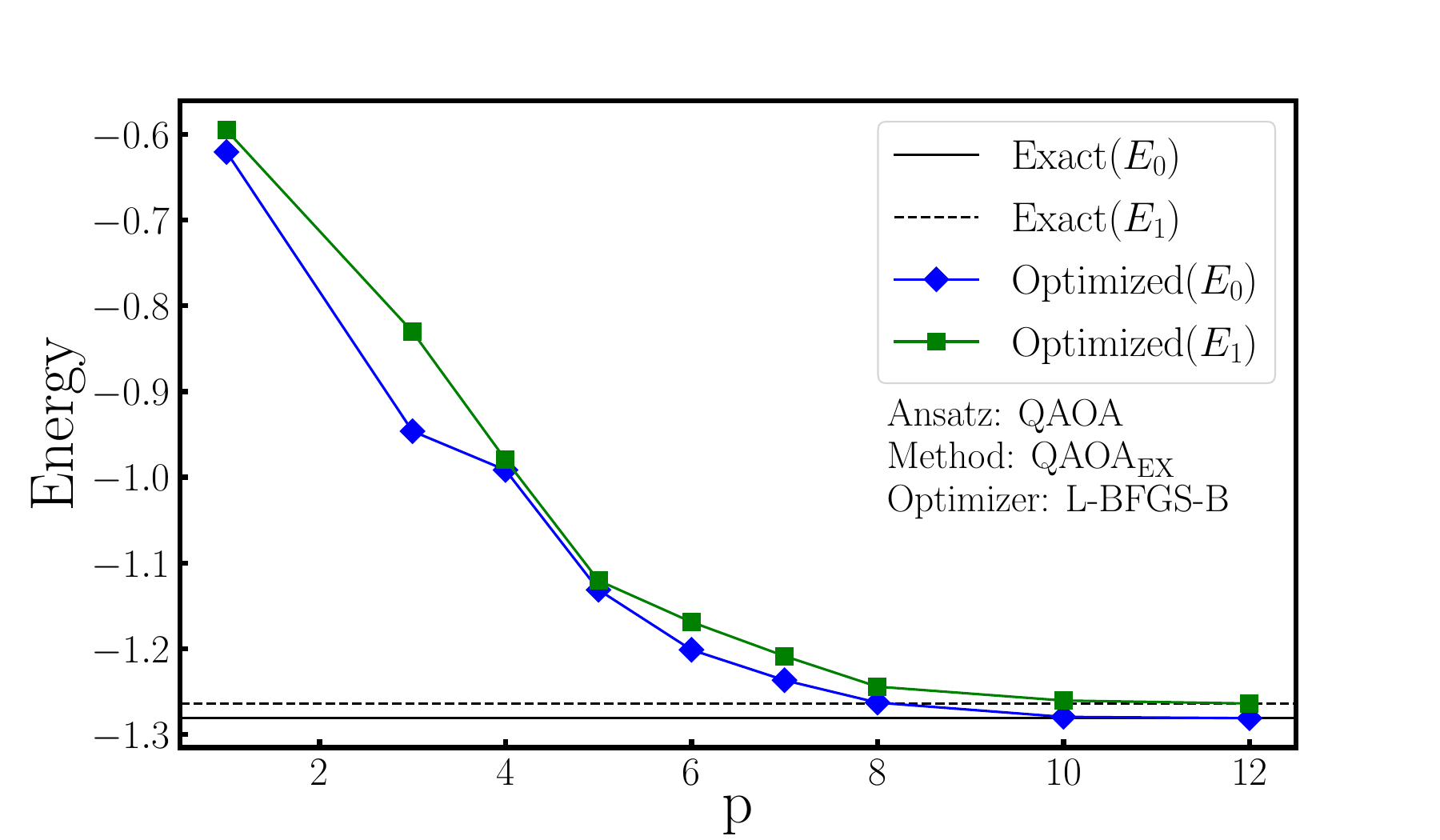}
\includegraphics[width=0.325\linewidth]{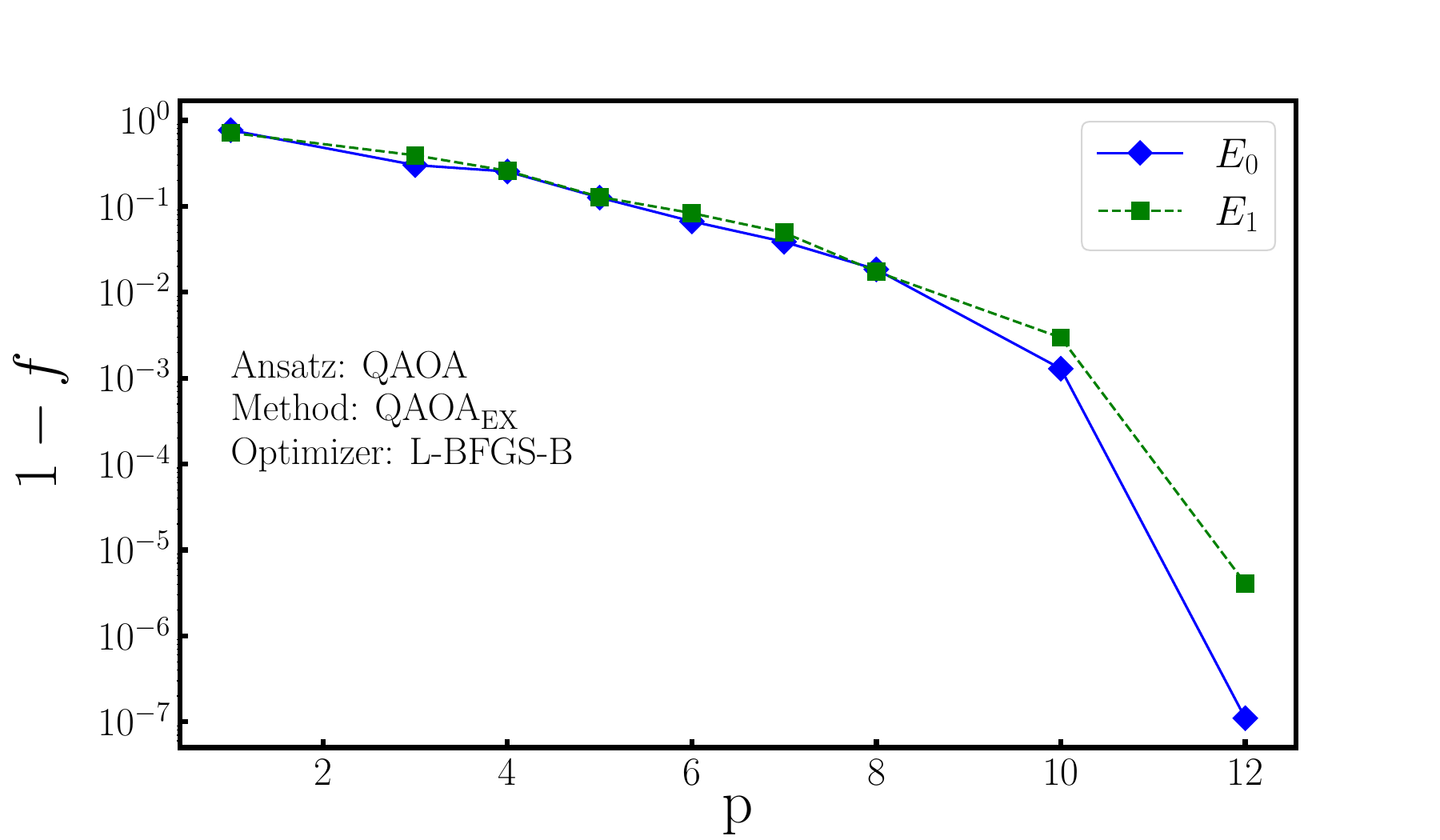}
\caption{\label{qaoa-res} Results for the energy and the fidelity as a function of
circuit depth for two different lattice sizes: the top column is the result for $2 \times  2$,
while the bottom column is for $2 \times 4$ lattice. The left panel shows the 
ground state and the first excited state energy as a function of circuit depth ($p$)
using the VQE algorithm ($\mathrm{QAOA}_\mathrm{shots}$); the middle panel shows the 
convergence of the ground state and the first excited state energy using the QAOA 
algorithm ($\mathrm{QAOA}_\mathrm{EX}$), and the right panel shows the in-fidelity with 
the circuit depth using the QAOA algorithm. Corresponding figures showing the 
convergence (restricted to our circuit depths) is shown in \cref{so3-latt2x6} 
\cref{sec:convg2t6}.
}
\end{center}
\end{figure*}
\end{widetext}

In \cref{qaoa-res}, we show the results obtained using the QAOA-inspired
symmetry resolved ans\"{a}tze for both the ground and the excited states. 
We note that while the energies themselves show agreements to better than 
0.1\%, the fidelities for the both the wavefunctions on the larger lattice 
are still several orders of magnitude larger than the smaller lattice. We 
point out that such considerations are important to be kept in mind while 
deciding the application for the quantum simulation methods. Clearly, when 
evaluating expectation values of local operators (such as order parameters), 
getting a few percent accuracy on the ground state is perhaps sufficient, 
while the computation of gaps in symmetry broken phase could be expensive. 
On the other hand, if the gaps have a $O(J)$ scaling, then the percent 
accuracy is enough. In summary, both methods seem to successfully capture 
the energies with good precision (better than 0.1\%) using 
at most linear growth in the number of layers of the VQE/VQD ans\"{a}tze 
as a function of system size.

\begin{figure}[!htb]
\begin{center}
\includegraphics[width=0.98\linewidth]{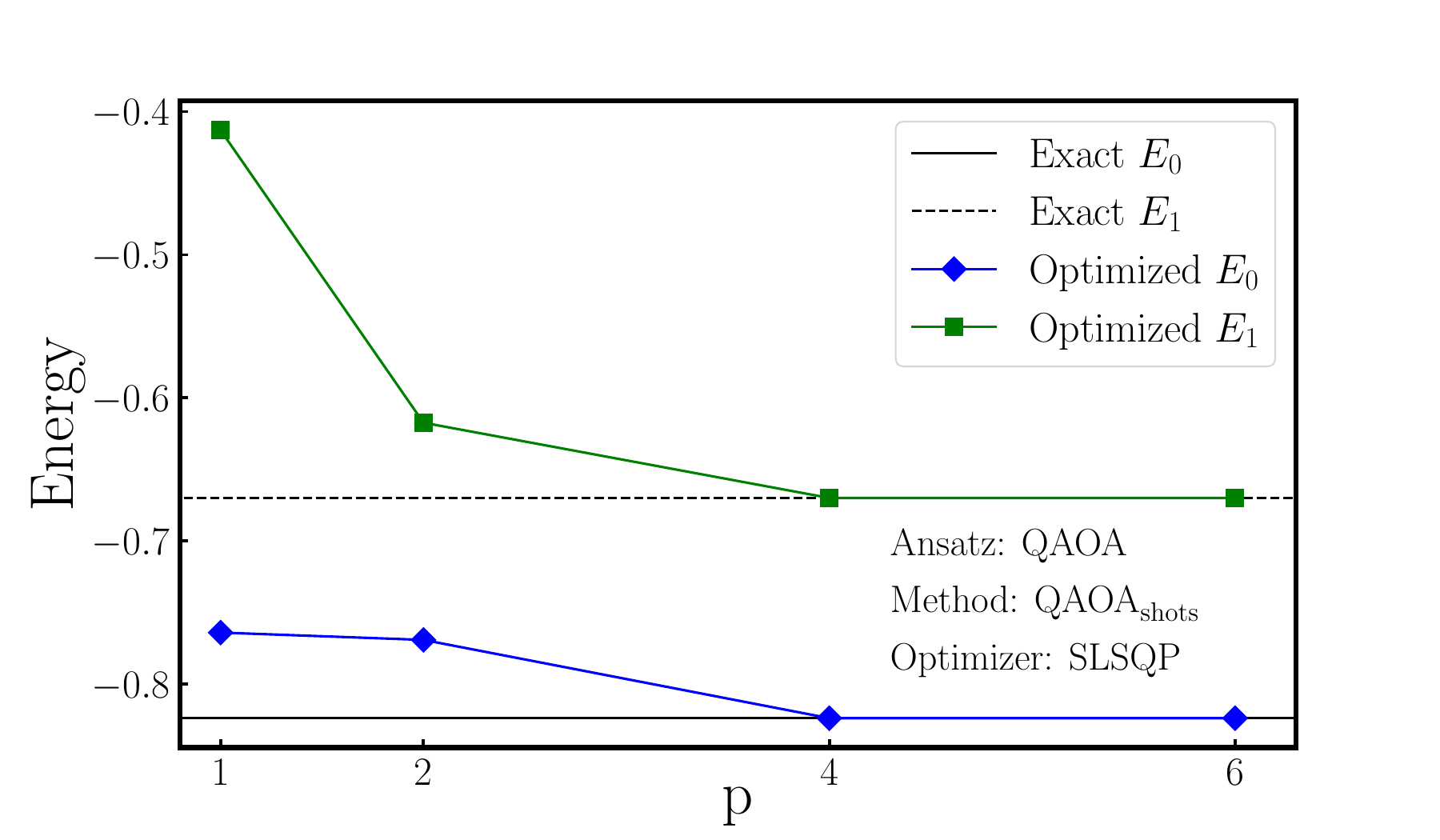}
\includegraphics[width=0.98\linewidth]{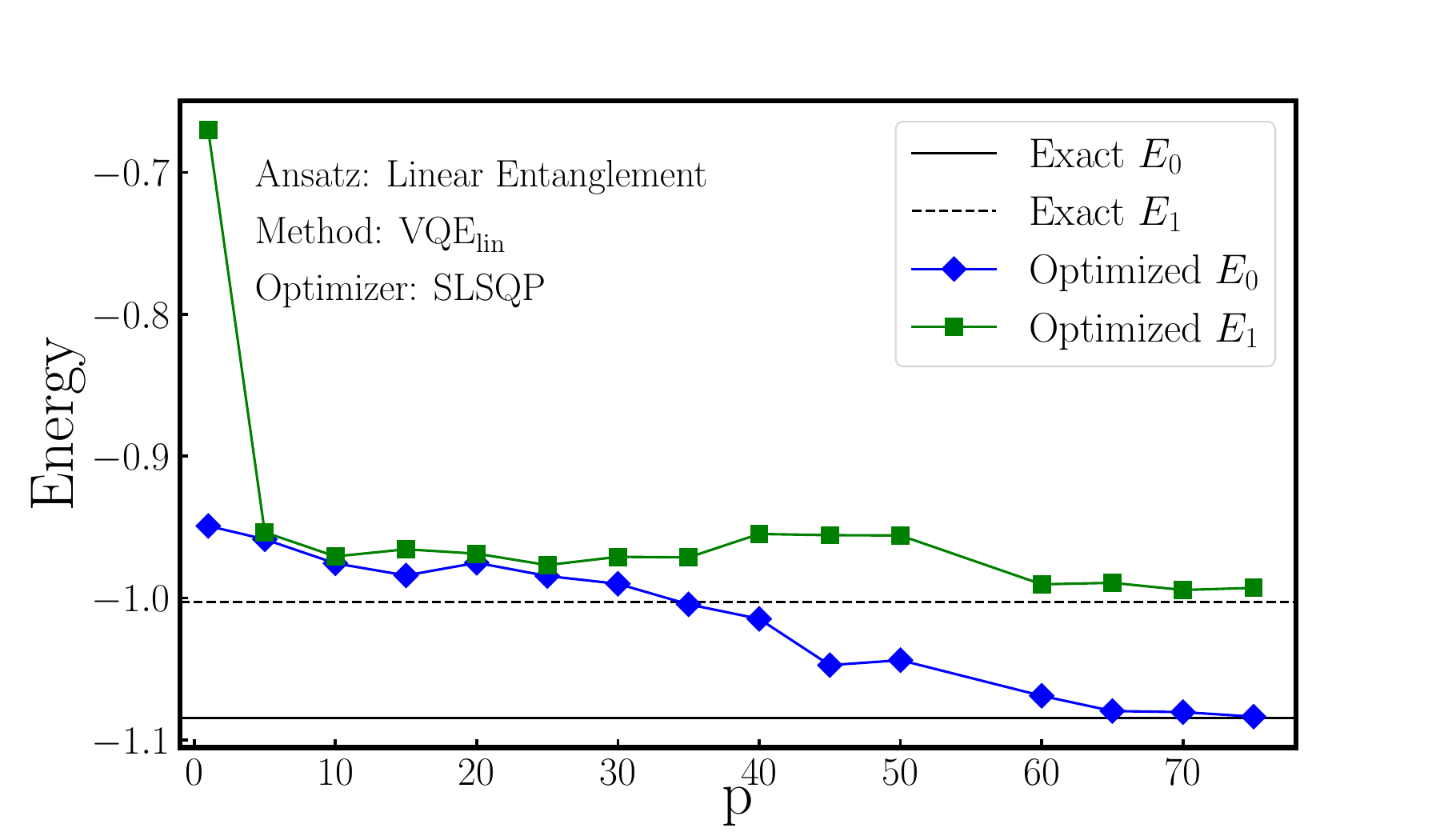}
\caption{\label{fig:twoDlatts} 
(Top): Convergence plots for the ground state and the 
first excited state for the $3 \times 2$ lattice, which share the same symmetry breaking 
patterns as the other lattices. The convergence is very fast for initial states chosen
in appropriate symmetry sectors. (bottom) For the $3 \times 3$ lattice, there is no
symmetry breaking between the ground and first excited state, and we required a circuit
depth of $600$ (75 layers each of linearly entangled ansatz, with CNOT gate of depth 8) 
for a decent convergence. The total number of terms in the Hamiltonian is 729, and the 
VQE/VQD values are $E_0 = -1.08364012$ (compare with the exact value $E_0 = -1.08475104$), and $E_1 = -1.0028875$ (against the exact value $E_1 = -0.99288576$).}
\end{center}
\end{figure}

At this point, we are in a good position to conjecture the method of choice
as the lattice becomes wider to a two-dimensional geometry from a ladder one. The next set of lattices are $3 \times 2$ and $3 \times 3$. The former lattice has the same symmetry breaking between the ground state and the first excited state as the ones with both even extents, and therefore, we can use the VQE algorithm
on the QAOA inspired starting ansatz for the ground state and the first excited states respectively, leading to very fast convergence, as illustrated in \cref{fig:twoDlatts} (top). However, the ground and the first excited states of the $3 \times 3$ lattice do not have any particular symmetry properties, and therefore require the VQD in conjunction with the VQE and large circuit depths. The details for the two cases are provided in the caption of \cref{fig:twoDlatts}.

\section{Benchmark computations for the Transverse Field Ising model}
\label{sec:results2}

In order to better understand the challenge of the resolving energy gaps with 
variational techniques, and in particular test our QAOA-inspired ans\"{a}tze 
in a different context, we also perform calculations for the transverse field Ising
model (TFIM), which is a paradigmatic model hosting a gapped and $Z_2$ broken phase,
separated by a second order phase transition \cite{PhysRevLett.25.443}. 
Moreover, this calculation contrasts the complexity involved in
the simulation of the fully dynamical non-Abelian SO(3) lattice gauge theory in the
last section. The Hamiltonian for TFIM is given by
\begin{equation}
    H = -J \sum_{\braket{i,j}} \sigma^3_i \sigma^3_j - h_x \sum_{i} \sigma^1_i\textcolor{blue}{,}
\end{equation}
with $J$ the interaction strength between adjacent spins, and $h_x$ the external 
magnetic field. We consider $|J|=1$ and explore three different regimes by varying 
the value of $h_x$. When $|h_x|<|J|$, the ground state breaks the spin-flip symmetry 
spontaneously (ferromagnetic phase). For a finite system, we expect the lowest mass gap 
to scale as $\exp(-\alpha V)$. When $|h_x|=1$, the system undergoes a quantum phase 
transition, and for $|h_x|>|J|$, the system is in a gapped phase (paramagnetic phase).

We use the QAOA algorithm to find the smallest mass gap for different lattice sizes, 
dividing the Hamiltonian for our variational ansatz into
\begin{equation}
\begin{aligned}
H_1 &= -J \sum_{\left\langle ij\right\rangle} \sigma^3_i \sigma^3_j , \qquad
H_2 = -h_x \sum_i \sigma^1_i.
\end{aligned}
\end{equation}
Note that here we set $H_1$ to be the interaction terms rather than the magnetic field 
terms which we used in the $SO(3)$ example, and indeed which are typically used for QAOA 
for the Ising model. The QAOA ansatz is then
\begin{equation}
\ket{GS}_{\mathrm{QAOA}} =\prod_{k=1}^p e^{iC_{1,k} H_1} e^{iC_{2,k} H_2} 
\ket{\psi_A},
\end{equation}
where to obtain the ground state we set 
$\ket{\psi_A} = 1/\sqrt{2} (\ket{\uparrow\uparrow... \uparrow} 
+ \ket{\downarrow\downarrow... \downarrow},$ the GHZ state. If we had built 
$\ket{\psi_A}$ using the transverse magnetic field terms instead (as is typically done), 
we would have $\ket{\psi_A} = \prod_i H_i\ket{\uparrow\uparrow... \uparrow}$, where 
$H_i(=1/\sqrt{2}(\sigma^1_i+\sigma^3_i))$ is the Hadamard operator acting on qubit $i$. 
The results for this other ansatz are given in the appendix in \cref{tfim-latt10}.

\begin{table*}[htbp]
\renewcommand{\arraystretch}{1.6}
\begin{tabular}{|c|c|c|c|c|cc|cc|cc|}
    \hline
    \textbf{Lattice} & \makecell{\textbf{N-terms} \\ \textbf{in} \(\mathcal{H}\)} & \textbf{N-qubits} & \multicolumn{2}{c|}{\makecell{\textbf{Circuit Depth} \\ $\textbf{QAOA}_{\mathrm{EX}}$}} & \multicolumn{2}{c|}{\makecell{\textbf{$h_x=0.5$} \\ ($E_1 - E_0$)}} & \multicolumn{2}{c|}{\makecell{\textbf{$h_x=1.0$} \\ ($E_1 - E_0$)}} & \multicolumn{2}{c|}{\makecell{\textbf{$h_x=1.5$} \\ ($E_1 - E_0$)}} \\
    \cline{4-11}
    & & & \textbf{CNOT} & \textbf{p} & \textbf{ED} & $\textbf{QAOA}_{\mathrm{EX}}$ & \textbf{ED} & $\textbf{QAOA}_{\mathrm{EX}}$ & \textbf{ED} & $\textbf{QAOA}_{\mathrm{EX}}$ \\
    \hline
    4  & 8  & 4  & 40  & 4 & 0.03549 & 0.03549 & 0.39782 & 0.39782 & 1.15446 & 1.15446 \\
    6  & 12 & 6  & 56  & 4 & 0.00689 & 0.00689 & 0.2633  & 0.2633  & 1.0523  & 1.0523  \\
    8  & 16 & 8  & 108 & 6 & 0.00146 & 0.00146 & 0.19698 & 0.19698 & 1.01945 & 1.01945 \\
    10 & 20 & 10 & 132 & 6 & 0.00032 & 0.0003  & 0.1574  & 0.1574  & 1.00757 & 1.00757 \\
    \hline
\end{tabular}
\caption{\label{TFIM-comp-TAB}Computational Resource and Algorithms Comparison for 1D-TFIM}
\end{table*}

\begin{figure}[h]
\begin{center}
\graphicspath{{../Notes/TFIM/}}
\includegraphics[width=1.05\linewidth]{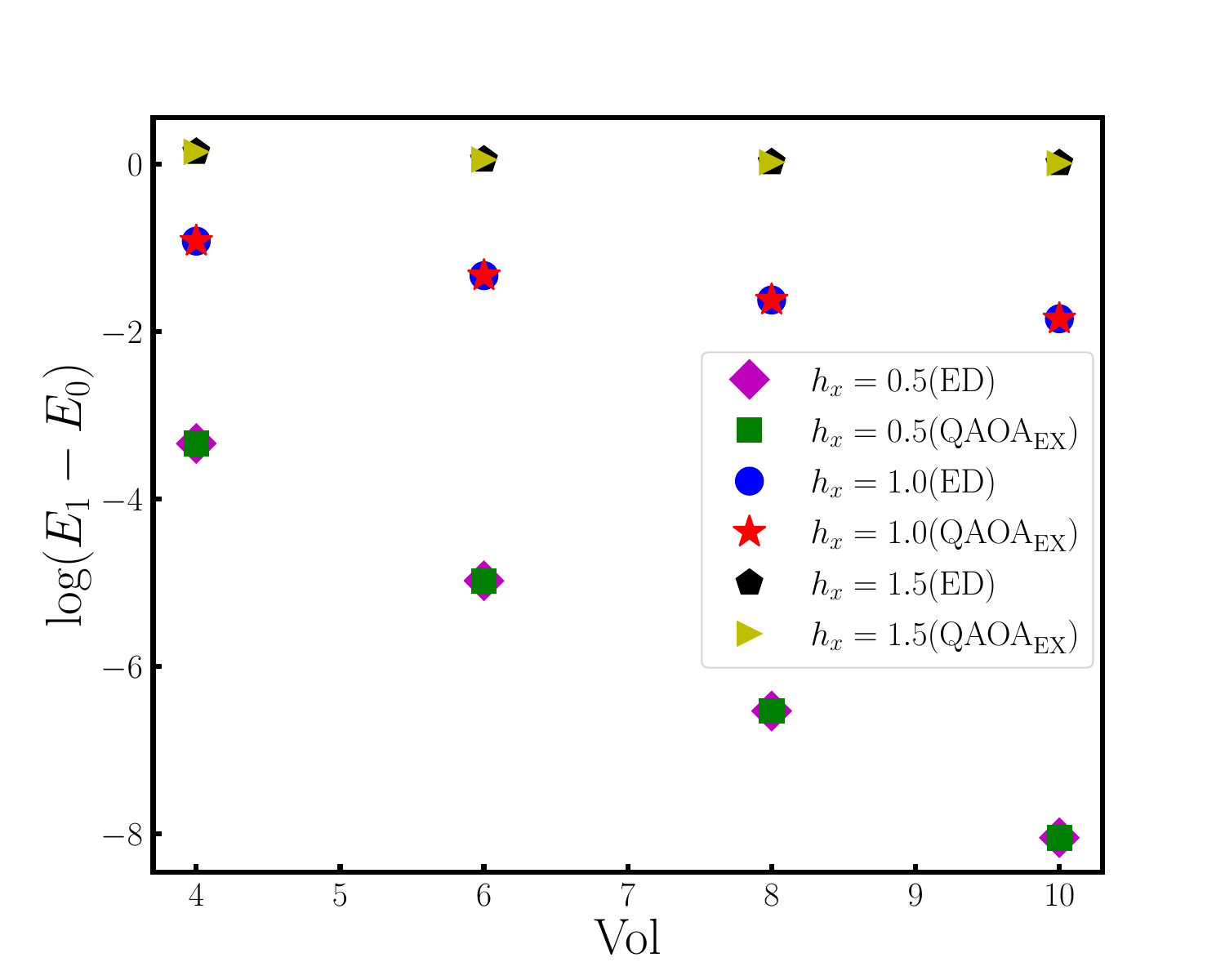}
\caption{\label{TFIM-qaoa}Plot of the energy difference for transverse field 
Ising model (TFIM) in 1-d.}
\end{center}
\end{figure}

The first excited state wavefunction of $H_1$ is antisymmetric under spin-flip symmetry. 
To compute the energy of the first excited state, we use the state 
$1/\sqrt{2} (\ket{\uparrow\uparrow... \uparrow} - \ket{\downarrow\downarrow... \downarrow})$ 
as the initial ansatz state $\ket{\psi_A}$. We compare both the ED results and the results 
from the quantum algorithm in \cref{TFIM-comp-TAB}. We point out that for the same number 
of qubits (8), the CNOT circuit depth for the TFIM is 56, while it is 5592 for the $SO(3)$ 
model. Moreover, the desired results in the TFIM are easily obtained with circuit depths 
of $p \simeq 6$, while at least double the circuit depth is necessary for the $SO(3)$ model. 
The energy gap of the TFIM model is shown in \cref{TFIM-qaoa}, where we can see that the 
QAOA-inspired algorithm accurately measures the energy gap in all regimes, matching the 
ED results. We show the performance of QAOA in \cref{tfim-latt10} in \cref{sec:tfim}, by    
plotting the ground state and first excited state energies, as well as the in-fidelity 
($1-f$) with circuit depth ($p$) for a 1-d lattice with 10 sites. While much work has 
already been done to compute the ground state of the 1D TFIM using QAOA 
\cite{Sun:2022zxi, PhysRevA.107.042418}, this extension of QAOA to compute the first 
excited state provides an additional proof of principle of the QAOA-inspired excited 
state ansatz that we have introduced in this paper.

\begin{figure}[h]
\centering
\includegraphics[width=0.98\linewidth]{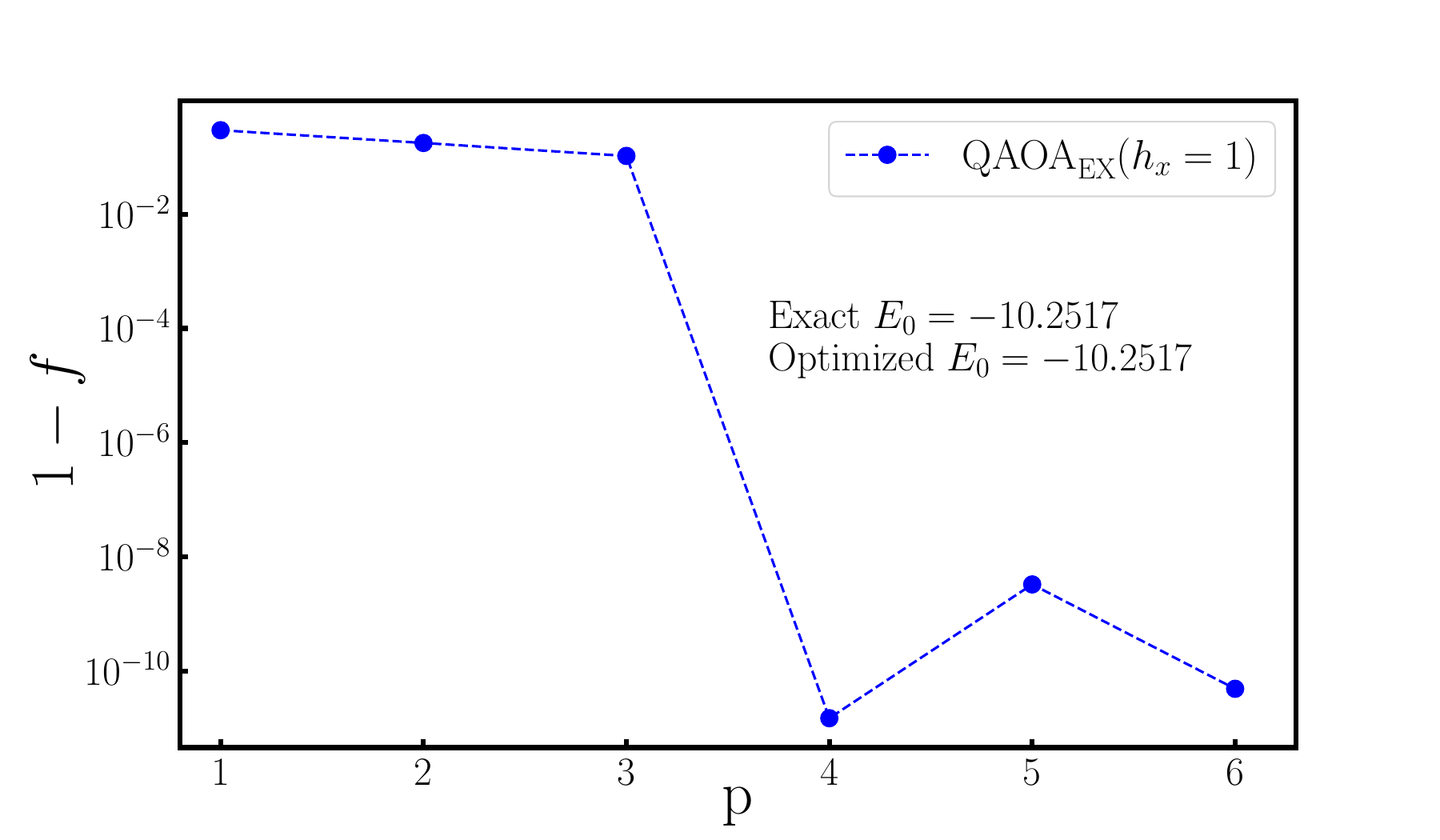}
\includegraphics[width=0.98\linewidth]{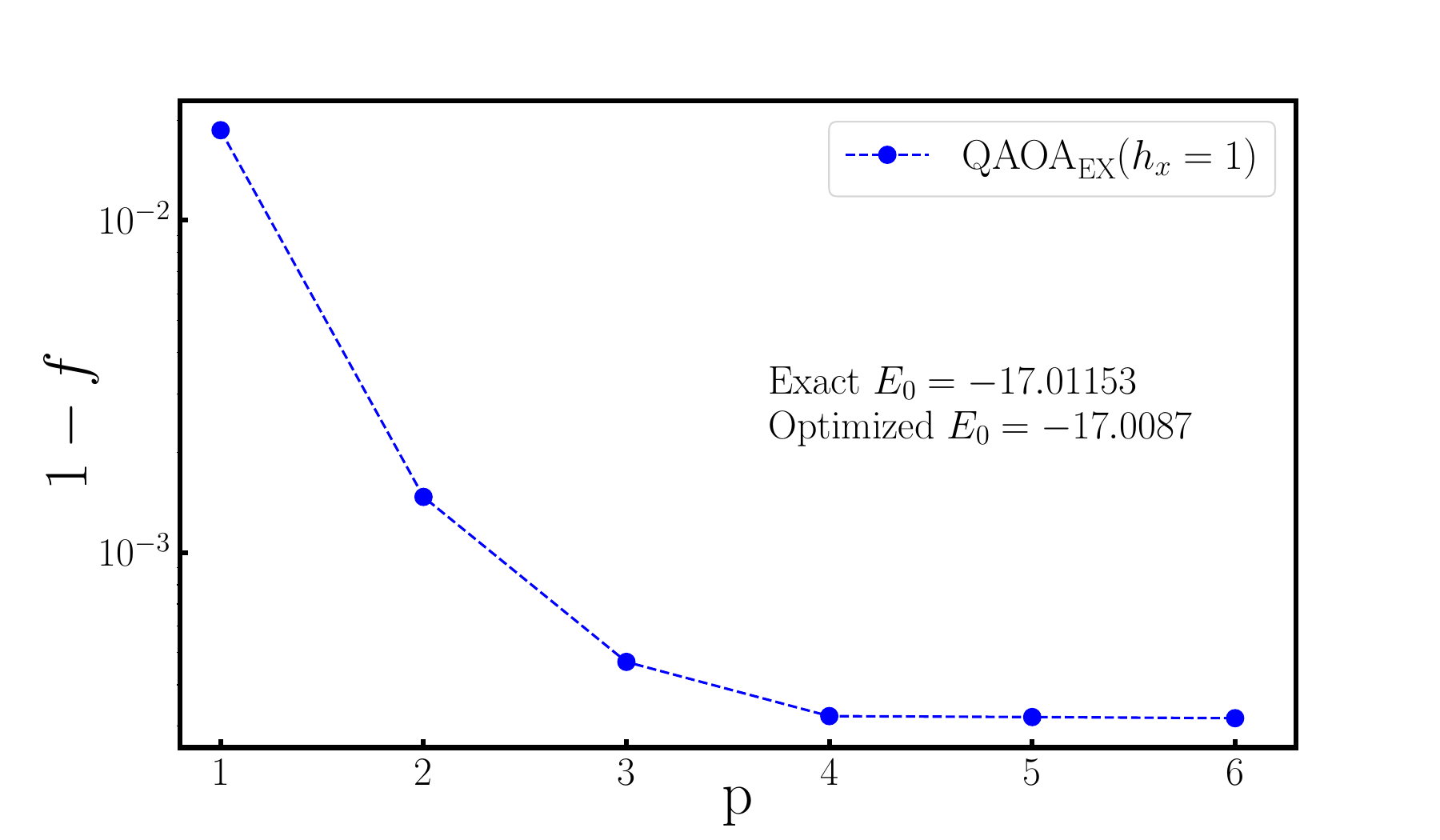}
\caption{\label{tfim-1d-2d}(Top): Plot of the in-fidelity of the ground state 
of TFIM at the critical point ($h_x=1)$ with circuit depth ($p$) on a 1D lattice 
with 8 sites. (Bottom): We plot the in-fidelity of the ground state on a $2 \times 4$ 
lattice. The plots demonstrate that for the same circuit depth, the 1D lattice 
performs much better than the 2D lattice.}
\end{figure}

 Due to the plaquette interaction, our $SO(3)$ model has the ingredients of a 2-d model.
 Thus, to have a fair comparison, we have also simulated the TFIM in 2-d using our proposed 
 methods. It becomes clear that the symmetry-resolved QAOA method does not work as well
 for the $2\times 4$ lattice compared to the comparable 1-d system. Specifically, we 
 compared the performance of the QAOA algorithm for the TFIM with periodic 
 boundary conditions at the critical point ($h_x=1$) on both a 1-d lattice with 8 sites and 
 a 2-d lattice of size $2\times4$. The number of qubits is the same (8) in both cases. 
 The performance is compared through the in-fidelity. As shown in \cref{tfim-1d-2d}, for 
 the ground state, the 1-d lattice achieves significantly better infidelity ($1-f$) at 
 lower circuit depths, while the $2\times4$ lattice requires much larger circuit depths 
 ($p$) to achieve a comparable in-fidelity. In fact, the same behaviour was also visible
 for the $SO(3)$ model in the context of the $2 \times 2$ and the $2 \times 4$ lattice. 
 Due to periodic boundary conditions, many terms cancel out in the Hamiltonian for the
 former lattice in contrast to the latter. The results in \cref{qaoa-res} clearly show
 the excellent convergence obtained for the former lattices, and the larger circuit
 depths required for a corresponding convergence on the $2 \times  4$ lattices. 

 The observed difference in convergence rate could perhaps be justified from general arguments
 about the entanglement structure for the ground states. Ground states of quantum systems
 interacting via a local Hamiltonian can exist in gapped, critical, or gapless phases. 
 The ground states of gapped phases are expected to have an area-law for the 
 entanglement entropy. This implies that the entanglement content of two-dimensional
 ground states is more than that of one dimensional counterparts, in particular if the 
 ground state of a gapless phase is under question. The observed results 
 indeed follow this general reasoning: to capture the ground state in two dimensions
 with the same precision as in one dimension, a larger circuit depth is necessary.
 
\section{Conclusion} \label{sec:conclusion}
 In this article, we have explored several theoretical and experimental
 aspects relevant for digital quantum simulation of non-Abelian lattice gauge theories
 in two spatial dimensions. Until now, most quantum simulations of gauge theories 
 have been restricted to just one spatial dimension, primarily due to the complexity 
 of managing large-dimensional Hilbert spaces associated with gauge fields and the 
 challenges of enforcing Gauss's law constraints.
 
 To get around these challenges in two spatial dimension, we consider a specific 
 quantum link model with $SO(3)$ gauge invariance in the absence of matter fields, 
 and, explicitly demonstrate the utility of defining and working
 in the gauge-invariant basis. Given that this model exhibits notable nuclear 
 physics phenomenology (such as binding, chiral symmetry breaking, and its 
 restoration at finite density—that are also observed in nature), it is worthwhile 
 to investigate the potential of quantum computing for studying this particular model.
 
 While working in the electric flux basis can be intuitive for understanding, 
 and exact gauge-invariant results can be obtained using classical computations, 
 implementing the problem on a quantum hardware does not always ensure that the 
 (non-Abelian) gauge invariance is exactly maintained. We demonstrated this using 
 the simplest possible plaquette with only two links (a bubble plaquette) on the 
 trapped-ion IonQ quantum computers. We observe that when employing variational 
 quantum algorithms like the VQE, the hardware generates an excess of states in 
 the electric flux beyond what is required to accurately reproduce the gauge-invariant 
 ground state. This is reflected in the ground state energy, which shows a deviation 
 of approximately $\sim 10 \%$ from the exact energy. However, by using a gauge-invariant 
 ansatz for the ground state, we have reproduced the wave function with a fidelity $\sim 1$. 

 Having justified the requirement of the gauge-invariant basis, we then formulated
 the variational ans\"atze directly in the gauge-invariant basis, and used different
 quantum algorithms such as the linearly connected VQE and the QAOA to explore the 
 ground state and the first excited state of this model. The key objective was to 
 determine the feasibility of using quantum algorithms in order to explore the 
 scenario of SSB of a discrete symmetry, when the finite volume mass gap closes 
 exponentially with increasing volume, which we show to be a challenging problem 
 even in the gauge-invariant basis. Our studies explore two different strategies 
 to combat this challenge: 
 the first being the VQE for the ground state, and then the VQD with converged ground state
 to extract the excited state, and the second is the QAOA with symmetry-resolved
 initial states. Our results resolve energies within $0.1\%$ using ans\"{a}tze with 
 layers that scale at most linearly with the system's volume. Further, 
 in order to have a fair comparison of the difficulty associated with simulating SSB 
 in non-Abelian gauge theories, we also simulate the paradigmatic TFIM which has a 
 $\mathbb{Z}_2$ SSB for 
 a range of parameters. In both the models, we demonstrate the necessity of using
 a symmetry-resolved variational ansatz in order to capture the ground state and
 first excited state with different symmetries.

 Another significant issue highlighted in our paper is the impact of higher than 
 one spatial dimensionality. While various variational quantum algorithms have 
 demonstrated remarkable success for quantum systems in one spatial dimension, 
 their performance degrades notably when applied to genuinely two-dimensional 
 systems. Although energy estimates remain accurate within a few percent, the 
 fidelities of the resulting ground state wavefunctions are orders of magnitude 
 lower. This phenomenon persists even for well-studied models like the paradigmatic 
 TFIM. Therefore, the development of more effective quantum algorithms for systems 
 in higher spatial dimensions remains an open and pressing challenge.

 Our results pave the way for various new investigations: the most immediate 
 is to include fermions in the problem, and explore how the quantum algorithms
 fare in the presence of fermions. This presents a significant challenge, as it 
 requires the development of efficient fermion-to-qubit encoding schemes that 
 maintain a high degree of locality, particularly in two-dimensional systems. 
 The phenomenology of the $SO(3)$ model with fermions
 is expected to be richer in two spatial dimensions as compared to the 
 previously studied one dimensional model because the magnetic field term can
 play a non-trivial role and generate more phases. We are currently investigating 
 these aspects in detail. In terms of quantum computing, the circuits developed
 here with considerable theoretical insights need to be implemented on actual 
 quantum hardware for the larger systems in order to understand their scaling.
 It is possible to place external charges in the pure gauge theory, and study
 the string breaking for an non-Abelian gauge theory on a quantum hardware. \\

{\bf Acknowledgments:} We would like to thank Aditya Banerjee, Shailesh Chandrasekharan, 
Arti Garg, Graham van Goffrier, Arnab Kundu, Marina Marinkovic, Nilmani Mathur, Indrakshi Roychowdhury, Arnab Sen, and Uwe-Jens Wiese for various illuminating conversations. Research of BC at the University of Southampton has been supported by the following research fellowship and grants - Leverhulme Trust (ECF-2019-223 G100820), STFC (Grant no. ST/X000583/1), STFC (Grant no. ST/W006251/1), and EPSRC (Grant no. EP/W032635/1). Research of EH at Perimeter Institute is supported in part by the Government of Canada through the Department of Innovation, Science and Industry Canada and by the Province of Ontario through the Ministry of Colleges and Universities. We acknowledge access to Ion Q through the AWS Cloud Credit for Research program and thank Sebastian Hassinger and Sebastian Stern for help with experiments.

\bibliography{apssamp}% Produces the bibliography via BibTeX.
\appendix
\vspace{20.0cm}
\section{Gauge transformation of the quantum link operators} \label{sec:gtrafo}
The commutation relations for ${L^a}_{x,y}$, ${R^a}_{x,y}$ and $O^{ab}_{x,y}$ are
\begin{equation}
\begin{split}
& [L^{a}_{x,y}, O^{bc}_{x,y}] = - t^{a}_{bd} O^{dc}_{x,y}\\
& [R^{a}_{x,y}, O^{bc}_{x,y}] =   O^{bd}_{x,y} t^{a}_{dc}
\end{split}
\end{equation}

The link operator $O^{ab}_{x,y}$ transforms under gauge transformation as 
\begin{align}
O^{ab}_{x,y} &\longrightarrow \exp{(-i(\alpha^{m}_x L^{m}_{x,y} + \alpha^{m}_y R^{m}_{x,y}))} O^{ab}_{x,y} \notag \\
&\qquad \exp{(i(\alpha^{m}_x L^{m}_{x,y} + \alpha^{m}_y R^{m}_{x,y}))} \notag \\
&= (1 - i\alpha^{m}_x L^{m}_{x,y} - i \alpha^{m}_y R^{m}_{x,y} - \mathcal{O}(\alpha^2)) O^{ab}_{x,y} \notag \\
&\qquad  (1 + i\alpha^{m}_x L^{m}_{x,y} + i \alpha^{m}_y R^{m}_{x,y} + \mathcal{O}(\alpha^2)) \notag \\
&= O^{ab}_{x,y} + i\alpha^{m}_x O^{ab}_{x,y} L^{m}_{x,y} + i\alpha^{m}_y O^{ab}_{x,y} R^{m}_{x,y} \notag \\
&\qquad - i\alpha^{m}_x L^{m}_{x,y} O^{ab}_{x,y} - i\alpha^{m}_y R^{m}_{x,y} O^{ab}_{x,y} + \mathcal{O}(\alpha^2) \notag \\
&= O^{ab}_{x,y} - i\alpha^{m}_x [L^{m}_{x,y}, O^{ab}_{x,y}] - i\alpha^{m}_y [R^{m}_{x,y}, O^{ab}_{x,y}] + \mathcal{O}(\alpha^2) \notag \\
&= O^{ab}_{x,y} + i\alpha^{m}_x {t^m}_{ac} O^{cb}_{x,y} - i\alpha^{m}_y  O^{ac}_{x,y} {t^m}_{cb} + \mathcal{O}(\alpha^2) \notag \\
&= (1 + i\alpha^{m}_x t^{m}_{ac} + \mathcal{O}(\alpha^2)) O^{cd}_{x,y} (1 - i\alpha^{m}_y t^{m}_{db} - \mathcal{O}(\alpha^2)) \notag \\
&= \big[\exp{(i\alpha^{m}_x {t^m})}\big]^{ac} O^{cd}_{x,y} \big[\exp{(-i\alpha^{m}_y {t^m})}\big]^{db}
\end{align}

\section{Expression of the Hamiltonian in the gauge-invariant basis} \label{sec:GIHam}
In this section, we show the relevant algebra to deduce the form of the Hamiltonian
in the gauge invariant basis. We refer the reader to \cref{fig:plaq-GIS} top right,
which shows four spins, labelled as $\sigma^b_{x,+\nu}$, $\sigma^a_{x,+\mu}$, $\sigma^b_{x,-\nu}$,
and $\sigma^a_{x,-\nu}$ on the four links starting from the twelve o'clock position
as we move in a clockwise fashion. The two gauge invariant states which can be defined
on the site $x$ are
\begin{align}  
 \ket{\psi_{1s}} &= \frac{1}{2}\left[ \ket{\uparrow \downarrow \uparrow \downarrow}
  - \ket{\uparrow \downarrow \downarrow \uparrow} - \ket{\downarrow \uparrow \uparrow \downarrow} 
  + \ket{\downarrow \uparrow \downarrow \uparrow} \right], \nonumber \\ 
 \ket{\psi_{2s}} &=  \frac{1}{2 \sqrt{3}}\left[-2 \ket{\uparrow \uparrow \downarrow \downarrow}
 -2 \ket{\downarrow \downarrow \uparrow \uparrow} \right. \\
 &+ \left. \ket{\uparrow\downarrow\uparrow\downarrow} 
 + \ket{\downarrow \uparrow \uparrow \downarrow} + \ket{\uparrow \downarrow \downarrow \uparrow}
 + \ket{\downarrow \uparrow \downarrow \uparrow} \right] \nonumber
\end{align}

 Next, note that there are four plaquettes which will touch the Gauss Law vertex in the four corners, 
and each corner will use a different pair of spins. The plaquette touching the top right uses the pair 
$\vec{\sigma}_{x,+\nu}/2 \cdot \vec{\sigma}_{x,+\mu}/2$, while the plaquette touching the bottom left
will use the pair $\vec{\sigma}_{x,-\mu}/2 \cdot \vec{\sigma}_{x,-\nu}/2$. The factors of 2 come from
the identification $S = \sigma/2$. To obtain the matrix representation
of these operators in the gauge invariant basis space, we need to evaluate the operators using the
generic expansion $\vec{S}_1 \cdot \vec{S}_2 = \frac{1}{2} (S^+_1 \cdot S^-_2 + S^-_1 \cdot S^+_2) 
+ S^z_1 \cdot S^z_2$ in the Hilbert space spanned by $\ket{\psi_{1s}}$ and $\ket{\psi_{2s}}$. 
For our four operators, we obtain:
\begin{align}
    \frac{\vec{\sigma}_{x,+\nu}}{2} \cdot \frac{\vec{\sigma}_{x,+\mu}}{2} &=
    \begin{pmatrix}
        0 & \frac{\sqrt{3}}{4} \\
        \frac{\sqrt{3}}{4} & -\frac{1}{2}
    \end{pmatrix} = \frac{1}{4}(\sqrt{3} \tau^1 - I + \tau^3),\\
    \frac{\vec{\sigma}_{x,-\mu}}{2} \cdot \frac{\vec{\sigma}_{x,+\nu}}{2} &=
    \begin{pmatrix}
        0 & -\frac{\sqrt{3}}{4} \\
        -\frac{\sqrt{3}}{4} & -\frac{1}{2}
    \end{pmatrix} \\&= \frac{1}{4}(-\sqrt{3} \tau^1 - I + \tau^3),\\
    \frac{\vec{\sigma}_{x,+\nu}}{2} \cdot \frac{\vec{\sigma}_{x,+\mu}}{2}
    &=\frac{\vec{\sigma}_{x,-\mu}}{2} \cdot \frac{\vec{\sigma}_{x,-\nu}}{2},\\
    \frac{\vec{\sigma}_{x,-\mu}}{2} \cdot \frac{\vec{\sigma}_{x,+\nu}}{2}
    &=\frac{\vec{\sigma}_{x,+\mu}}{2} \cdot \frac{\vec{\sigma}_{x,-\nu}}{2}.
\end{align}
Here $\tau^i$ operates in the $2 \times 2$ gauge invariant space and $I$ is
the $2 \times 2$ identity matrix. The plaquette term (by symmetry) then picks up
these two types of corner operators, and has the form as given in \cref{eq16}
in the main text.

\section{Hamiltonian decomposition for $2\times4$ lattice} \label{sec:trot}
The Hamiltonian for the $2\times4$ lattice was decomposed into twenty-four pieces for 
implementation with QAOA (each of these terms operate in the gauge-invariant subspace
represented by the $\tau^i$ operators):
\begin{equation*}
\begin{aligned}
H_1 &= \frac{1}{2\cdot4^3}\sum_{x=1}^8 \tau^3_x,\\
H_2 &=  -\frac{1}{2\cdot4^3}\sum_{x\neq y} \tau^3_x \tau^3_y
    + \frac{1}{2\cdot4^3}\sum_{x\neq y\neq z} \tau^3_x\tau^3_y\tau^3_z \\
   &-\frac{1}{2\cdot4^3}(\tau^3_1 \tau^3_2 (\tau^3_3\tau^3_4 + \tau^3_7\tau^3_8) 
    + \tau^3_5 \tau^3_6 (\tau^3_3\tau^3_4 + \tau^3_7\tau^3_8)), \\
H_3 &= -\frac{3}{2\cdot 4^3} \tau^1_3 \tau^1_4 (-\tau^3_1\tau^3_2 
     - \tau^3_5\tau^3_6 + \tau^3_1 + \tau^3_2 + \tau^3_5 + \tau^3_6 - 2I),\\
H_4 &= -\frac{3}{2\cdot 4^3} \tau^1_2 \tau^1_4 (-\tau^3_1\tau^3_3 + \tau^3_1 + \tau^3_3 - I), \\
H_5 &= -\frac{3}{2\cdot 4^3} \tau^1_1 \tau^1_4 (\tau^3_2\tau^3_3 - \tau^3_2 - \tau^3_3 + I), \\
H_6 &= -\frac{3}{2\cdot 4^3} \tau^1_2 \tau^1_3 (\tau^3_1\tau^3_4 - \tau^3_1 - \tau^3_4 + I), \\
H_7 &= -\frac{3}{2\cdot 4^3} \tau^1_1 \tau^1_3 (-\tau^3_2\tau^3_4 + \tau^3_2 + \tau^3_4 - I), \\
H_8 &= -\frac{3}{2\cdot 4^3} \tau^1_1 \tau^1_2 (-\tau^3_3\tau^3_4 
     -\tau^3_7\tau^3_8 + \tau^3_3 + \tau^3_4 + \tau^3_7 + \tau^3_8 - 2I),\\
H_9 &= -\frac{3}{2\cdot 4^3} \tau^1_5 \tau^1_6 (-\tau^3_3\tau^3_4 
     -\tau^3_7\tau^3_8 + \tau^3_3 + \tau^3_4 + \tau^3_7 + \tau^3_8 - 2I),\\
H_{10} &= -\frac{3}{2\cdot 4^3} \tau^1_4 \tau^1_6 (-\tau^3_3\tau^3_5 + \tau^3_3 + \tau^3_5 - I), \\
H_{11} &= -\frac{3}{2\cdot 4^3} \tau^1_3 \tau^1_6 (\tau^3_4\tau^3_5 - \tau^3_4 - \tau^3_5 + I),
\end{aligned}
\end{equation*}

\begin{equation*}
\begin{aligned}H_{12} &= -\frac{3}{2\cdot 4^3} \tau^1_3 \tau^1_5 (-\tau^3_4\tau^3_6 + \tau^3_4 + \tau^3_6 - I), \\
H_{13} &= -\frac{3}{2\cdot 4^3} \tau^1_4 \tau^1_5 (\tau^3_3\tau^3_6 - \tau^3_3 - \tau^3_6 + I), \\
H_{14} &= -\frac{3}{2\cdot 4^3} \tau^1_5 \tau^1_7 (-\tau^3_6\tau^3_8 + \tau^3_6 + \tau^3_8 - I), \\
H_{15} &= -\frac{3}{2\cdot 4^3} \tau^1_6 \tau^1_7 (\tau^3_5\tau^3_8 - \tau^3_5 - \tau^3_8 + I), \\
H_{16} &= -\frac{3}{2\cdot 4^3} \tau^1_5 \tau^1_8 (\tau^3_6\tau^3_7 - \tau^3_6 - \tau^3_7 + I), \\
H_{17} &= -\frac{3}{2\cdot 4^3} \tau^1_7 \tau^1_8 (-\tau^3_5\tau^3_6 
        - \tau^3_1\tau^3_2 + \tau^3_1 + \tau^3_2  + \tau^3_5 + \tau^3_6 - 2I),\\
H_{18} &= -\frac{3}{2\cdot 4^3} \tau^1_2 \tau^1_8 (-\tau^3_1\tau^3_7 + \tau^3_1 + \tau^3_7 - I), \\
H_{19} &= -\frac{3}{2\cdot 4^3} \tau^1_1 \tau^1_8 (\tau^3_2\tau^3_7 - \tau^3_2 - \tau^3_7 + I), \\
H_{20} &= -\frac{3}{2\cdot 4^3} \tau^1_2 \tau^1_7 (\tau^3_1\tau^3_8 - \tau^3_1 - \tau^3_8 + I), \\
H_{21} &= -\frac{3}{2\cdot 4^3} \tau^1_1 \tau^1_7 (-\tau^3_2\tau^3_8 + \tau^3_2 + \tau^3_8 - I), \\
H_{23} &= -\frac{3}{2\cdot 4^3} \tau^1_6 \tau^1_8 (-\tau^3_5\tau^3_7 + \tau^3_5 + \tau^3_7 - I), \\
H_{24} &= -\frac{3^2}{2\cdot 4^3}( \tau^1_1 \tau^1_2 (\tau^1_3\tau^1_4 + \tau^1_7\tau^1_8) 
        + \tau^1_5 \tau^1_6 (\tau^1_3\tau^1_4 + \tau^1_7\tau^1_8)). 
\end{aligned}
\end{equation*}

\section{Energy convergence for the $SO(3)$ model on a $2 \times 6$ lattice} \label{sec:convg2t6}
\begin{figure}
   \centering
   \includegraphics[scale=0.25, trim=0 0 0 0.1]{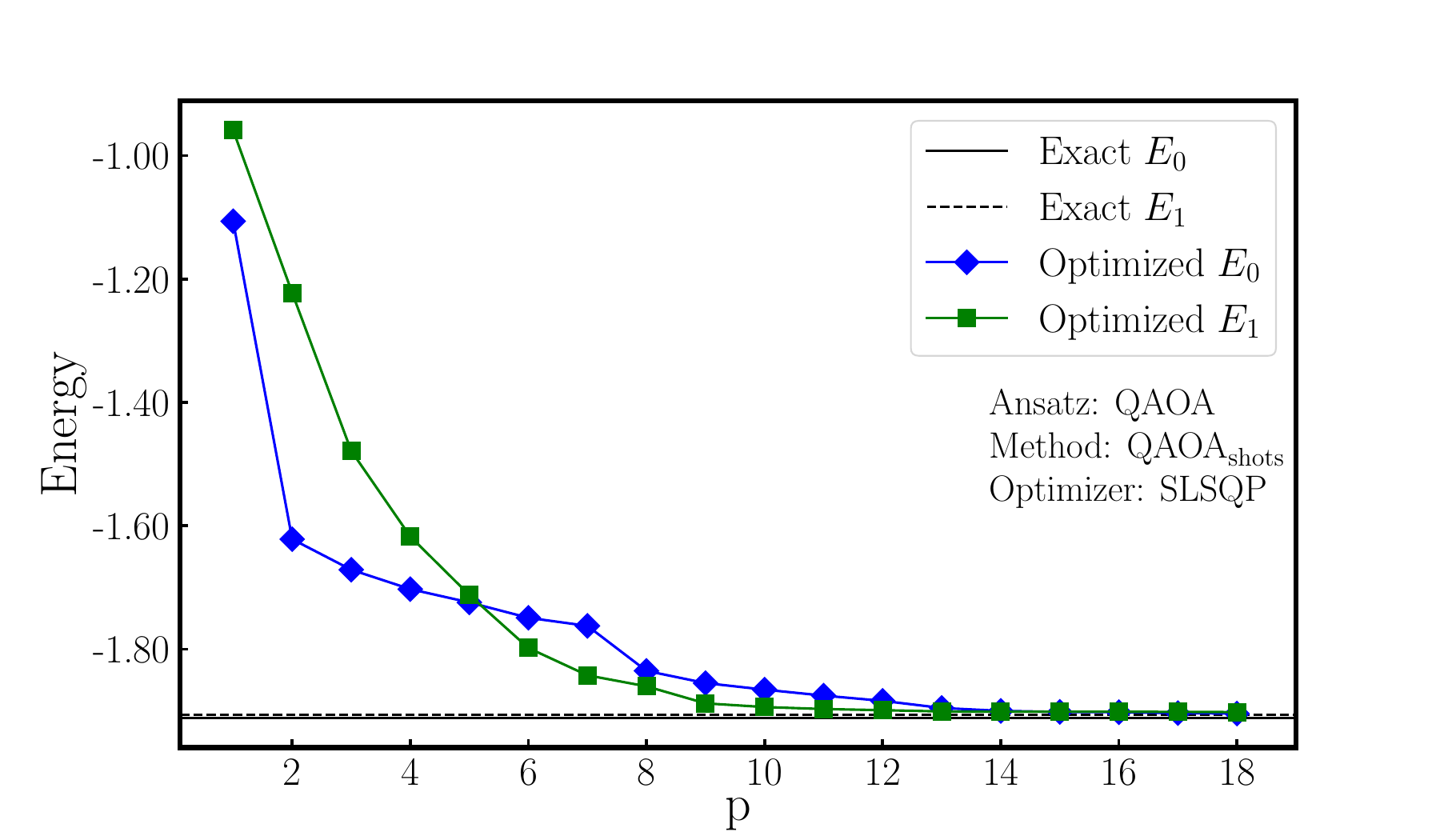}
   \includegraphics[scale=0.25, trim=0 0 0 3]{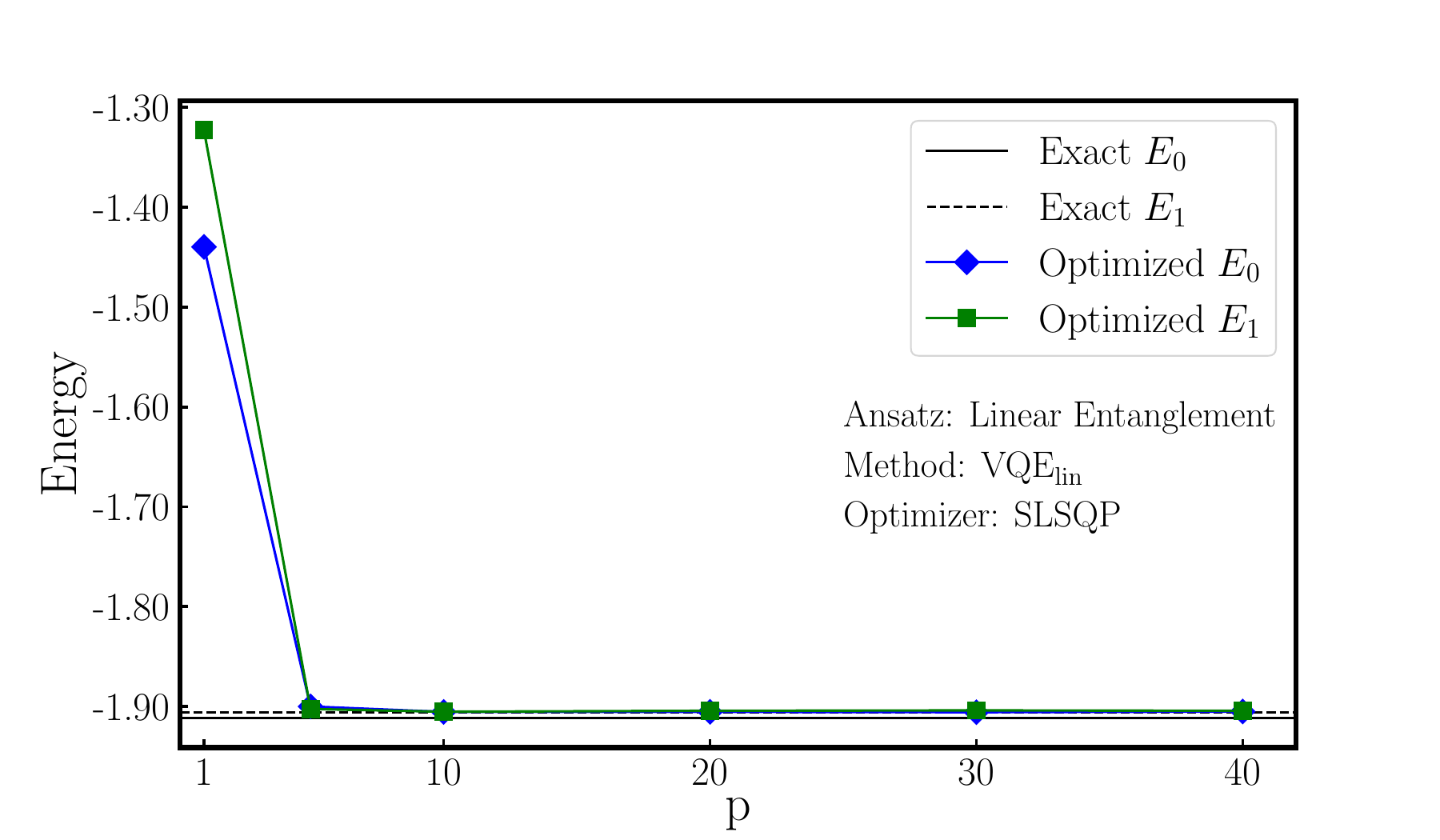}
   \caption{\label{so3-latt2x6}(Top): The optimized ground state and first excited state 
   energies are shown as a function of circuit depth ($p$) on a $2\times6$ lattice using 
   the VQE algorithm ($\mathrm{QAOA}_\mathrm{shots}$). (Bottom): The energy convergence 
   results are presented for the VQD method ($\mathrm{VQE}_\mathrm{lin}$).}
\end{figure}

In this section, we show the energy convergence for both the ground state and 
the first excited state energy as a function of circuit depth ($p$) for the $SO(3)$ 
model on a $2\times6$ lattice, using VQE and VQD methods in \cref{so3-latt2x6}.

\section{The results for 1-d TFIM} \label{sec:tfim}
\begin{figure}[H]
\includegraphics[scale=0.27, trim =0 0 0 3.35]{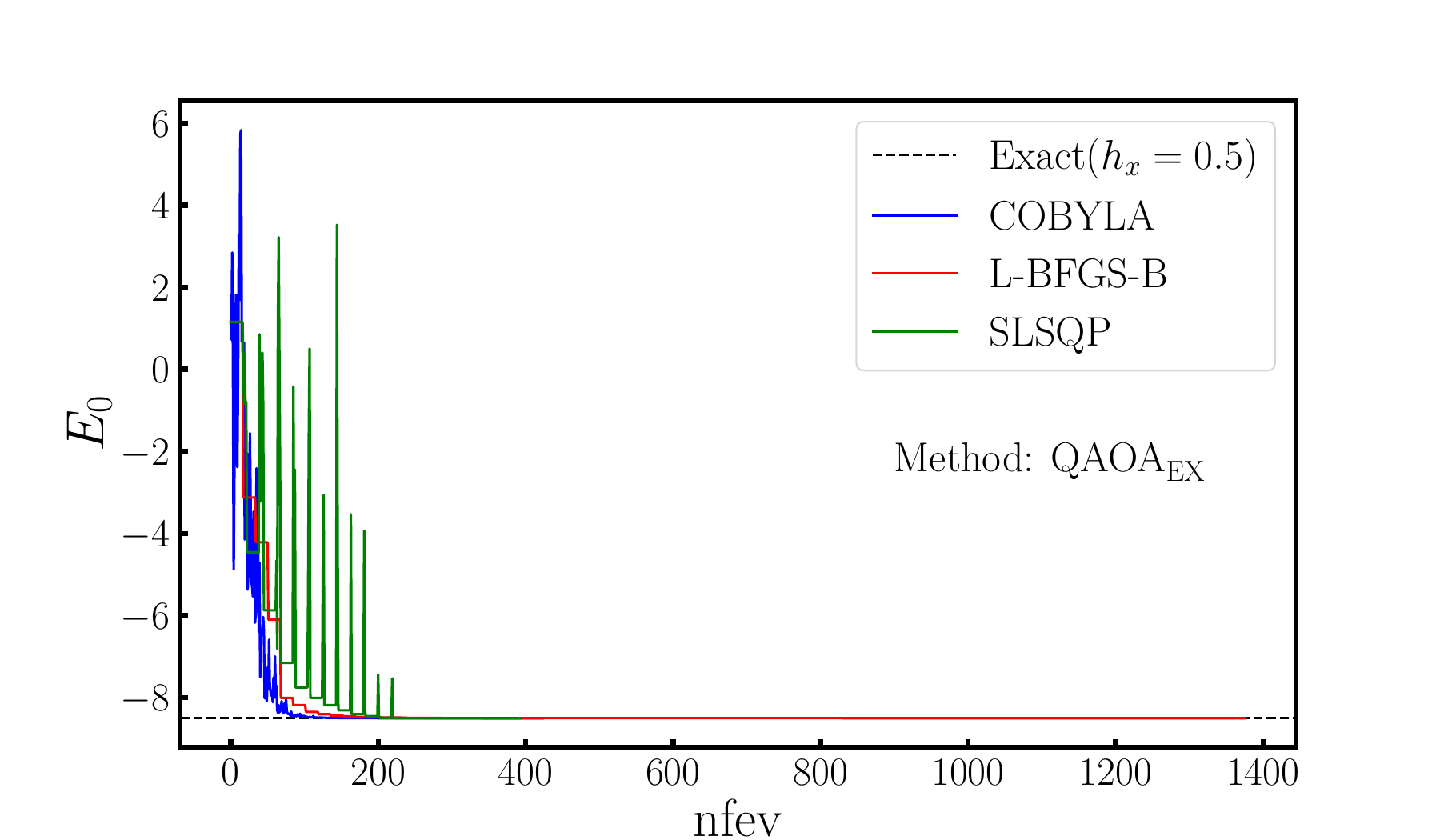}
\caption{\label{tfim-latt8-comp} Plot of the optimized ground state energy using the QAOA method ($\mathrm{QAOA}_{\mathrm{EX}}$) against the number of function evaluations for a lattice with 8 sites at $h_x=0.5$, comparing three different optimizers. The optimizer L-BFGS-B took more iterations to converge than the others but achieved a very low in-fidelity of about $10^{-10}$. In contrast, the COBYLA reached an in-fidelity of around $10^{-6}$, and the SLSQP optimizer had an in-fidelity of approximately $10^{-4}$.}
\end{figure}

\begin{figure*}[t]
\centering
\includegraphics[width=0.325\linewidth]{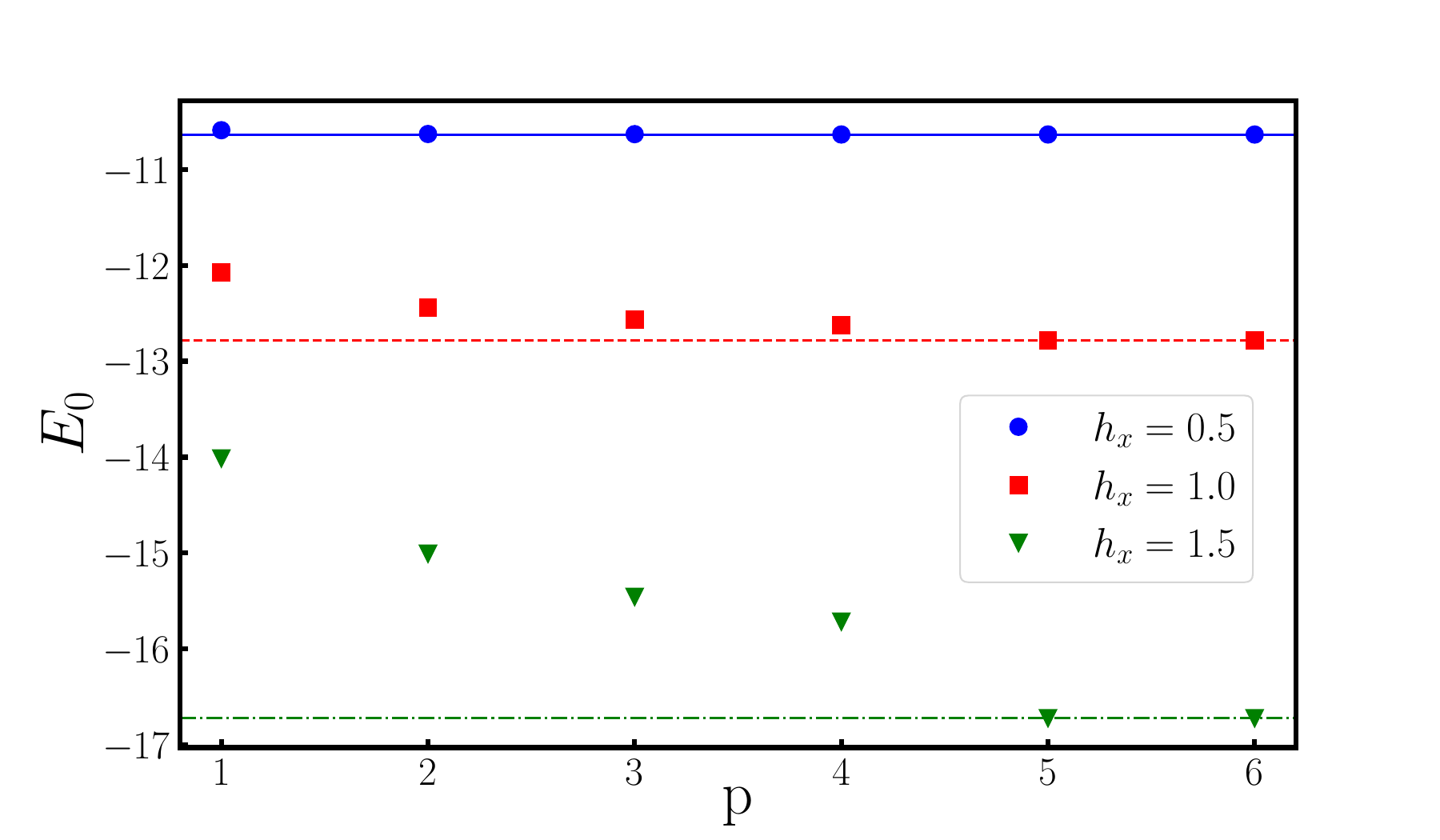}
\includegraphics[width=0.325\linewidth]{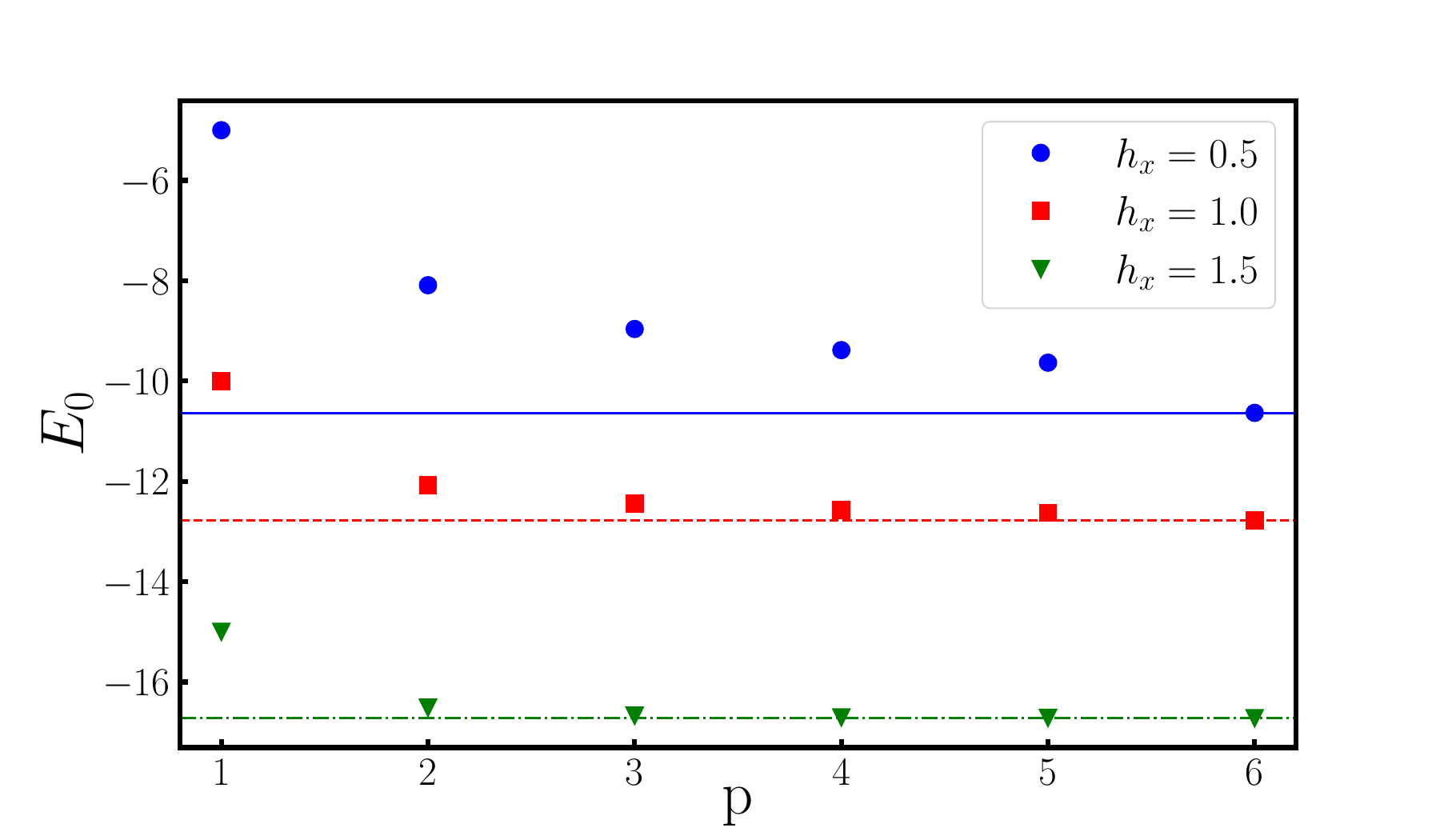}
\includegraphics[width=0.325\linewidth]{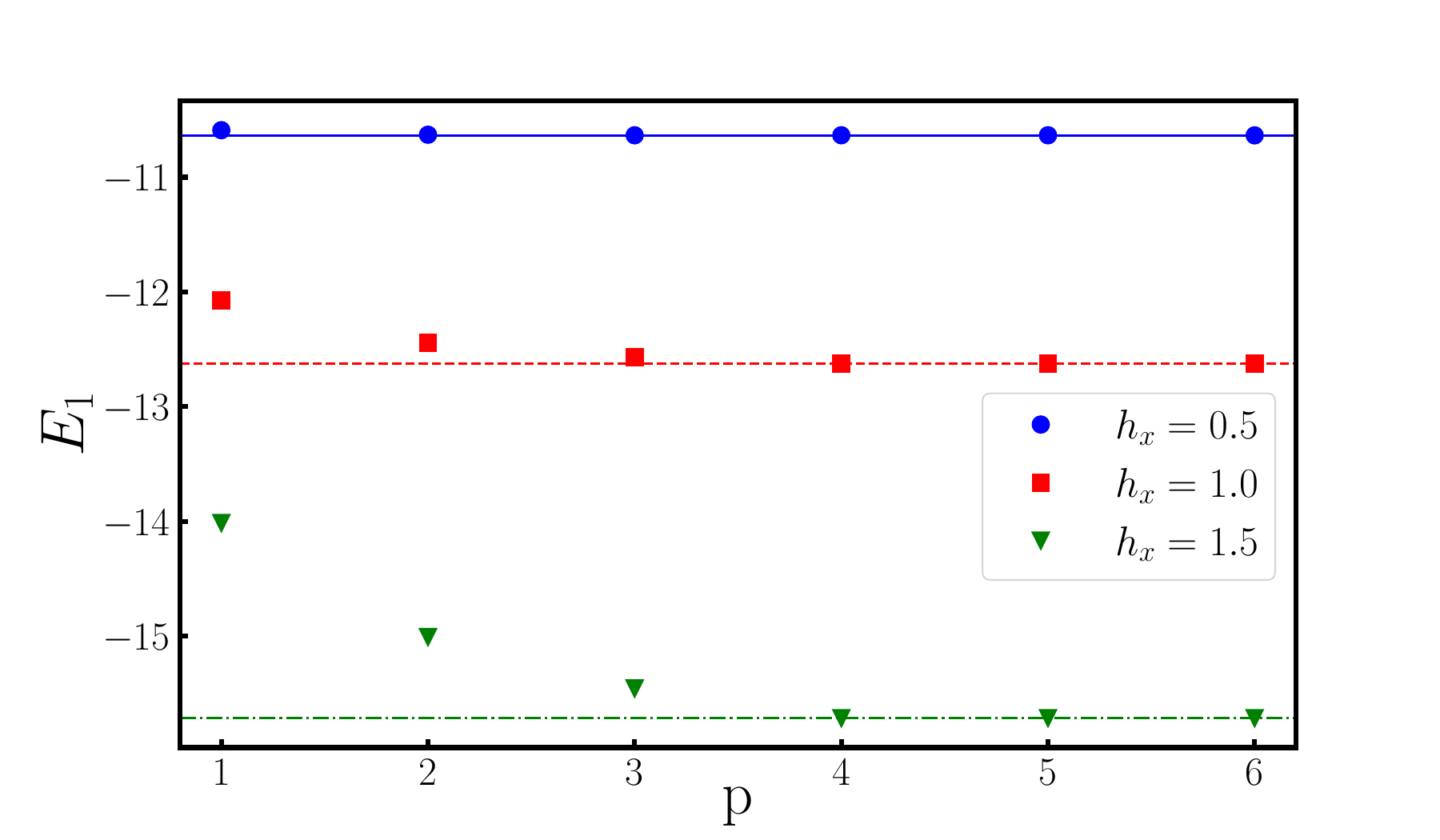}
\includegraphics[width=0.325\linewidth]{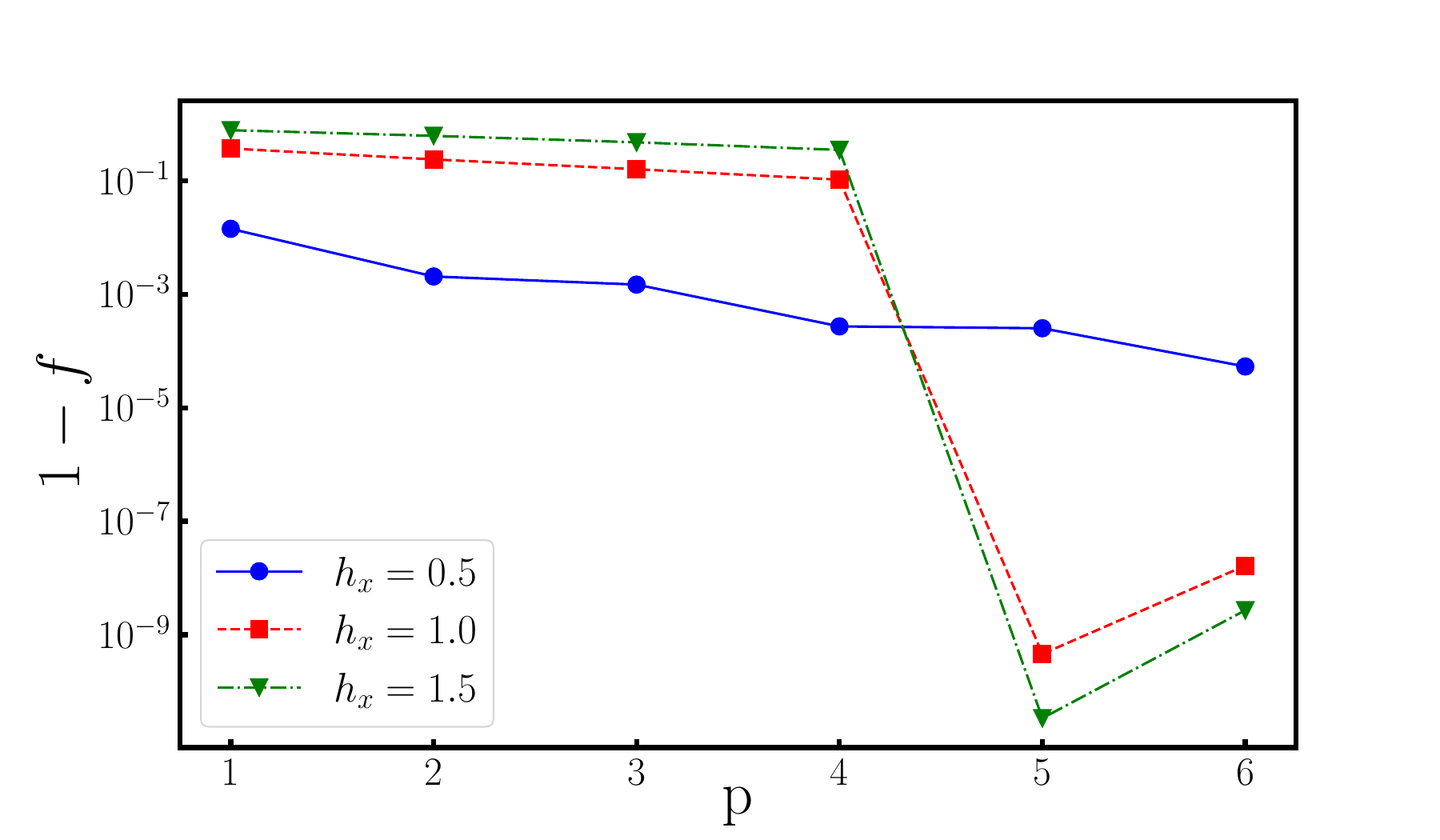}
\includegraphics[width=0.325\linewidth]{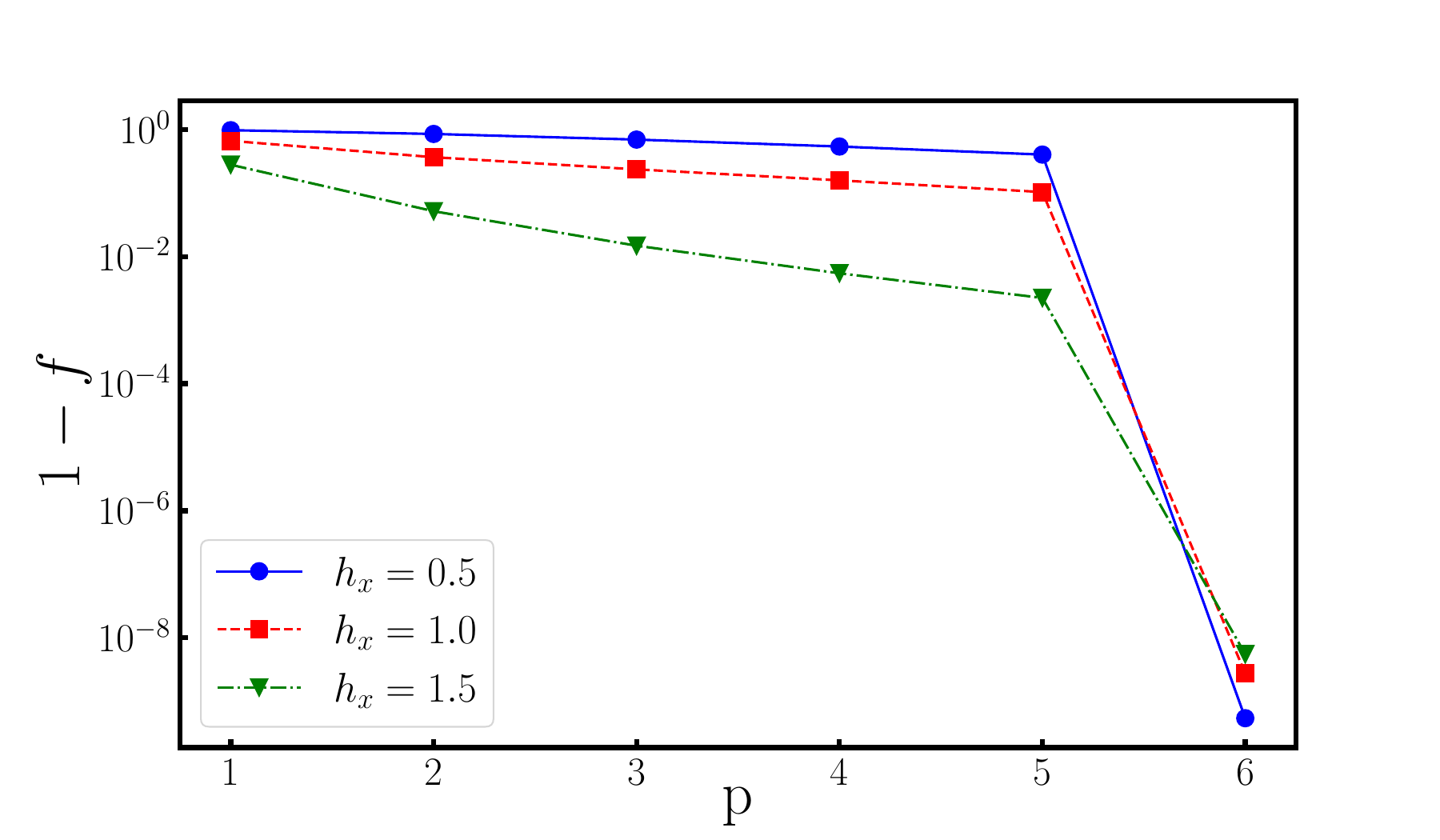}
\includegraphics[width=0.325\linewidth]{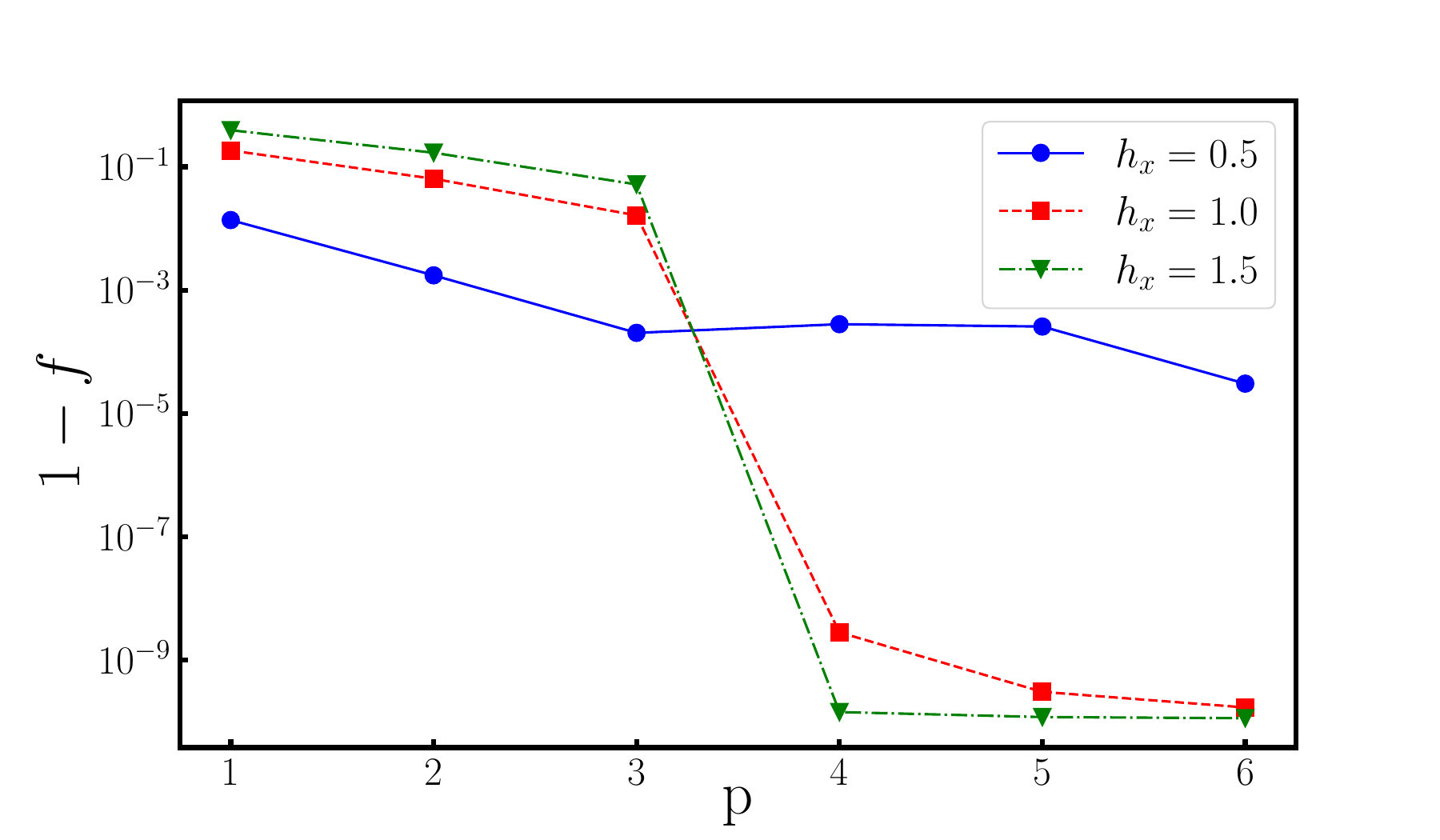}
\caption{\label{tfim-latt10} (First row): We display both the ground state and the 
first excited state energy as a function of circuit depth ($p$) on a 1-d lattice 
with 10 sites. In the first row, the leftmost figure shows the results for the
ground state with the GHZ state as an initial state, the middle figure also 
shows the ground state results when one starts with an eigenstate of the 
$H_2$ term as the initial state, while the rightmost figure presents the performance 
for the first excited state. The second row shows the corresponding in-fidelity 
for each of the cases in the first row.}
\end{figure*}

This section collects the results of the ground and the first excited state energy 
for the 1-d TFIM in three different regimes for various lattice sizes using the QAOA 
method. To see the performance of QAOA we plot both energies and in-fidelity with the 
circuit depth ($p$) for a 1-d lattice with 10 sites in \cref{tfim-latt10}. In addition,
we show an example of how the minimization proceeds for a given circuit depth and an 
initial state in \cref{tfim-latt8-comp} for several different classical optimizers 
(for $h_x = 0.5$ and $L=8$) and compare among them.

%\begin{widetext}
%\end{widetext}
\end{document}